\documentclass[preprint,showpacs,preprintnumbers,bibnotes,amsmath,amssymb,aps]{revtex4-1}
\usepackage{dcolumn}
\usepackage{bm}
\usepackage{amssymb}
\usepackage{graphicx}

\def\araa{ARA\&A}
\def\apj{ApJ}
\def\apjl{ApJL}
\def\apjs{ApJS}

\def\apss{Ap\&SS}
\def\aap{A\&A}

\def\mnras{MNRAS}

\def\prd{Phys.\ Rev.\ D}

\def\pasj{PASJ}

\def\nat{Nature}

\begin{document}

    
\title{Coherent Network Analysis of Gravitational Waves from Three-Dimensional
 Core-Collapse Supernova Models}
\author{Kazuhiro Hayama$^{1,2}$, Takami Kuroda$^{3}$,  Kei Kotake$^{4,5}$, and 
 Tomoya Takiwaki$^{6}$}
\affiliation{$^1$Department of Physics, Osaka City University, Sugimoto 3-3-138, Sumiyoshi, Osaka, 558-8585, Japan }
\affiliation{$^5$Gravitational Wave Project Office, National Astronomical Observatory of Japan, 
2-21-1, Osawa, Mitaka, Tokyo, 181-8588, Japan}
\affiliation{$^3$Department of Physics, University of Basel,
Klingelbergstrasse 82, 4056 Basel, Switzerland}
\affiliation{$^4$Department of Applied Physics, Fukuoka University,
 8-19-1, Jonan, Nanakuma, Fukuoka, 814-0180, Japan}
\affiliation{$^5$Center for Computational Astrophysics, National Astronomical Observatory of Japan, 
2-21-1, Osawa, Mitaka, Tokyo, 181-8588, Japan}
\affiliation{$^6$Astrophysical Big Bang Laboratory, RIKEN, Saitama, 
351-0198, Japan}

\begin{abstract}
Using predictions from three-dimensional (3D) hydrodynamics 
simulations of core-collapse supernovae (CCSNe), we present a coherent 
network analysis to detection, reconstruction, and the source localization 
of the gravitational-wave (GW) signals. 
We use the {\tt RIDGE} pipeline for the analysis, in which the network 
 of LIGO Hanford, LIGO Livingston, VIRGO, and KAGRA is considered.
By combining with a GW spectrogram analysis, we show that several important 
hydrodynamics features in the original waveforms persist in the 
waveforms of the reconstructed signals. The characteristic excess in
 the spectrograms originates not only from rotating core-collapse, bounce and 
the subsequent ring down of the proto-neutron star (PNS) as previously 
identified, but also from the formation of magnetohydrodynamics jets and 
non-axisymmetric instabilities in the vicinity of the PNS. 
 Regarding the GW signals emitted near at the rotating core bounce, the 
horizon distance extends up to $\sim$ 18 
kpc for the most rapidly rotating 3D model in this work.
Following the rotating core bounce,
the dominant source of the GW emission shifts to the
 non-axisymmetric instabilities.
The horizon distances extend maximally up 
to $\sim$ 40 kpc seen from the spin axis.
 With an increasing number of 3D models trending towards explosion 
recently, 
 our results suggest that in addition to the best studied GW signals due to 
rotating core-collapse and bounce, the time is ripe to consider
 how we can do science from GWs of CCSNe much more seriously than before.
 Particularly the quasi-periodic signals due to the 
non-axisymmetric instabilities and the detectability 
should deserve further investigation to elucidate the inner-working of the 
rapidly rotating CCSNe. 
\end{abstract}

\maketitle

\section{Introduction}
\label{introduction}

Significant progress has been made in the development of an international
 network of gravitational wave (GW) detectors.
 Although the first 
 detection has not been accomplished yet, the non-detection has
 already yielded scientific results setting upper bounds to a rich variety of 
 astrophysical GW sources (e.g., \cite{abbot09,abadie10,abbot10,abadie12,asai13,asai14}).
 The second generation detectors such as Advanced LIGO \cite{harry},
 Advanced VIRGO \cite{advv}, and KAGRA \cite{aso13,somiya}, will be on line in the 
coming years. The possibility to construct the third generation detectors is 
also recently being proposed \cite{punturo,fair}.
 At such a high level of precision, these advanced detectors are
 enough sensitive to many compact objects, including binary neutron star
(black hole) systems (e.g., \cite{schutz,faber,duez}),
 neutron star normal mode oscillations (e.g., \cite{nils1}), 
rotating neutron star mountains (e.g., \cite{horowitz}), and core-collapse 
supernova (CCSN) explosions (e.g., \cite{kota06,ott_rev,fryer11} for recent reviews), on the final of which 
we focus in this work.
 
According to the Einstein's theory of general relativity (e.g., \cite{shap83}), no GWs can be emitted if gravitational collapse of the 
stellar core proceeds perfectly spherically symmetric. 
To produce GWs, the gravitational collapse should 
proceed aspherically and dynamically.
Gathered over the last decades, observational evidence 
from electromagnetic-wave observations, e.g., of
 ejecta morphologies, spatial distributions of
 nucleosynthetic yields (as recently discovered by the NuSTAR observations 
of Cas A \cite{casA}) and natal kick of pulsars has pointed towards
 CCSNe indeed being generally aspherical (i.e., multi-dimensional (multi-D), 
e.g., \cite{wang01,wang02,maeda08,tanaka} and references therein). 
Unfortunately, however, these electromagnetic 
signatures are rather indirect to probe the inner-working
  because they can only provide an image of optically 
thin regions far away from the central core. 

Much more direct information is carried
 away by neutrinos and GWs. The detection of neutrinos from SN1987A 
 paved the way for the neutrino astronomy \cite{hirata1987,bionta1987}.
 Even though there were just two dozen neutrino events from SN1987A (which 
are not enough to say something solid about the multi-D feature), these 
events have been studied extensively (yielding $\sim$ 500 papers) and
 have allowed us to 
have a confidence that our basic picture of the supernova 
 physics is correct (e.g., \cite{Sato-and-Suzuki}, see
 \cite{raffelt_review} for a recent review). 
In propagating the stellar envelope,  
SN neutrinos produced deep inside the core are
 influenced (at least) by the well-known Mikheyev-Smirnov-Wolfenstein effect
 (e.g., \cite{mikheev1986,kneller}).
Therefore, GWs are primary 
observables, which carry us a direct episode of the supernova 
 engine. 

From a theoretical point of view, clarifying what makes the dynamics of the 
central engine deviate from spherical symmetry is essential 
also in understanding the yet uncertain CCSN mechanism. As a result of continuing efforts 
for decades, theory and neutrino radiation-hydrodynamics simulations 
are now converging to a point that multi-D hydrodynamic matter motions play a 
crucial role in facilitating the neutrino mechanism, which is the most favoured scenario
 to trigger explosions (e.g., \cite{kota06,Janka12,Kotake12_ptep,tony14} 
for recent reviews). 
The neutrino mechanism \cite{bethe85,bethe} 
 requires convection and the standing-accretion-shock instability (SASI) to increase the 
neutrino heating efficiency in the gain region where net energy absorption is 
positive. For canonical massive stars heavier than $\sim 10 M_{\odot}$,
 the neutrino mechanism fails in spherical symmetry (1D) 
\cite{rampp00,lieb01,thom03,sumi05}. 
A number of two-dimensional (2D) simulations with spectral neutrino
 transport now report successful neutrino-driven 
models that are trending towards explosion
\cite{marek,Bmueller12a,bruenn13,suwa},
 whereas the first such three-dimensional (3D)
 simulations \cite{takiwaki12,Hanke13,takiwaki14}
 have reported explosions only for a light progenitor model.

Another candidate mechanism is the magnetohydrodynamic 
(MHD) mechanism
 \cite{Bisnovatyi76,Leblanc70,Kotake06,Burrows07,Takiwaki09,Takiwaki11}.  
Rapid rotation of precollapse iron cores
 is preconditioned for this mechanism, 
because it relies on the extraction of 
rotational free energy of the core by means of 
the field-wrapping and magnetorotational instability (e.g., 
\cite{Balbus98,Obergaulinger09,Masada12} and references therein). 
Such rapid rotation is likely to obtain $\sim$ 1\% of massive star 
population (e.g., \cite{Woosley06}). Minor as they may be, the MHD explosions 
are receiving great attention
 as a possible relevance to magnetars and collapsars
(e.g., \cite{Metzger11,MacFadyen01,Harikae09_a,Harikae10}), 
which are hypothetically linked to the formation of long-duration 
gamma-ray bursts (e.g., \cite{Meszaros06} for a review).

Keeping step with these advances in the CCSN theory and modelling, 
considerable progress in understanding the GW emission processes has been 
made simultaneously (e.g., \cite{ott_rev,Kotake13,fryer11} for recent reviews). 
In the MHD mechanism, rapid rotation of the precollapse core leads to
significant rotational flattening of the collapsing and bouncing core,
leading to a theoretically best-studied, the so-called type I waveform of 
the bounce signals. The waveform is characterized by sharp spikes
 at bounce followed by a subsequent ring-down 
phase \citep{Dimmelmeier07,Ott07_prl,Ott07_cqg}. After bounce, 
a large variety of the emission processes have been proposed, including
convective motions in the proto-neutron star (PNS) 
and in the region behind the stalled shock
\citep{Burrows96,EMuller97,Fryer04_apjl,EMuller04}, the SASI
(e.g., \citep{Marek09,Kotake07,Kotake09,kotake_ray,Murphy09}), 
non-axisymmetric rotational instabilities \cite{Ott05,Scheidegger08,Scheidegger10,Kuroda14}, anisotropy in neutrino emission \cite{EMuller04,Kotake09,EMuller12,BMuller13}, and pulsations of the PNS \cite{ott_new}.

If we were able to associate the above GW signatures with the proposed 
explosion mechanisms (basically either the neutrino or MHD mechanism), 
then the GW signals, if successfully detected, should help confirm the 
 mechanisms. To this end, one must extract a real GW signal buried in 
detector noises
 and determine the waveform characters by matching somehow 
to signal predictions obtained from the multi-D CCSN simulations mentioned 
above.  

The most established method is matched filtering (see \cite{schutz} 
for review) as is done when looking for compact binary coalescence
 signals (e.g., \cite{rasio}). 
%
 However, such a template based search is not suitable for the GW 
signals from CCSNe. This is because the waveforms, except for the bounce signals in rapidly rotating cores, are all affected
 by turbulence in the postbounce phase, which is governed by the non-linear hydrodynamics (e.g., \cite{ott_rev,Kotake13}). 
Hence, the waveforms are of stochastic 
 nature \cite{Kotake09} and impossible to predict {\it a priori}. 
To detect such signals from the 
 next nearby event and extract the information of the explosion
 physics, one needs to construct a suitable analysis method to 
 signal extraction, reconstruction, and model selection,  which is able to deal with the stochastic GW nature.

Considering GW signals from CCSN simulations into signal
 detection/reconstruction and parameter estimation (of the supernova 
physics) was pioneered by \citet{brady}.
The authors introduced a Gram-Schmidt method to parameterize the 
 bounce GW signals \cite{zweg} using a small set of orthonormal basis 
vectors that represent characteristic features common to all the waveforms. 
More efficient method to derive the basis vectors was introduced
by \cite{heng09} with principal component analysis (PCA). \citet{roever2009}
 combined the PCA with the Bayesian inference to recover the bounce 
GW signals by \cite{dimm08} and obtained excellent waveform 
reconstructions.
Going step a further, \citet{logue12} developed a Bayesian model selection
 framework to tell the proposed explosion mechanisms (MHD, neutrino, 
 or acoustic mechanism) apart in the presence of detector noises. They pointed out 
that the Bayesian approach could identify any of the candidate mechanisms 
with high confidence for CCSN events at distances of up to $\sim 2$ kpc. 
 
The above work has demonstrated that the PCA is indeed a powerful tool to 
extract robust waveform features of the bounce signals 
 in rapidly rotating core-collapse.
 However as already mentioned by \cite{heng09,roever2009,logue12,engels14}, one of the disadvantage is 
 that only from the PCs it is not easy to directly extract 
the physical parameters of the central core (such as the rotational
 parameters in this case). To get around the difficulty, 
\citet{engels14} recently presented a multivariate regression model,
 by which several important parameters
 to determine the bounce signals (i.e., in the context of the MHD mechanism) 
were shown to be nicely extractable. Yet, as the authors mentioned,
 their current regression model cannot 
 deal with the stochastic waveforms, which are inherent to the neutrino mechanism, probably the most canonical way
 to blow up massive stars.

There have been other approaches to detection and reconstruction
 of the GW signals from stellar core collapse. The discipline 
 stems from the work by \citet{tinto}, in which a maximum likelihood 
 approach was introduced to reconstruct the time evolution of the burst GW 
signals.
More efficient methods for inferring the incident GW signals 
have been proposed so far including the Tikhonov regularization 
scheme by \cite{rakhmanov,hayama07} and a maximum entropy approach
 \cite{summerscales08}. \citet{summerscales08} successfully reconstructed 
 the injected waveforms by \cite{ott04} using data from two detectors without 
any a priori knowledge of the signal shape. 
Extending the Tikhonov regularization of \cite{rakhmanov}, \citet{hayama07} 
added new time domain data conditioning to the analysis pipeline, 
creating complete stand-alone
 coherent network analysis pipeline called {\tt RIDGE}. Using the pipeline,
 they explored the possibility whether one could infer the degree of
 differential rotation from a small set of waveforms by \cite{kotake04}, 
which is more recently reexamined by \cite{abdika} with a more 
complete set of waveforms.

 Joining in these efforts,
we present a coherent network analysis for detection and 
waveform reconstruction of the GW predictions 
obtained from our 3D CCSN simulations.
The network we consider in this work consists of the 4 km LIGO Hanford
 (H), LIGO Livingston (L), VIRGO (V), and KAGRA (K) interferometers
 \cite{fligo,advv,somiya}.
One of the advantage using such world-wide detector networks 
is that both of the GW polarisations ($h_{+}$ and $h_{\times}$) can be 
reconstructed, furthermore permitting the source position on the sky map
 to be determined. In most of the work mentioned above, 
  the employed CCSN models are 
limited to 2D, which can produce only linearly polarized signals, 
 and a single detector has been often 
considered for simplicity \cite{roever2009,logue12}. 
By performing Monte Carlo simulations using the RIDGE pipeline 
\cite{hayama07}, we discuss detectability of 
the gravitational waveforms from our 3D models and 
discuss to what extent information
 about the CCSN engine could be extracted from successful
GW detection of the future nearby CCSN event.

This paper is structured as follows. In Section
\ref{CCSN},  after we shortly review the candidate CCSN mechanisms,
 we summarize the individual gravitational waveforms that we employ in this 
work. Then we discuss the detectability of the GW signals in a most 
prevalent way, that is, by comparing the 
 root-sum-square (rss) waveform amplitudes with the sensitivity curves of various
 GW interferometer detectors. 
Main results of this work are given and discussed 
 in Section \ref{results}. We summarize our results and discuss their 
implications in Section \ref{summary}.

\section{Gravitational-Wave Signatures and their Optimistic Detectability}
\label{CCSN}
In this study, we consider the neutrino mechanism and the MHD mechanism
 for CCSN explosions and describe their characteristic GW signatures in 
 the following sections. Regarding the neutrino mechanism,
 we take the model waveforms from 3D simulations by 
\citet{Kotake09,kotake11}, which we refer to them as {\tt KK+09} and 
{\tt KK+11} waveforms (e.g., top panel in Figure 1), respectively. 
For the waveforms in the context
 of the MHD mechanism, we use the waveforms from 3D models 
by \citet{Kuroda14} (e.g., middle panels in Figure 1)
and 2D models by \citet{Takiwaki11} (e.g., bottom panels 
in Figure 1) ({\tt KTK14} and {\tt TK11} for short below, respectively). 
 In Appendix A, the waveform 
properties of the {\tt KK+09}, {\tt KK+11}, {\tt KTK14}, and {\tt TK11} 
catalogues are summarized with the numerical methods and initial
 conditions, respectively. Validities and variations
 of the model waveforms are discussed elsewhere in the following.
 
\begin{figure}[htbp]
 \includegraphics[width=8.0cm]{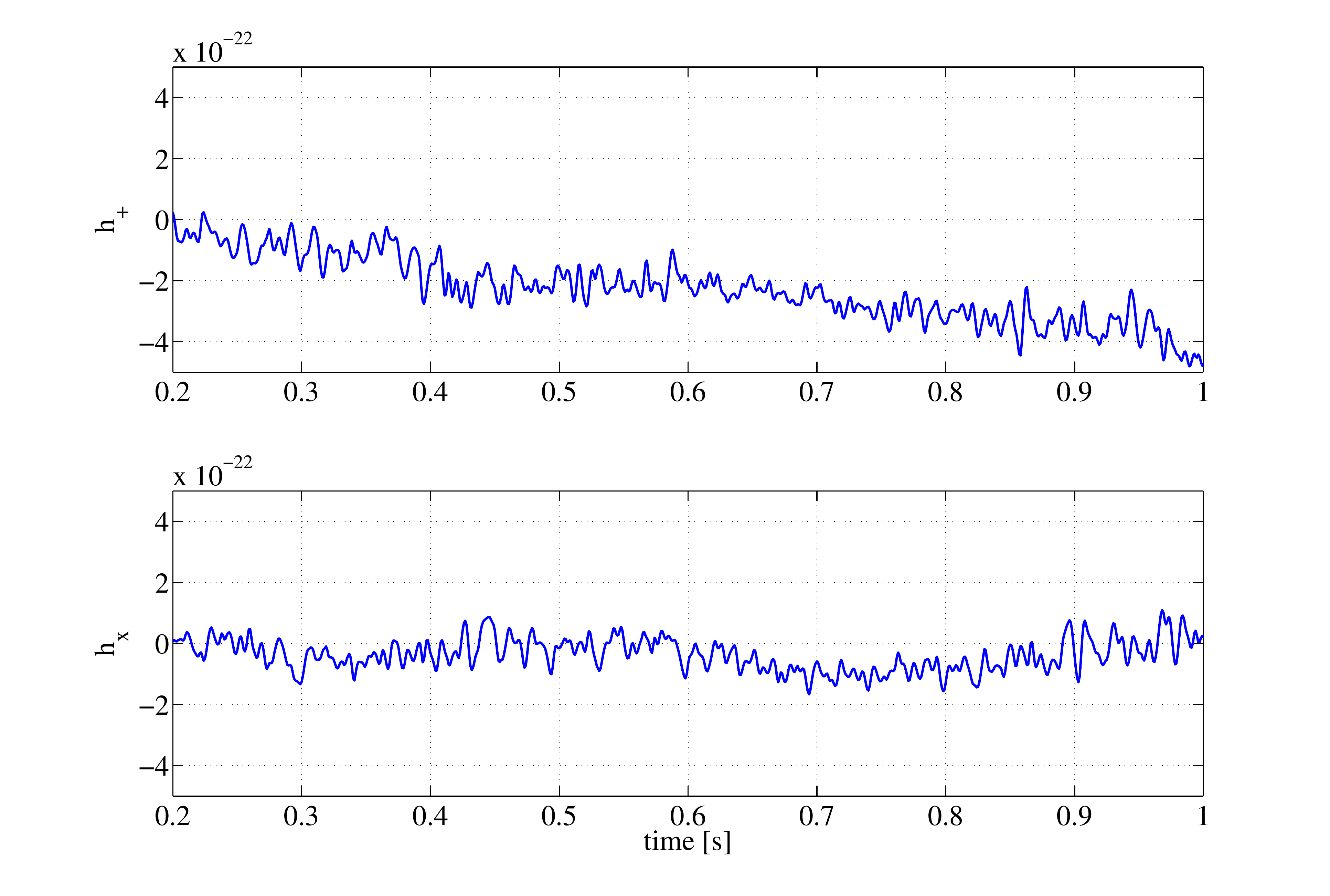}
 \includegraphics[width=5.0cm]{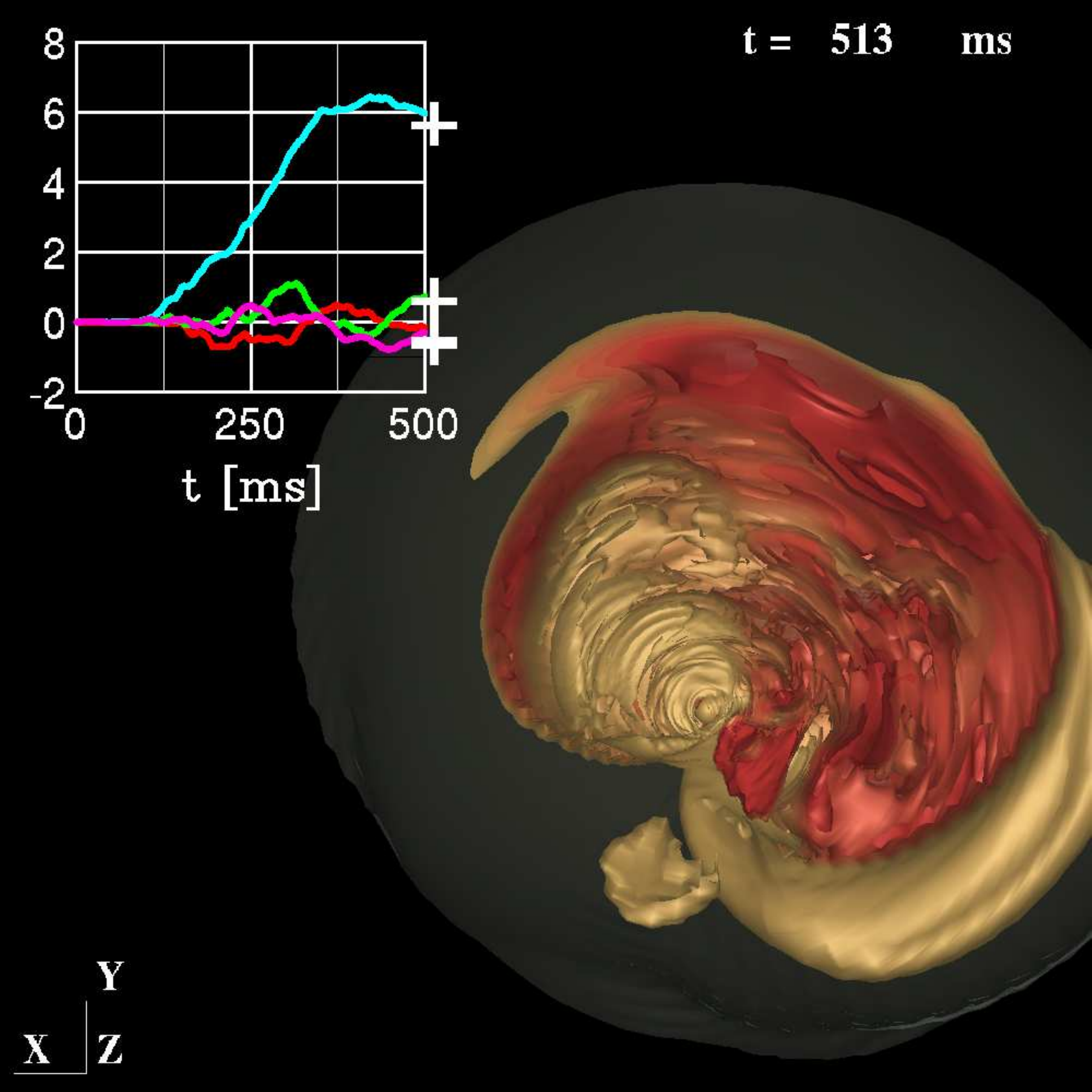}
  \includegraphics[width=8cm]{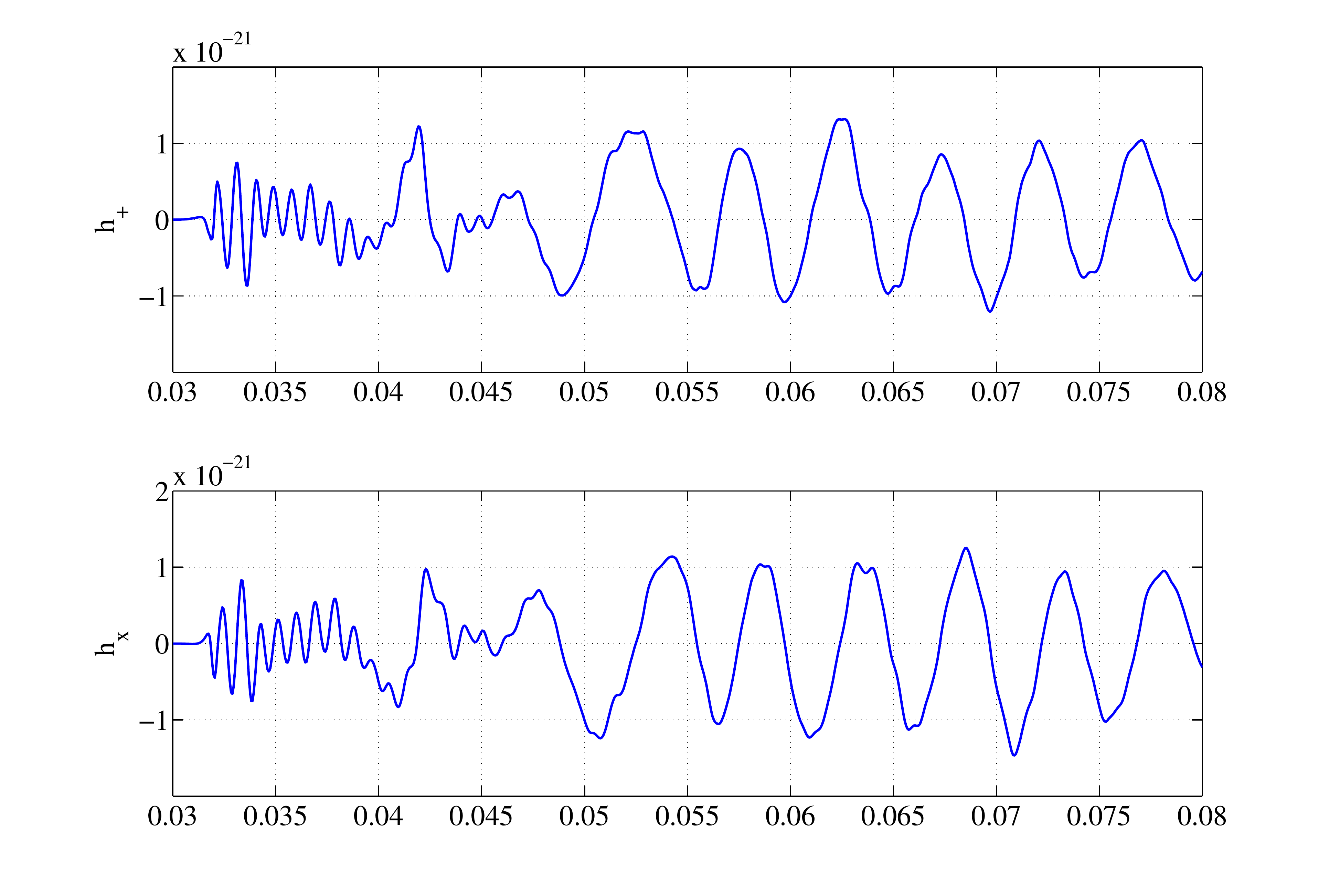}
 \includegraphics[width=5cm]{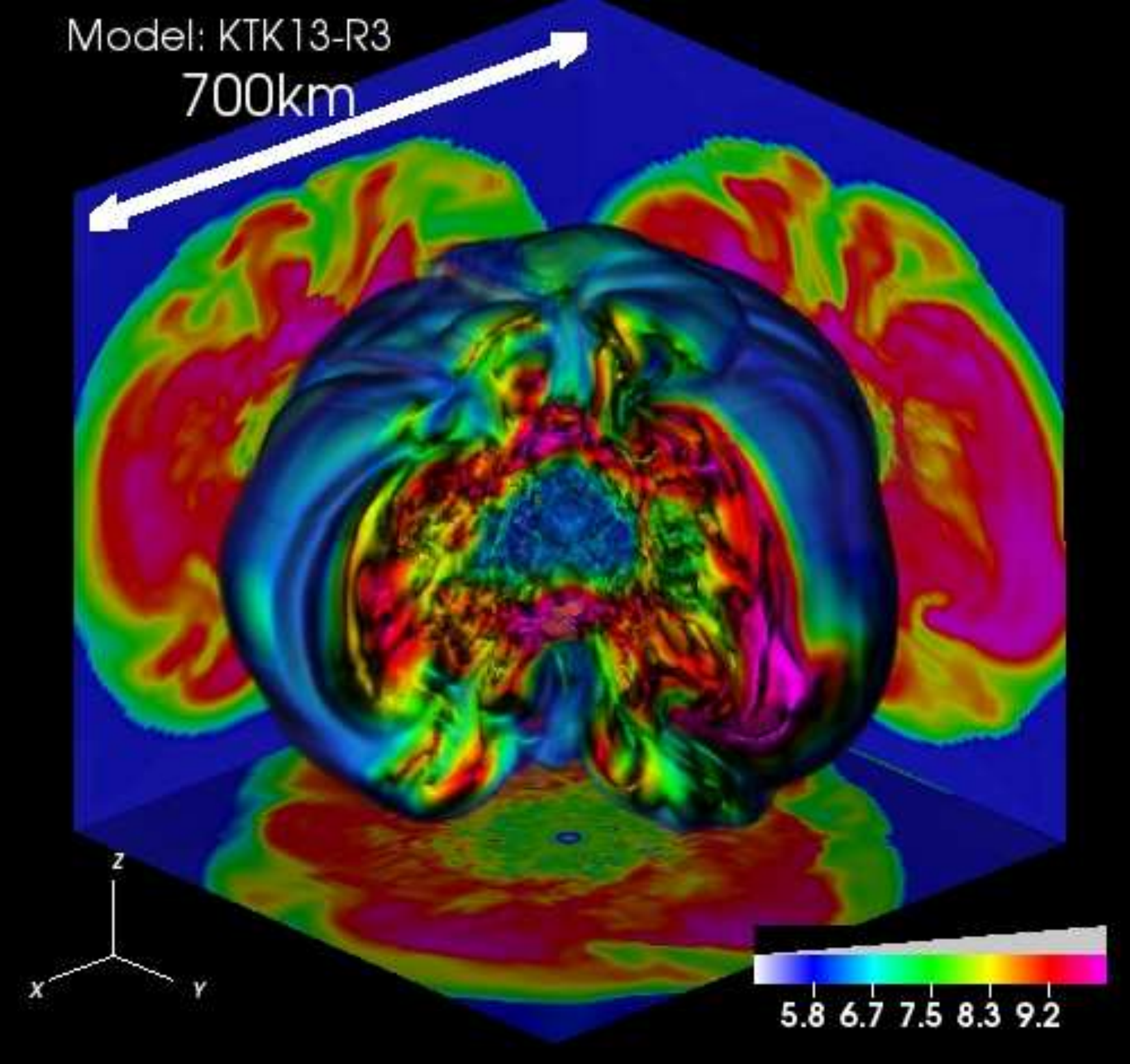}
 \includegraphics[width=8cm]{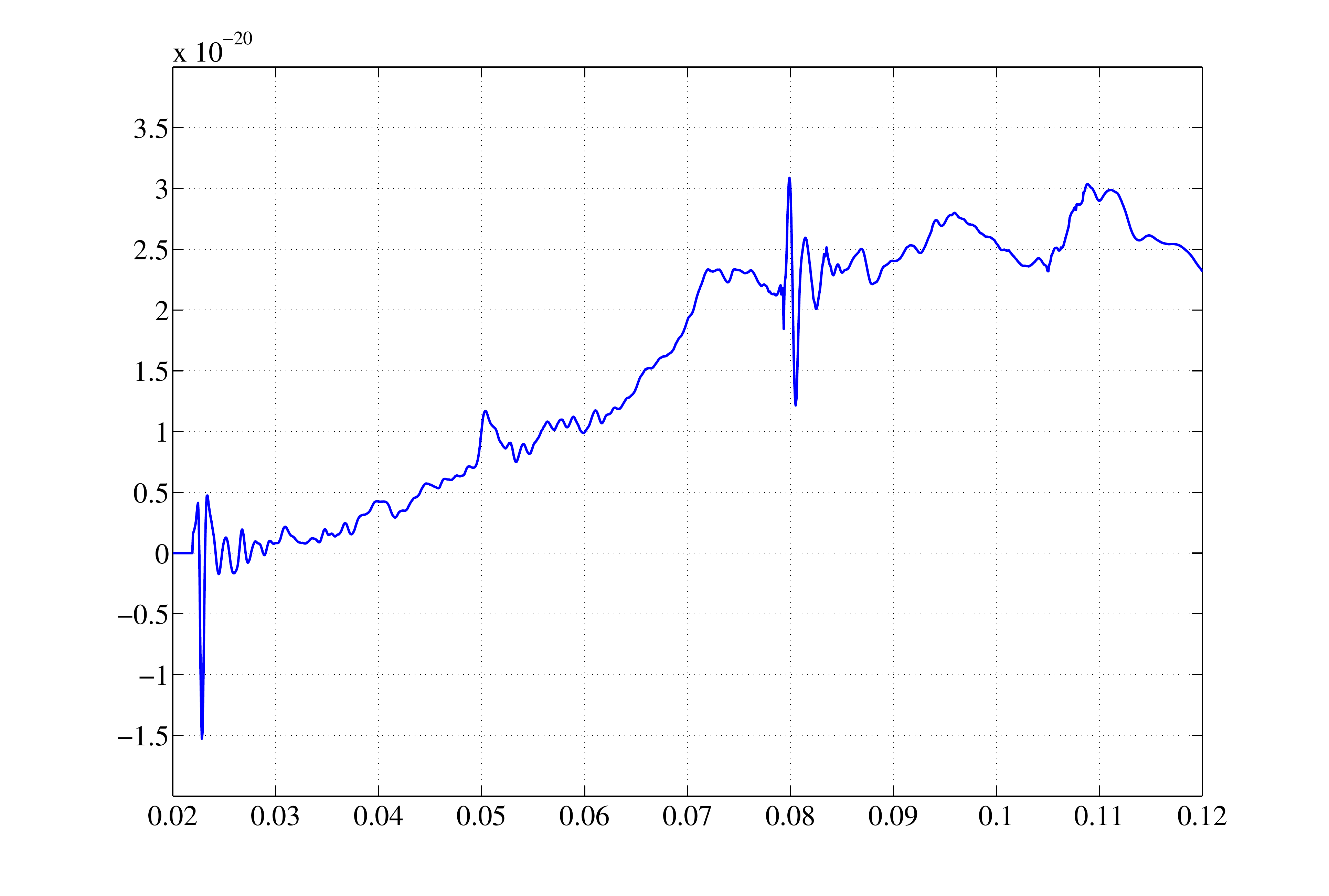}
 \includegraphics[width=5cm]{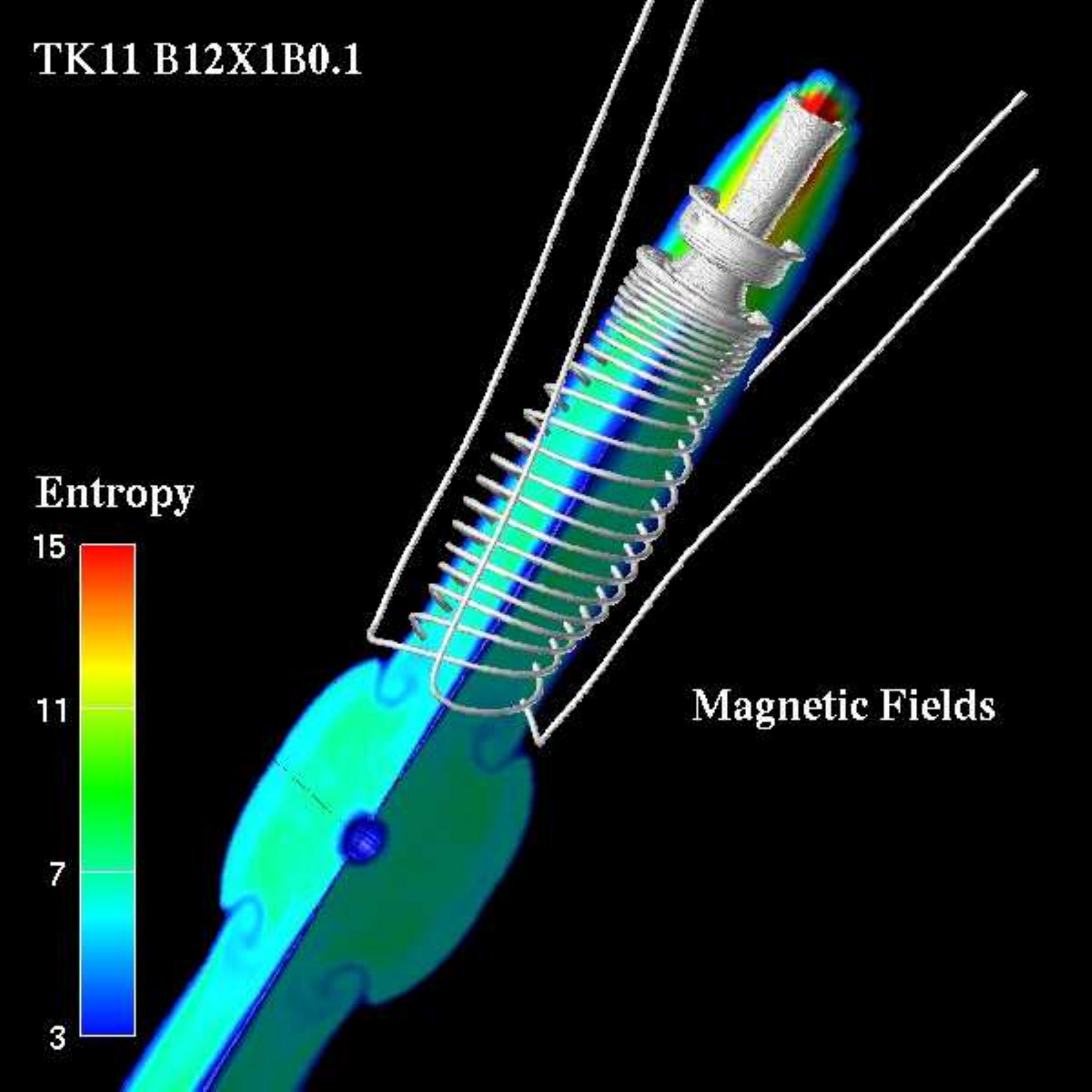}
\caption{GW signal predictions (left panels) for a Galactic event 
 (at a distance of 10 kpc) and the blast morphologies (right
 panels) for the neutrino mechanism (top panels from {\tt KK+11}) and the 3D general-relativistic (GR)
 models that exhibit non-axisymmetric rotational 
instability (middle panels from {\tt KTK14}) and jet-like
 explosion (bottom panels from {\tt TK11}) possibly associated with 
 the MHD mechanism. All the waveforms 
(left panels) come from quadrupole matter motions, whereas the inset 
of the top right panel shows the waveform only from 
anisotropic neutrino emission (see Appendix A for more details).
Not to make the plots messy, waveforms 
seen only from the polar
 direction (with respect to the computational domain) are shown 
 for the 3D models (top left and middle left panels) and waveforms 
 seen only from the equatorial plane for the 2D model (bottom left panel).
 The time ($t_{\rm sim}$) is measured from the epoch 
 when simulations are started. The polarization of the GWs is indicated 
 by "$+$" and "$\times$" (see Appendix A for more details).   }
\label{f1}
\end{figure} 

\subsection{Model Predictions versus Sensitivity Curves}


\subsubsection{Characteristic frequency and root-sum-square of GW signals}
\begin{figure}[hbtp]
\begin{center}
\includegraphics[width=0.8\linewidth]{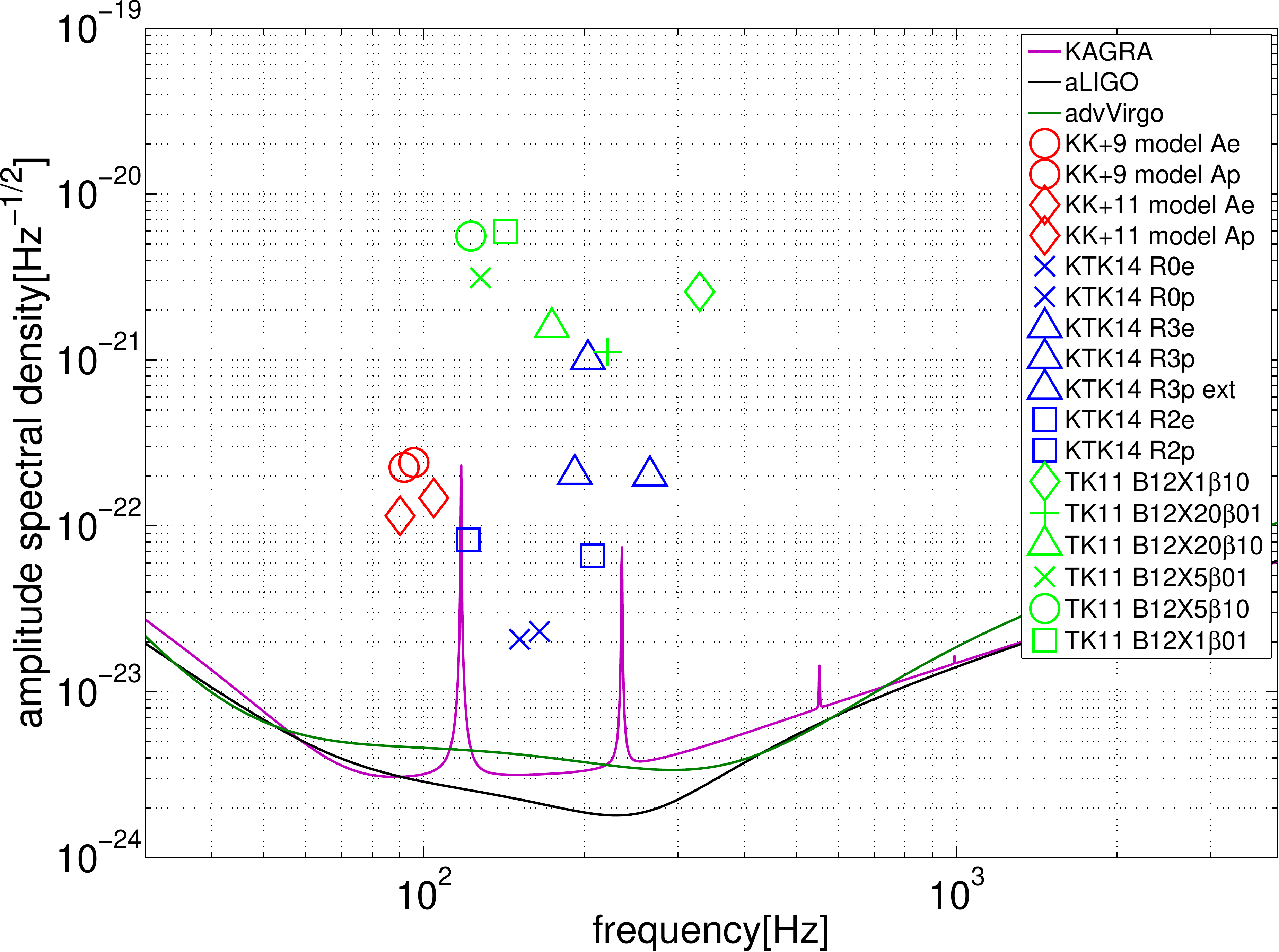}
\caption{Location of GW signal predictions for all the waveform 
 catalogues of {\tt KK+} (circles and diamonds colored by red), {\tt KTK14}
 (circles, triangles, cross, and squares colored by blue), and 
{\tt TK11} (triangles, squares, crosses colored by green) in the 
 $h_{\rm rss}$-$f_c$ plane relative to the sensitivity curves 
of KAGRA, advanced LIGO (labelled as aLIGO), and advanced VIGRO 
(labelled as advVirgo). The source is located at a distance of 10 kpc.}
\label{f2}
\end{center}
\end{figure}

Figure \ref{f2} shows the (frequency-integrated)
 root-sum-square (rss) strain amplitude (${h}_{\rm rss}$) from the 
{\tt KK+, KTK14}, and {\tt TK11} catalogues
against the characteristic frequency ($f_c$) 
relative to the sensitivity curves of KAGRA, advanced LIGO (labelled as 
 aLIGO), and advanced VIGRO (labelled as advVirgo). 
The sources are assumed to be 
located at $10$ kpc from the Earth and optimally 
oriented to the design sensitivity of the 
detectors. Following \cite{flanagan}, 
we calculate the spectral density ${h}_{\rm rss}$ [${\rm Hz}^{-1/2}$] from 
 the Fourier transform of the wave signals 
(${h}_{+,\times}(f) = \int_{-\infty}^{\infty} e^{2 \pi i f t} h_{+,\times}(t)\, dt$) 
as ${h}_{\rm rss} = \sqrt{ \sum_{A = +, \times}^{} |\tilde{h}_{A}(f)|^2}$,
 and the (detector-dependent) characteristic frequency as,
\begin{equation}
f_c =\left(\int^{\infty}_0 \frac{\sum_{A} \tilde{h}_A(f)
\tilde{h}_A^*(f)}{S_\mathrm{n}(f)}\,f\mathrm{d}f\right)/ 
\left(\int^{\infty}_0\frac{\sum_A^{} \tilde{h}_A(f)\tilde{h}_A^*(f)}{S_\mathrm{n}(f)}\,\mathrm{d}f\right),
\label{fc}
\end{equation}
where $S_\mathrm{n}(f)$ is the detector noise power spectral density 
in units of ${\rm Hz}^{-1/2}$.  Having the same unit as 
the strain equivalent spectrum density, ${h}_{\mathrm{rss}}$ is often
used in burst GW searches to compare the signal strength with 
the detector sensitivity.

In the $h_{\rm rss}$--$f_{\rm c}$ plane (Figure \ref{f2}), 
the {\tt KK+} catalogues (symbols colored by red) that can 
 be associated with the neutrino mechanism are rather localized to a small
 region (e.g., $f_{\rm c} \sim 100$Hz and 
$h_{\mathrm{rss}}$ $\sim$ $1-2\times10^{-22}\mathrm{[Hz^{-1/2}]}$ )
 for the 3D models without or with rotation 
(labelled as {\tt KK+09} or {\tt KK+11}) either seen from equator 
or pole (with the model name ending with {\tt e} or {\tt p}, such as 
{\tt KK+09Ae} or {\tt KK+09Ap}). This is because the 
initial rotation rate assumed in the {\tt KK+11} models is not enough rapid, 
as is consistent with outcomes of recent stellar evolutionary calculations 
\cite{hege05}, to affect the quadrupole matter motions 
in the postshock region. 

Regarding the waveforms that can be associated with the MHD mechanism,
 the {\tt KTK14} models (symbols colored by blue) and the {\tt TK11} models
 (symbols colored by green)
 are in the range of  $f_c = 100 - 300$Hz,
 $h_{\mathrm{rss}}$ $\sim$ $2-100\times10^{-23}\,\mathrm{[Hz^{-1/2}]}$
 and of  $100$ - $300$Hz, $h_{\mathrm{rss}}$ $\sim$ $1-6\times10^{-21}
\,\mathrm{[Hz^{-1/2}]}$, respectively.

For the {\tt KTK14} models, it can be seen that the rss amplitudes 
 become higher for models with larger initial angular momentum
 (model R3 (blue triangle) followed in order by R2, R1 
(not shown in the plot), and R0 (non-rotating)). Note that for model "R3p ext"
 we manually extrapolate the quasi-periodic gravitational waveform 
(middle panel in Figure \ref{f1}) up to 
 1 s after bounce, assuming that 
the non-axisymmetric instability observed in the limited simulation time 
(until $\sim$ 60 ms after bounce) persists afterwards
(with the mean oscillation period during the simulation time) up to 
 1 s after bounce.
  This ad-hoc model has the maximum amplitude among the {\tt KTK14}
 catalogue (the highest amplitude among the blue triangles){\footnote{
 Remembering that 3D-GR simulations are still computationally very 
 expensive, we here consider such optimistic model just as a guide to 
infer a (very crude) upper limit of the GW signals 
 expected from very rapidly rotating 3D-GR models with non-axisymmetric instabilities}}.

The wave amplitudes for {\tt TK11} are generally higher than 
 those for {\tt KTK14} simply because the assumed initial rotation rates
  are generally higher for the {\tt TK11} catalogue.
As is well known from previous studies (e.g., \cite{ott_rev}),
 the characteristic frequency 
($f_c$) becomes generally lower for models with larger initial angular 
momentum (e.g., compare R3 with R2 and B12X1$\beta$10 with 
B12X20$\beta$01). This is because the central density ($\rho_c$) 
in the vicinity of PNS becomes
 generally lower for models with larger initial angular momentum
 due to the stronger centrifugal forces. This 
makes the dynamical timescale $t_{\rm dyn}
 \sim (G \rho_c)^{-1/2}$ (with $G$ being the gravitational constant)
 longer and the typical frequency ($f_c \propto 1/t_{\rm dyn}$) lower.

As can be seen from Figure \ref{f2},  the signal predictions taken in this work
 are above the sensitivity curves of the advanced 
detectors for a Galactic event. And it is also worth mentioning that 
the characteristic frequency for all the models
 is in the range of $100$ - $400$Hz, which is close to the 
 highest sensitivity domain of the advanced detectors.
In the next section, we proceed to 
 discuss the detectability more quantitatively 
by calculating the Signal-to-Noise Ratio (SNR).

\begin{table}[hbt]
\caption{Optimal SNR as a function of distance to the source 
for several representative models for GW emission 
in the context of the MHD mechanism ({\tt KTK14} and
 {\tt TK11}) and 
 the neutrino mechanism ({\tt KK+}).
Theoretical noise power spectral density for KAGRA is used. 
As a threshold to claim detection, we take the SNR of 8.
\label{table1}}
\begin{ruledtabular}
\begin{tabular}{lrr}
\textrm{Model}&
\textrm{SNR at 10 kpc}&
\textrm{Distance at SNR=8 [kpc]} \\
\colrule
  KTK14 R0e & 7.35 & 9.1875 \\
    KTK14 R0p & 7.55 & 9.4375 \\
     KTK14 R2e & 21.62 & 27.0250 \\
    KTK14 R2p & 22.88 & 28.6 \\ 
   KTK14 R3e & 58.65 & 73.3125 \\
    KTK14 R3p & 73.93 & 92.4125 \\
    KTK14 R3p ext & 360.97 & 451.2125 \\ \hline
      TK11 B12X1$\beta$10 & 386.54 & 483.175 \\
    TK11 B12X20$\beta$01 & 78.38 & 97.975 \\
    TK11 B12X20$\beta$10 & 191.75 & 239.6875 \\
    TK11 B12X5$\beta$01 & 248.63 & 310.7875 \\
    TK11 B12X5$\beta$10 & 306.2 & 382.75 \\
    TK11 B12X1$\beta$01 & 327.64 & 409.55 \\ \hline
KK+9 Ae & 13.88 & 17.35 \\
    KK+9 Ap & 13.28 & 16.6 \\
    KK+11 Ae & 9.43 & 11.7875 \\ 
    KK+11 Ap & 10.38 & 12.975 \\ 
\end{tabular}
\end{ruledtabular}
 \end{table}


\subsubsection{Optimal horizon distances for KAGRA}

Before presenting a multiple detector analysis from the 
 next sections, we briefly compute matched-filtering SNR in this section.
By taking KAGRA as an example detector and assuming perfect orientation,
we compute the SNR as 
$\mathrm{SNR} = \frac{h_c}{\sqrt{f_cS_\mathrm{n}(f_c)}}$,
where $f_c$ is the characteristic frequency (equation (\ref{fc})) and
 $h_c$ is the characteristic strain amplitude~\cite{Thorne:1987},
\begin{equation}
h_c \equiv \left(3\int^{\infty}_0\frac{S_\mathrm{n}(f_c)}{S_\mathrm{n}(f)}\sum_A\tilde{h}_A(f)\tilde{h}_A^*(f)f\mathrm{d}f\right)^{1/2}.
\end{equation}
 For computing the SNR, we employ the theoretical
 noise spectral densities $S_{\rm n}$ of KAGRA from
 \footnote{http://gwcenter.icrr.u-tokyo.ac.jp/en/researcher/parameter}.

Table \ref{table1} summarizes the optimal SNR at 
 a distance of 10 kpc and the distance at which the SNR is 
at 8 for the representative models, respectively. 
To claim detection, a SNR significantly
 greater than unity and probably in the range of 8 - 13 would be needed
 \cite{flanagan,schutz}.
 We optimistically take the threshold as 8, by which the so-called 
 horizon distance is conventionally defined,
 which is the maximum distance at which the GW signals from an optimally 
oriented and optimally located source could be detected.
 It should be also mentioned that the detection distances
 with realistic noise (rather than idealised Gaussian noise considered
 here) can be significantly worse,
 which remains to be investigated more in detail.

The horizon distance of the {\tt KTK14} waveforms
 is in the range of $9$ - $450$ kpc and it becomes longer 
for models with larger initial angular momentum
 and longest $\sim 450$ kpc for the
 extrapolated waveform (model "R3p ext"). For the non-rotating 
(R0) model, the detection distance is much the same
either seen from pole (model R0p) or equator (model R0e) 
(with respect to the source coordinate system), both of which closely reach
 to $\lesssim 10$kpc for the SNR = 8 threshold. 
As one would guess easily, the pole to equator asymmetry in the horizon 
distance becomes remarkable for models with
 larger initial angular momentum (compare model R3p with R3e. 
Seen from the polar direction (e.g., parallel to the rotational axis), 
the SNR becomes generally higher because the more efficient GW
 emission is associated with the violent growth of
 the non-axisymmetric instabilities in the {\tt KTK14} models.

For 2D models that produce explosions by the 
 MHD mechanism. 
 the detection distance extends from $\sim 100$ to  $\sim 480$kpc,
 depending on the initial rotation rates (bigger for more rapidly
 rotating models). These {\tt TK11} models provide the most 
 distant horizons in this work. Due to the absence of such rapid 
rotation, the detection distance 
 becomes much smaller ($12$ - $17$kpc) for the {\tt KK+} models.
 As one would expect, 
the pole to equator asymmetry in the horizon distance is only weak
 in the {\tt KK+} waveforms.

Having discussed the SNR and the optimistic
detectability with a very idealized situation (i.e., 
  a single detector for an optimally 
oriented and optimally located source), 
we shall turn to a more realistic situation, in which multiple detectors 
 are used for an arbitrary oriented source. 




\section{Results}
\label{results}

In section II, we have discussed an ideal detection limit 
 with the design sensitivity of KAGRA. In reality the performance of the 
detection strongly depends on the antenna pattern functions of a network of 
the multiple detectors. In this section we study the performance 
using the {\tt RIDGE} pipeline which 
takes full advantage of the global network of 
 currently working and future interferometers 
(LIGO Hanford (H), LIGO Livingston (V), VIRGO (V), and KAGRA (K)), 
resulting in enhanced detection efficiency (see 
Appendix B and \cite{hayama07,hayama2013} for more details).
\begin{figure}[hbtp]
\begin{center}
\includegraphics[width=0.4\linewidth]{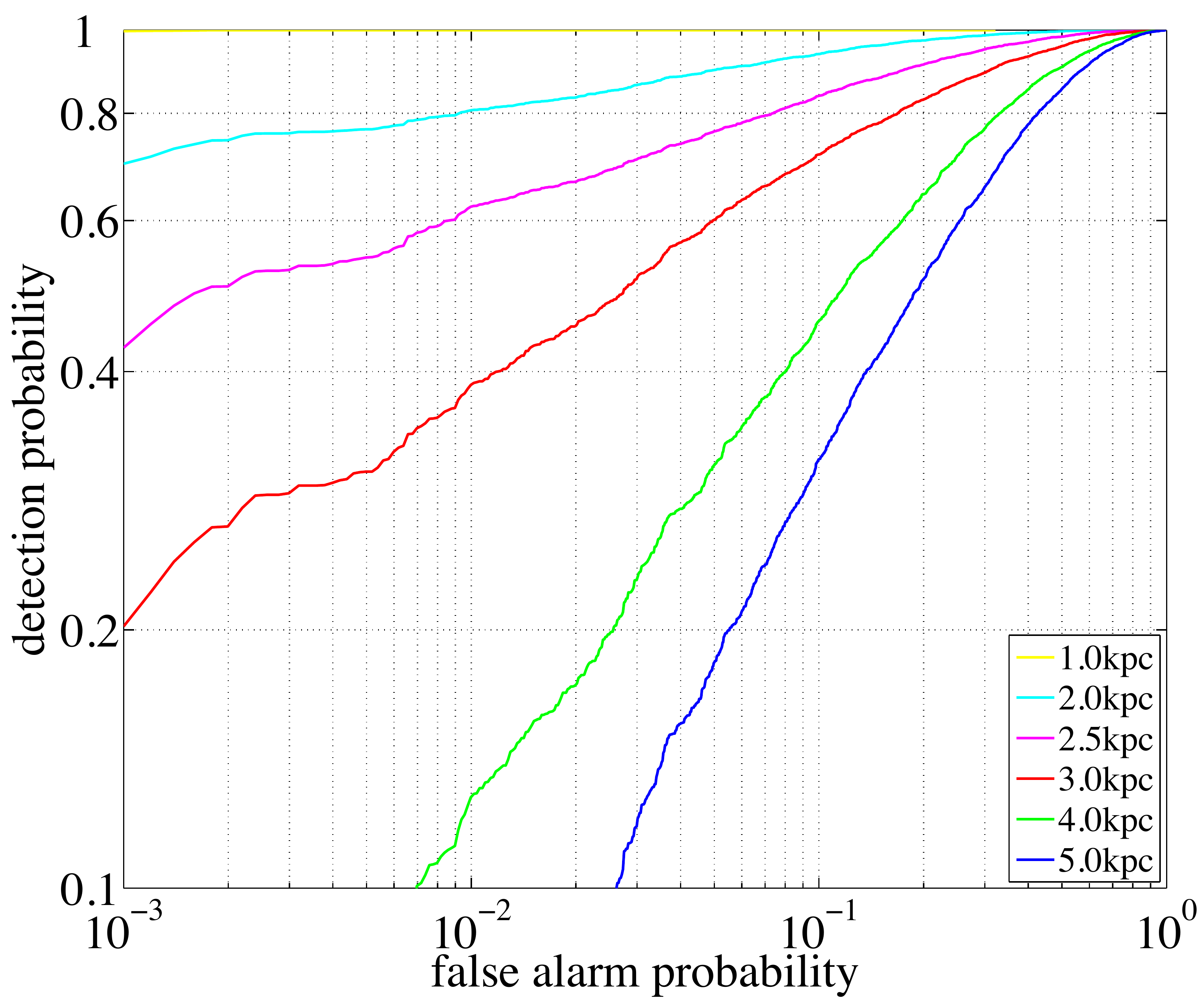}
\includegraphics[width=0.4\linewidth]{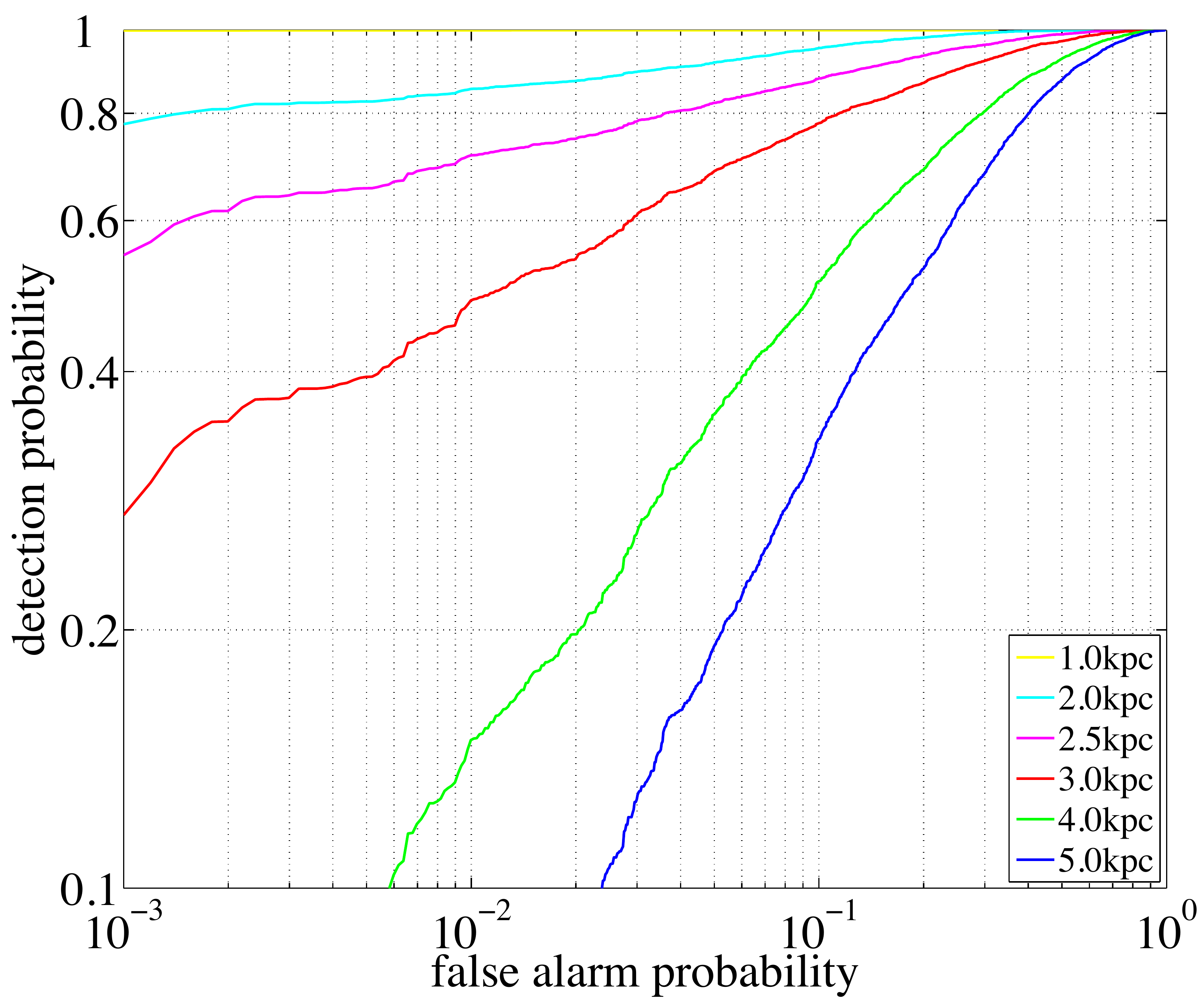}
\includegraphics[width=0.4\linewidth]{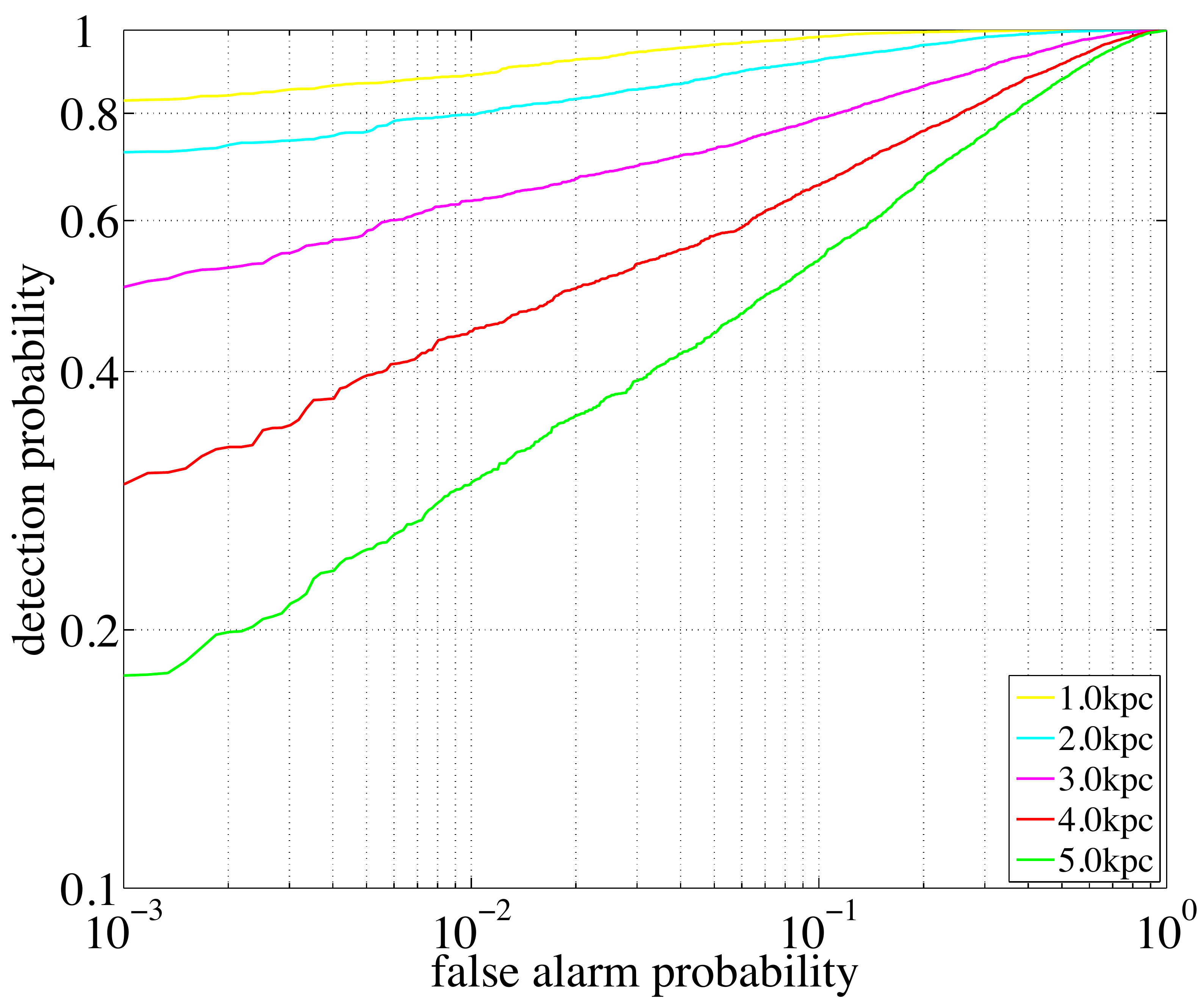}
\includegraphics[width=0.4\linewidth]{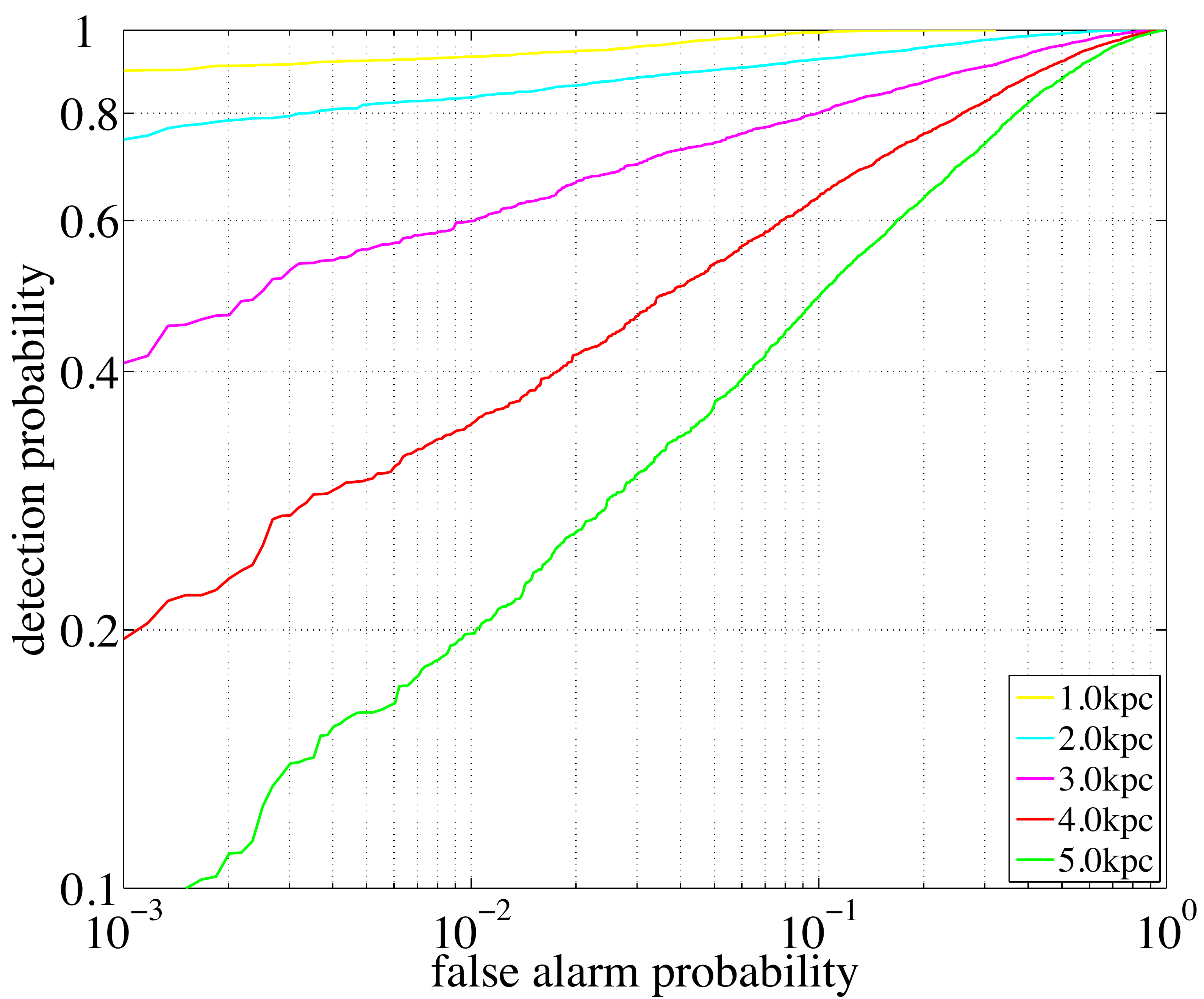}
\caption{ROC curves of non-rotating 3D models 
from the {\tt KK+09} (top panels) and {\tt KTK14} catalogues (bottom panels)
 for polar (left panels) and equatorial (right panels) observer at 
different source distances. }
\label{f5}
\end{center}
\end{figure}

\subsection{Detection efficiency}
We first focus on the receiver operating 
characteristic (ROC) curves, which are useful to see how the detection efficiency ($y$ axis) changes 
with the false alarm probability ($x$ axis) for sources at different 
 distances.
 
Figure \ref{f5} is for the non-rotating 3D models 
from the {\tt KK+09} (top panels) and {\tt KTK14} catalogues (bottom panels). Without rotation, the detection efficiency
 is much the same either seen from pole (left panels) or equator (right panels).
 If we set the false alarm probability of 0.01 and the detection 
probability of $0.5$ as a threshold of the detection, the 
detection distance of the non-rotating model 
 is $2.5 - 3$ kpc from the {\tt KK+09} catalogue and $\sim 3 - 4$
 kpc from the {\tt KTK14} catalogue,
 respectively. It is noted that the detection
 horizon becomes smaller by a factor of $\sim 5$ compared to that for 
the most optimal situation (shown in Table \ref{table1}). 

Figure \ref{f6} is for 3D models with rapid rotation 
(i.e., the {\tt KTK14} catalogue). As already mentioned in Appendix A,
  the growth of non-axisymmetric instabilities was clearly observed 
 in model R3 (the most rapidly models in {\tt KTK14}, top panels
 in Figure \ref{f6}) in the vicinity of the equatorial plane. Due to this, 
the horizon distance of model R3
 (top panels) is bigger seen from 
 pole ($ \sim 60$ kpc, left panel) compared to that seen from 
equator ($ \sim 25$ kpc, right panel). As similar to the non-rotating models 
(Figure \ref{f5}), the horizon distances for the rapidly rotating models 
become up to a factor of $\sim$ 3 smaller 
compared to the most optimal case (e.g., Table 3, $\sim 92$ kpc for
 pole and $\sim 73$ kpc for equator for model R3). Figure \ref{f7} shows
the horizon distance of our 2D MHD model (B12X1$\beta$10, {\tt TK11}) which 
has the biggest optimistic SNR in our catalogues. 
The horizon distance changes from the optimal estimation of $\sim 483$ kpc 
(Table 3) to $\sim 150$ kpc in the more realistic situation.
 These 2D models as well as the 3D rapidly rotating models 
have pronounced GW peaks either associated with core bounce or with 
 the non-axisymmetric instabilities. As a result, the reduction in the horizon
 distance relative to the most idealized situation becomes smaller 
 than for the non-rotating models.

\begin{figure}[hbtp]
\begin{center}
\includegraphics[width=0.4\linewidth]{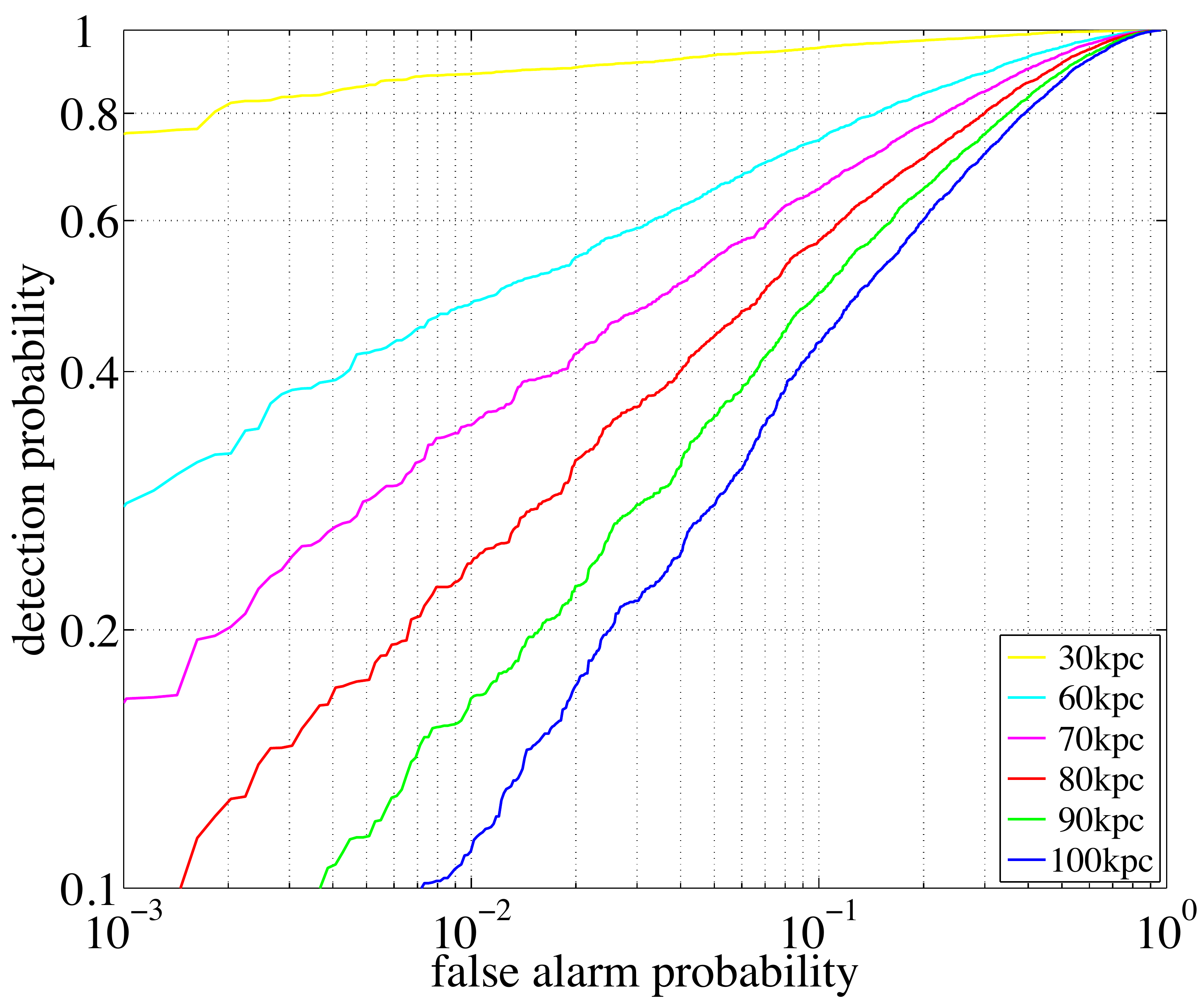}
\includegraphics[width=0.4\linewidth]{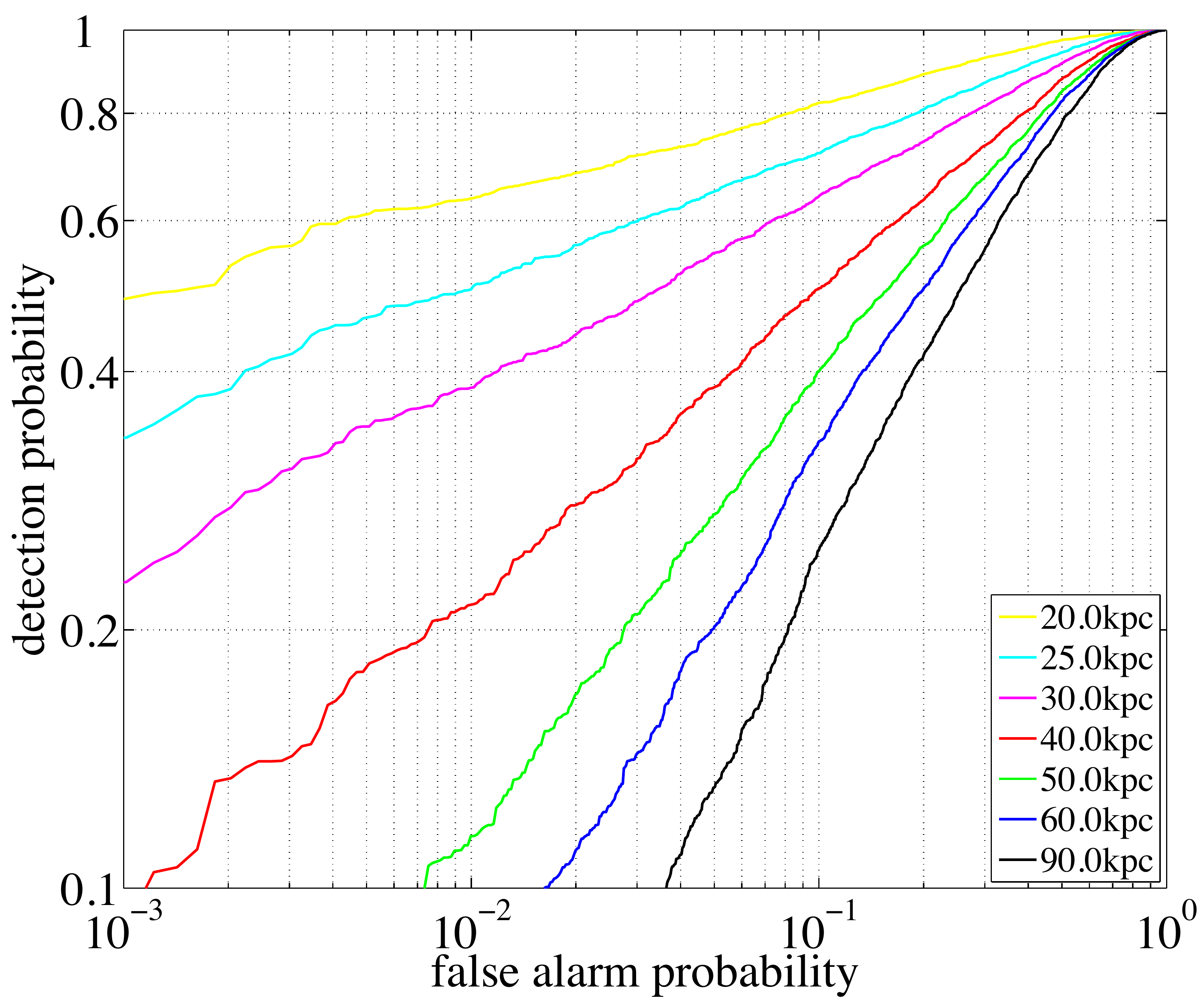}
\includegraphics[width=0.4\linewidth]{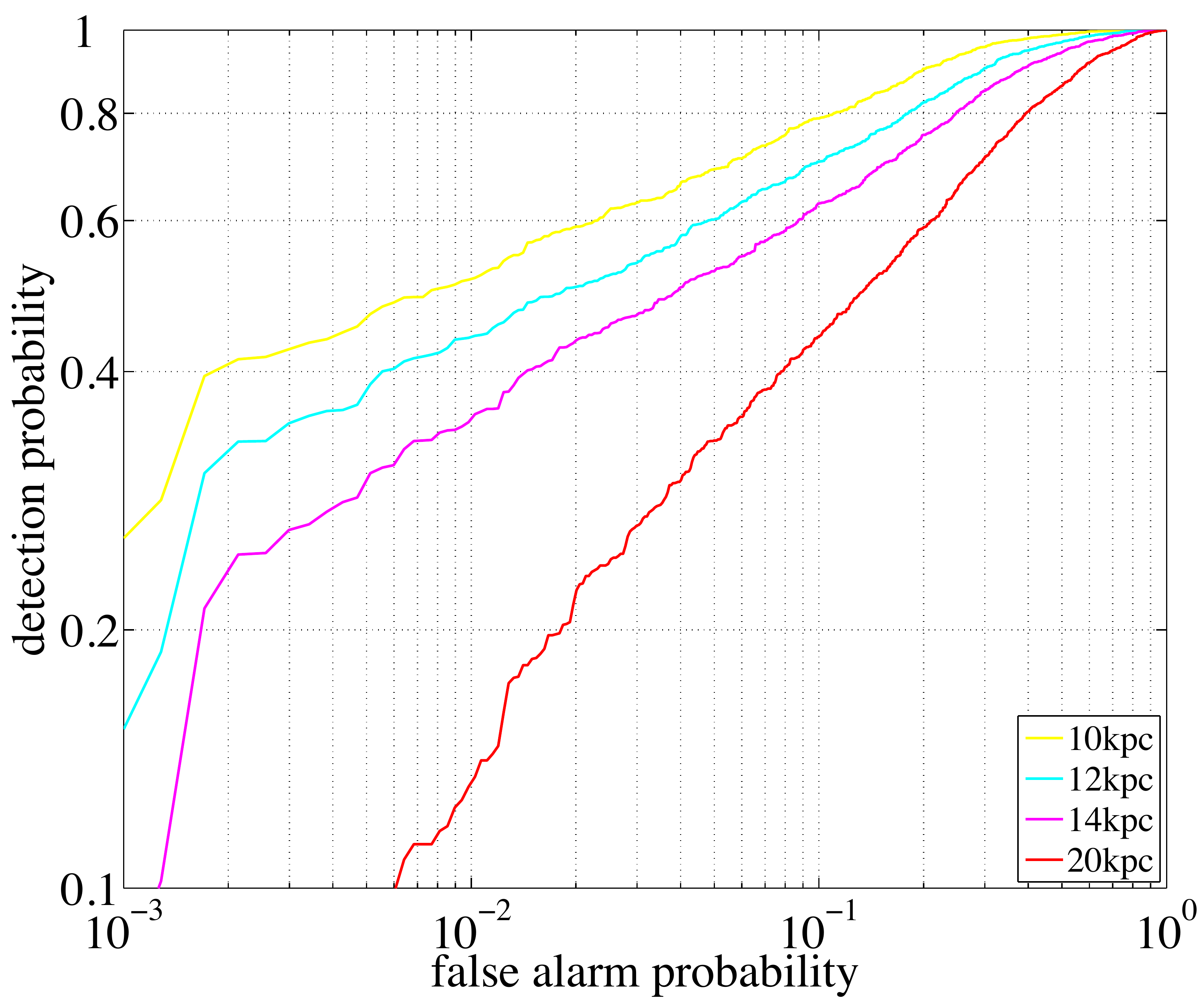}
\includegraphics[width=0.4\linewidth]{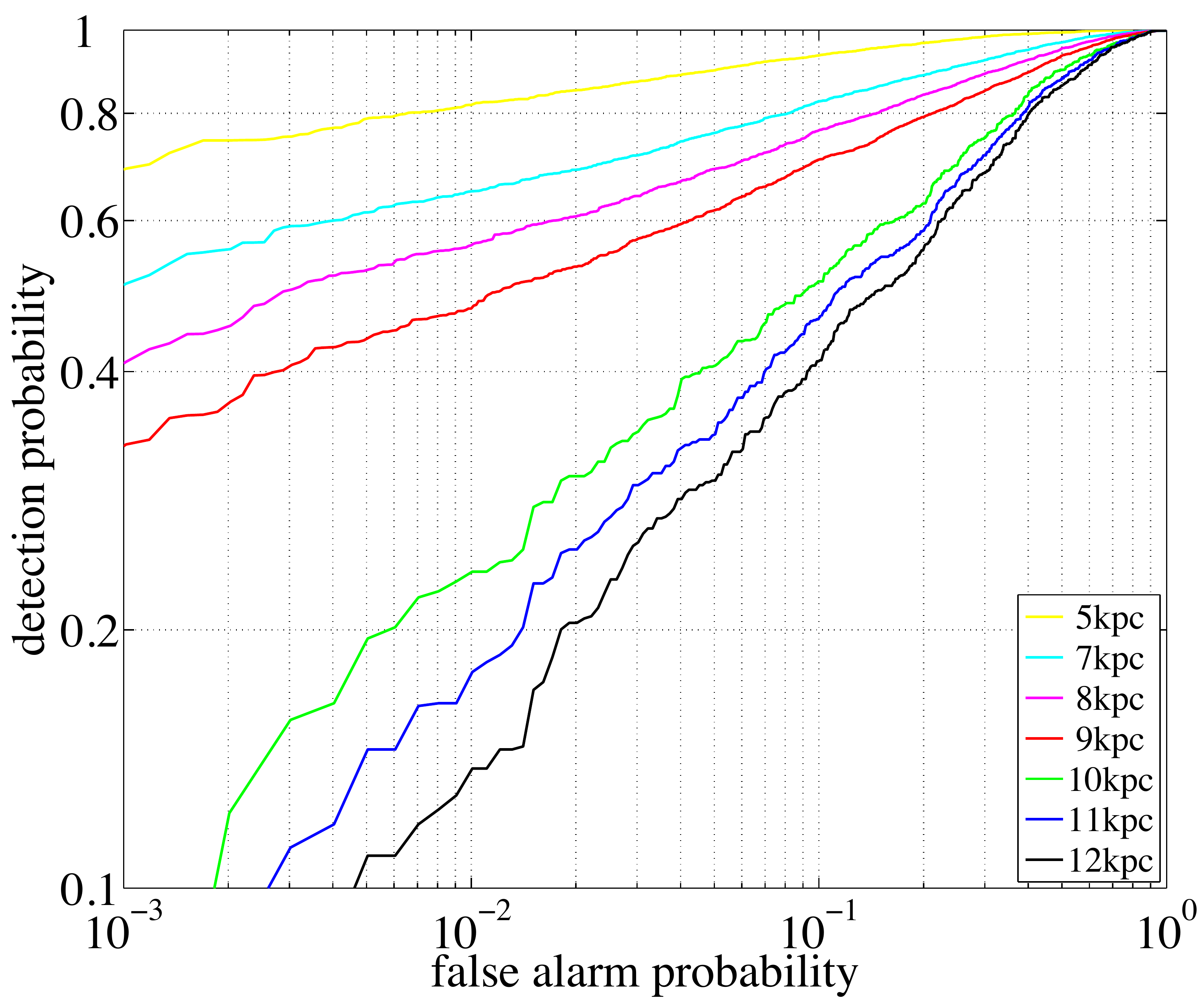}
\caption{Same as Figure \ref{f5} but for 3D models with rotation from 
 the {\tt KTK14} waveforms (model R3 (top panels) and R2 (bottom panels).
}
\label{f6}
\end{center}
\end{figure}

\begin{figure}[hbtp]
\begin{center}
\includegraphics[width=0.4\linewidth]{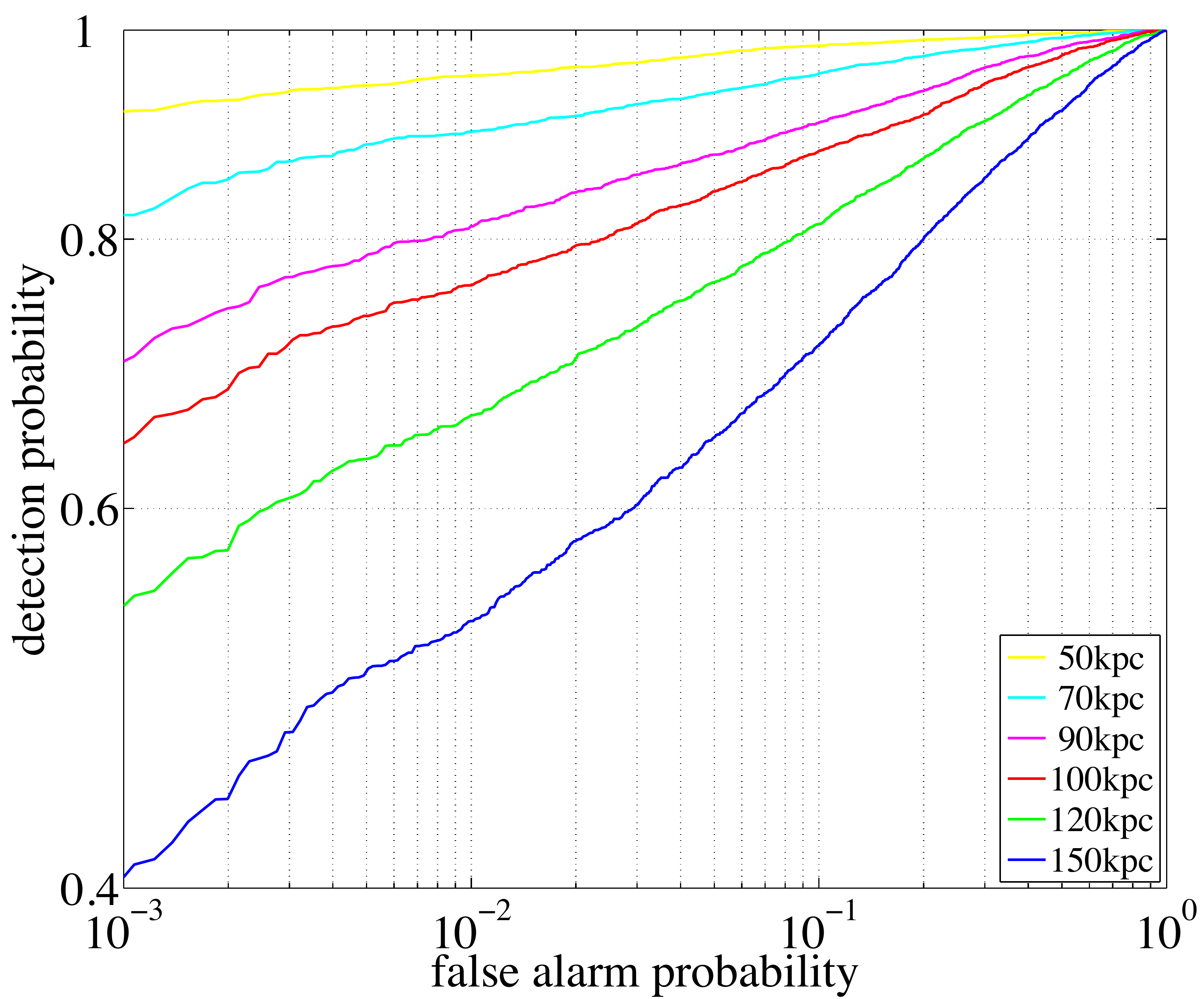}
\caption{Same as Figure \ref{f5} but for the 2D MHD model 
(B12X1$\beta$10) of the {\tt TK11} waveform that possesses the biggest SNR 
(e.g., Table 1) in our catalogue.} 
\label{f7}
\end{center}
\end{figure}

\begin{figure}[hbtp]
\begin{center}
\includegraphics[width=0.43\linewidth]{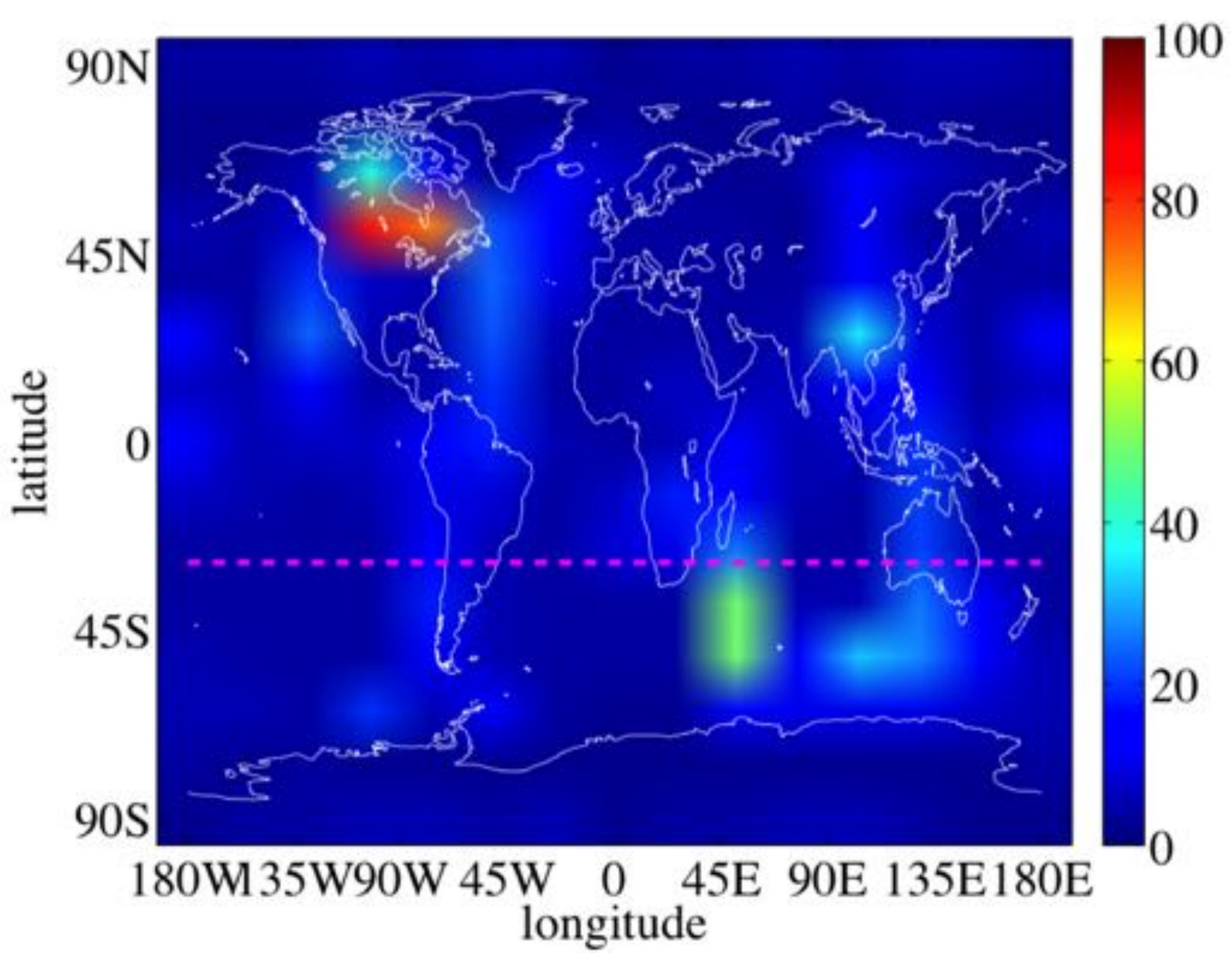}
\includegraphics[width=0.43\linewidth]{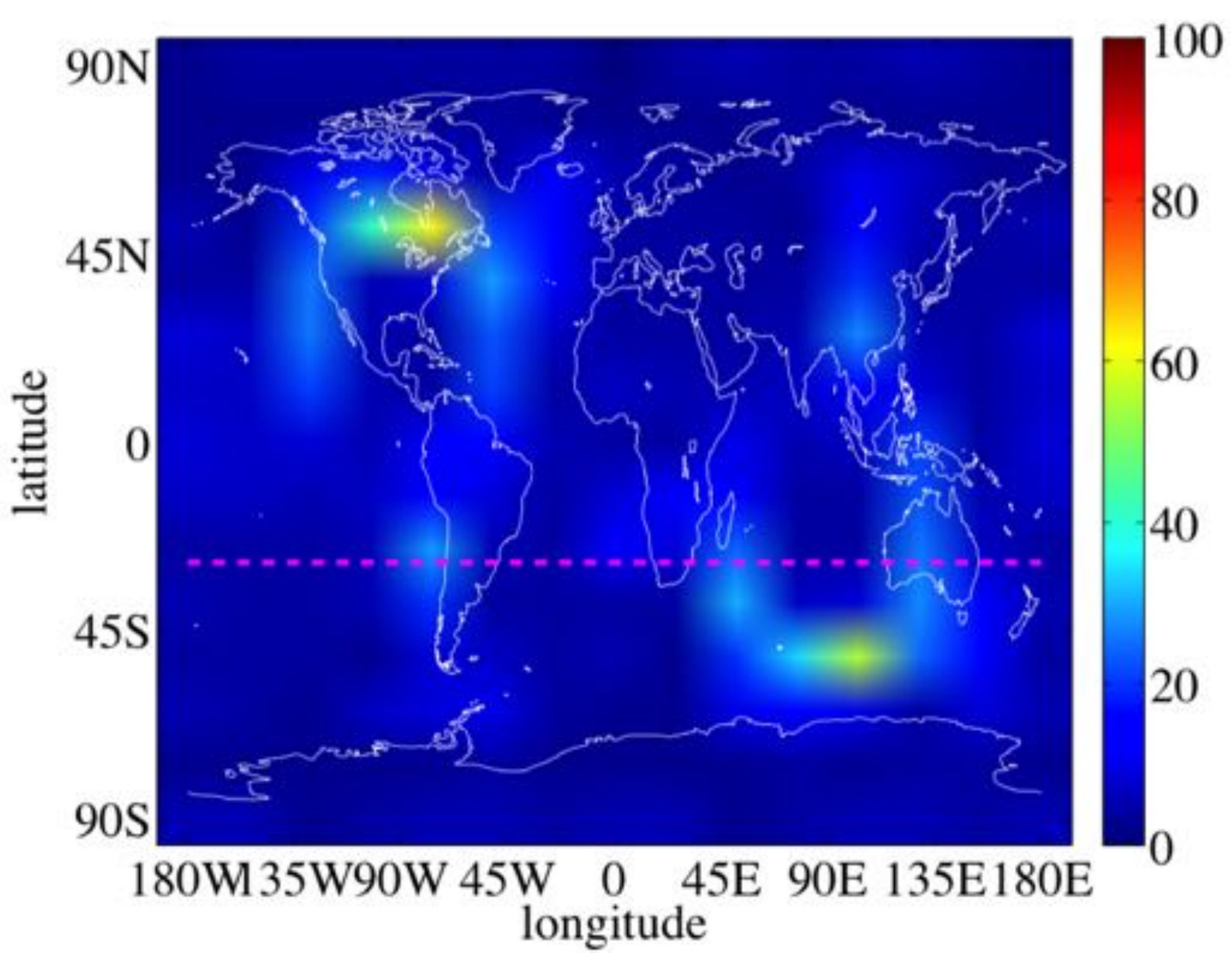}
\caption{Plots of the skymap from the network analysis of H-L-V-K for 
the 3D non-rotating (top panels) from the {\tt KK+09} waveform
 for the equatorial (left) or polar (right) observer,
 respectively. The horizontal yellow line corresponds to the 
 dairy motion of the Galactic center in the skymap. 
 Note that for the advanced detectors such as advanced LIGO, 
advanced Virgo and KAGRA, the Galactic Center is a sky direction with high 
probability of the detection of GWs from the CCSNe.
The $x$ and $y$ axis is the longitude and latitude, 
respectively. Note in this panel that the distance to the source 
 is set as 2 kpc (see text for more details).}
\label{f8}
\end{center}
\end{figure}

\begin{figure}[hbtp]
\begin{center}
\includegraphics[width=0.43\linewidth]{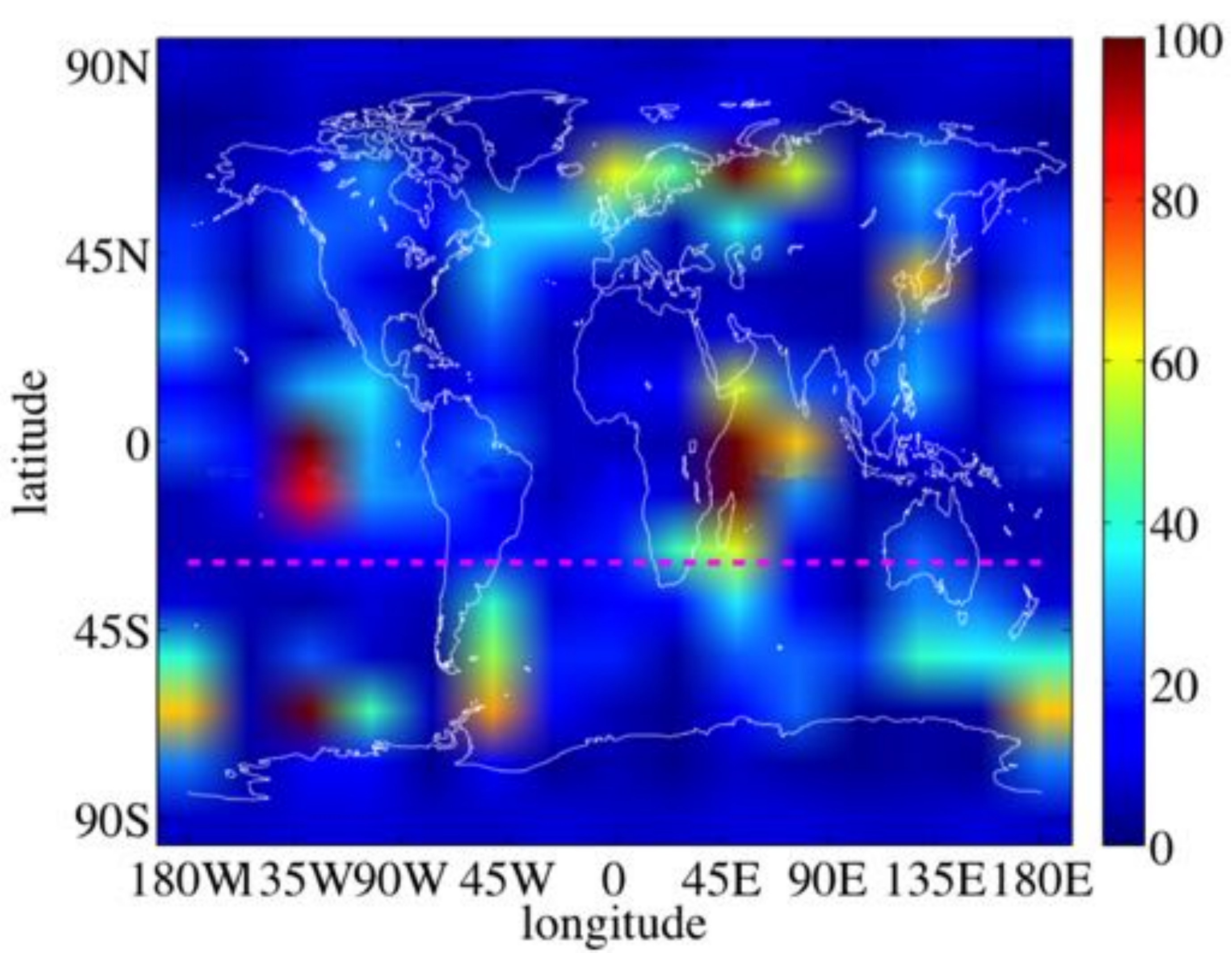}
\includegraphics[width=0.43\linewidth]{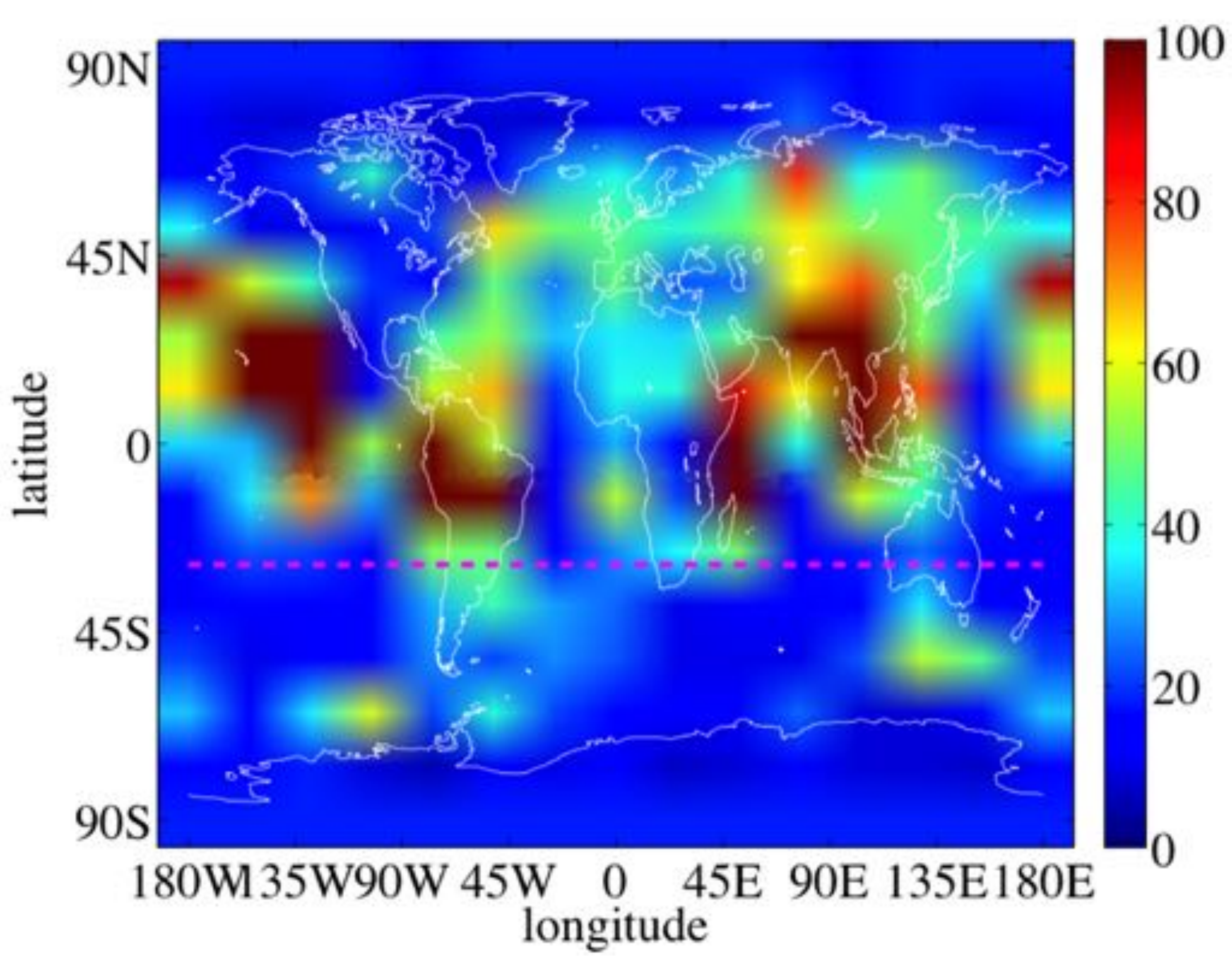}
\caption{Same as Figure \ref{f8} but for 3D models with rapid rotation from 
 the {\tt KTK14} waveforms (model R3) either seen from equator (left) or 
from pole (right).
Note in this panel that the distance to the source 
 is set as 10 kpc.}
\label{f9}
\end{center}
\end{figure}

\begin{figure}[hbtp]
\begin{center}
\includegraphics[width=0.45\linewidth]{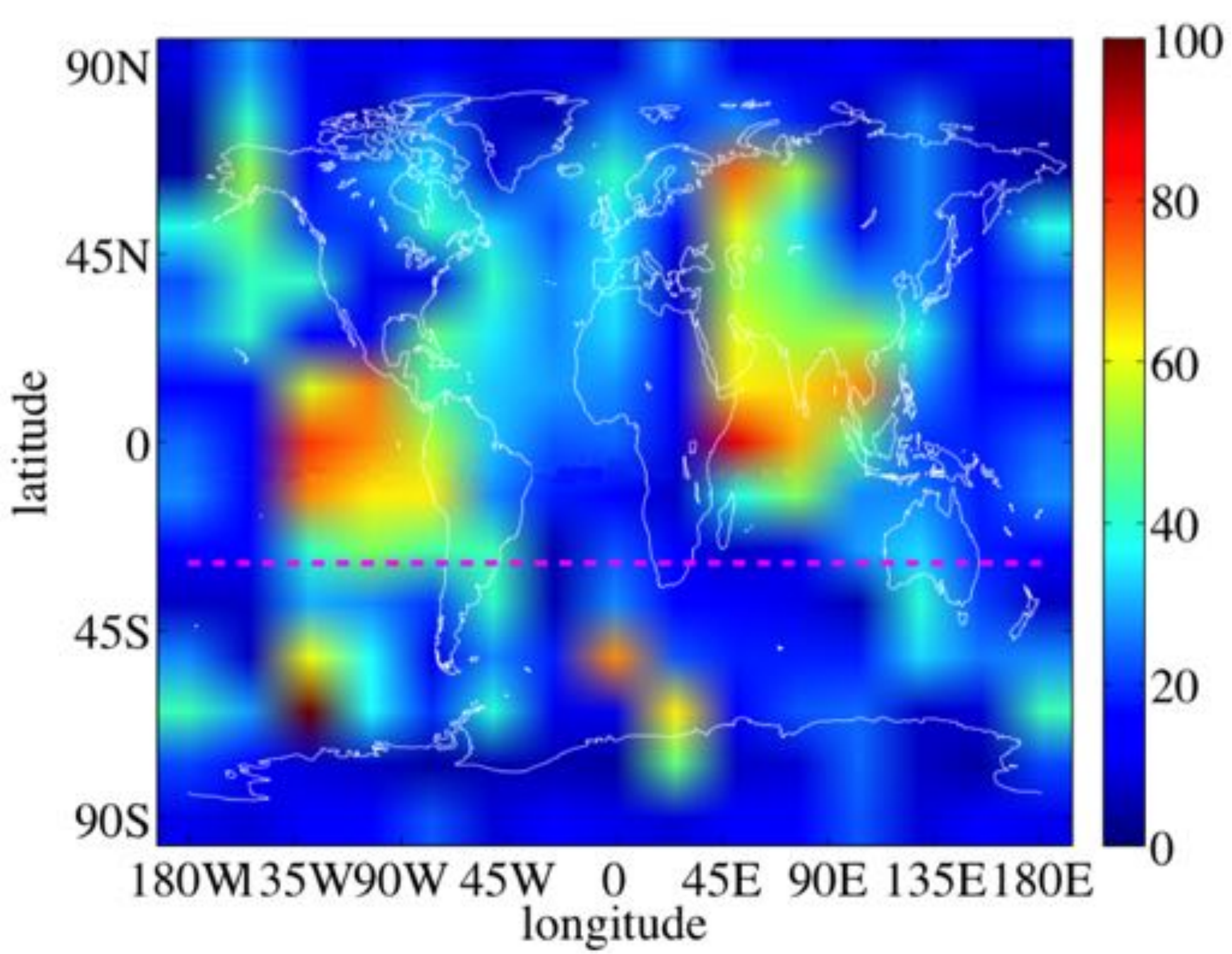}
\includegraphics[width=0.44\linewidth]{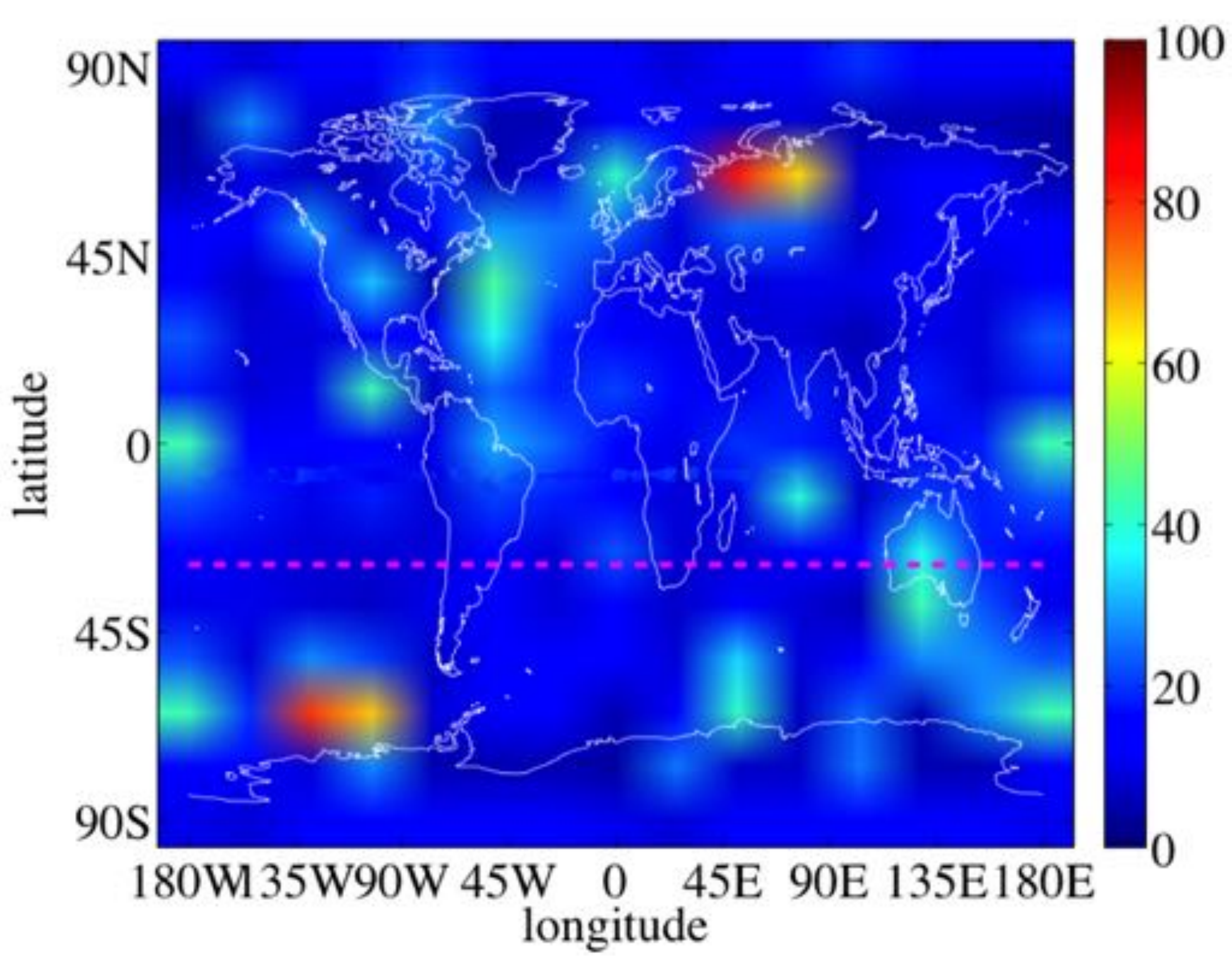}
\caption{Same as Figure \ref{f8} but for 2D MHD models of 
the {\tt TK11} waveforms (models B12X5$\beta$01 (left panel)
 and B12X5$\beta$10 (right panel). 
Note in this panel that the distance to the source 
 is set as 50 kpc.}
\label{f10}
\end{center}
\end{figure}

\subsection{Position reconstruction}
Based on the Monte Carlo simulations using the {\tt RIDGE} (e.g., 
Appendix B),  
we discuss the signal reconstruction of the sky location in this section. 

Following the method in \cite{klimenko08,klimenko11}, 
we inject all our model waveforms onto the simulated detector data 
streams (with the Gaussian noise) in a wide range of SNRs with the coordinates uniformly
 distributed in the sky. These signals are injected 20 times every
 3 s, 0.2 s, and 1 s for the {\tt KK+}, {\tt KTK14}, and {\tt TK11} 
waveforms, respectively. The duration of each injection is different,
 reflecting the different simulation timescale. For each injected event, 
 the skymap ($\boldsymbol{S}(\theta,\phi)$, Appendix B) 
is calculated with the angular resolution of $d \Omega = 4 \times 4$ square
 degrees. To quantify the accuracy of the skymap localization for 
 a single injection, we calculate an error region: total area of 
all pixels in the sky which satisfy the condition $\boldsymbol{S}(\theta,\phi)
 \geq \boldsymbol{S}(\theta_i,\phi_i)$, where 
$\boldsymbol{S}(\theta_i,\phi_i)$ is the injection sky location.
In the following, we choose a threshold of 50\%
 (which contains 50\% of injections) to estimate the error regions 
(namely, we take the 50\% CL error regions).

Figures \ref{f8}, \ref{f9}, and \ref{f10} show the distributions of the error 
regions as a function of latitude ($\theta$) and longitude ($\phi$) 
for the {\tt KK+09}, {\tt KTK14}, and
 {\tt TK11} waveforms, respectively. 
Note in these figures that 
the distance to the source ($d$) is set differently 
as $d = 2$ kpc, 10 kpc, and 50 kpc each for 
Figure \ref{f8}, \ref{f9}, and \ref{f10}, respectively, by which the 
 optimal SNR exceeds 8 (e.g., Table \ref{table1}) bearing in mind that 
 signal detection should be preconditioned for the localization of the 
source. 

In the skymaps, the color scale corresponds to the area of the error regions, 
so that the smaller values (bluish regions) or high values (reddish regions)
 correspond to good or bad accuracy regarding the 
 skymap reconstruction. Comparing Figure \ref{f8} with Figures 
\ref{f9} and \ref{f10},
 one can see from the area of bluish regions that the accuracy is 
 biggest for the {\tt KK+09}
 waveform (Figure \ref{f8}) at $d = 2$ kpc, which is followed roughly in
 order by the {\tt TK11} at $d = 50$ kpc (Figure \ref{f10}) and 
the {\tt KTK14} waveform at $d = 10$ kpc (Figure \ref{f9}).
The accuracy of the skymap reconstruction depends on many ingredients, 
such as the signal strength and duration, waveform morphology, etc.
For our three sets of the signals, the source localization 
(positioned at the optimal distance) turns out to be most 
accurately determined for
 the {\tt KK+09} waveform, which may be rather counter-intuitive 
due to the absence of the distinct peaks in the waveform. However,
 this is mainly due to the longer simulation time compared to the other
 two waveforms. This kind of ambiguity is typical for the least constrained
 unmodeled search and networks with only four spatially separated detectors (e.g., 
 \cite{klimenko11}). 

For the {\tt KK+} waveforms, the distribution and the area of the 
error regions are almost similar 
between the 3D model without or with rotation or either seen from equator 
 or pole.  As mentioned, this is due to the assumed initial small rotation 
rate. The region, which we call as {\it hot spots}: the source localization 
there is not good (colored by red or yellow 
in the skymap), is confined in small clusters in this case. On the 
 other hand, the hot spots are distributed over a large area and split into
 more smaller clusters in the {\tt KTK14} waveforms (Figure \ref{f9}).


The {\tt TK11} signals (Figure \ref{f10})
from 2D MHD models were assumed to be positioned at $d = 50$ kpc and 
 seen from the equator of the source. Comparing the left (model
 B12X5$\beta$01) with right panel (model
 B12X5$\beta$10),
 the area of the error regions is shown to be smaller 
 for the model with the larger initial 
 rotation rates (right panel). This feature is clearly seen in the rest of 
 the 2D models with rapid rotation. 

In this study, the number of the examined multi-D models is rather limited 
(17) and it is truly far from comprehensive. 
 At this stage, we are only able to discuss how well we can 
 localize the GW signals for the limited set of the CCSN models. 
To seek for some systematic trend, we summarize in Table \ref{table2} 
the range of the solid 
angle within which the (given) position reconstruction can be done. 
 It is shown in the table that the accuracy of the sky position 
reconstruction is generally higher for models with rapid rotation 
(e.g., model R3 of {\tt KTK14} 
 and models from {\tt TK11}) than for the non-rotating models (e.g., 
model R0 of {\tt KTK14} and models from {\tt KK+9}). 
Finally we point out that the
 configuration of the H-L-V-K network is fine because the 
sky position of the Galactic center (horizontal
 yellow line in the skymap) does not always coincide with the hot 
 spots as shown in Figures \ref{f8} - \ref{f10}.

\begin{table}[hbt]
\caption{Fraction of the solid angle relative to the whole sky ($4\pi$),
 within which the position reconstruction can be done by each of the 
 angular resolution ($ d\Omega = 5 \times 5, 10 \times 10$ square degrees,
 and so on).
\label{table2}}
\begin{ruledtabular}
\begin{tabular}{lrrrrr}
\textrm{Model}&
\textrm{5 deg}&
\textrm{10 deg}&
\textrm{20 deg}&
\textrm{25 deg}&
\textrm{30 deg}  \\
\colrule
 KK+9 Ae & 0.2304 & 0.3926 & 0.5832 & 0.6472 & 0.6899\\
 KK+9 Ap & 0.3414 & 0.5249 & 0.7084 & 0.7468 & 0.7952\\\hline
 KTK14 R0e & 0.6558 & 0.8250 & 0.9289 & 0.9488 & 0.9545\\
 KTK14 R0p & 0.5533 & 0.7411 & 0.8734 & 0.9004 & 0.9232\\
 KTK14 R2e & 0.1422 & 0.3272 & 0.5306 & 0.5832 & 0.6430\\
 KTK14 R2p & 0.4893 & 0.7212 & 0.8535 & 0.8962 & 0.9260\\
 KTK14 R3e & 0.6714 & 0.8492 & 0.9644 & 0.9844 & 0.9957\\
 KTK14 R3p & 0.6330 & 0.8805 & 0.9758 & 0.9872 & 0.9986\\ \hline
 TK11 B12X1$\beta$01 & 0.5007 & 0.7112 & 0.8321 & 0.8606 & 0.8834\\
 TK11 B12X1$\beta$10 & 0.4979 & 0.7084 & 0.8450 & 0.8691 & 0.8890\\
\end{tabular}
\end{ruledtabular}
 \end{table}


\begin{figure}[hbtp]
\begin{center}
\includegraphics[width=0.5\linewidth]{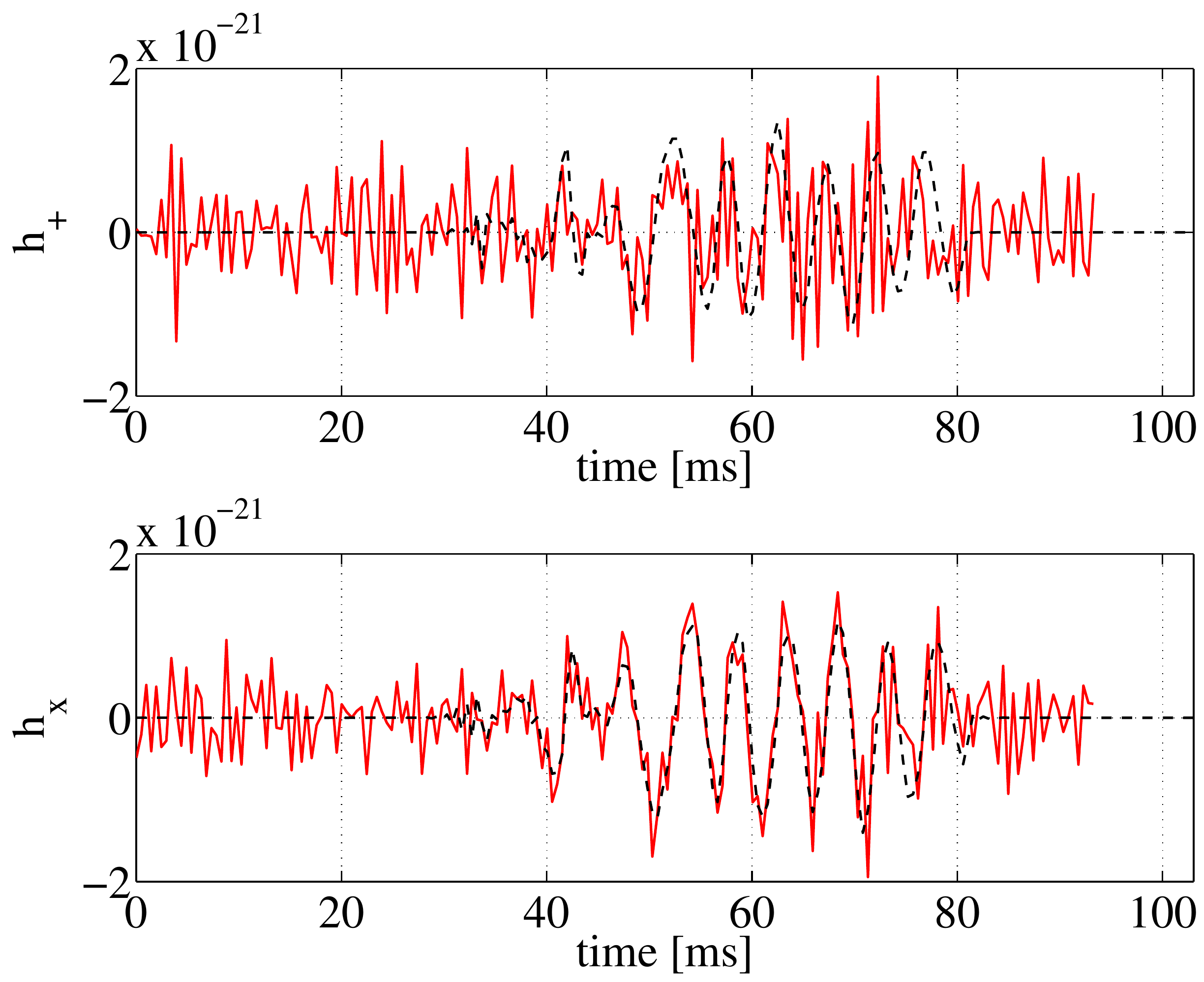}
\includegraphics[width=0.45\linewidth]{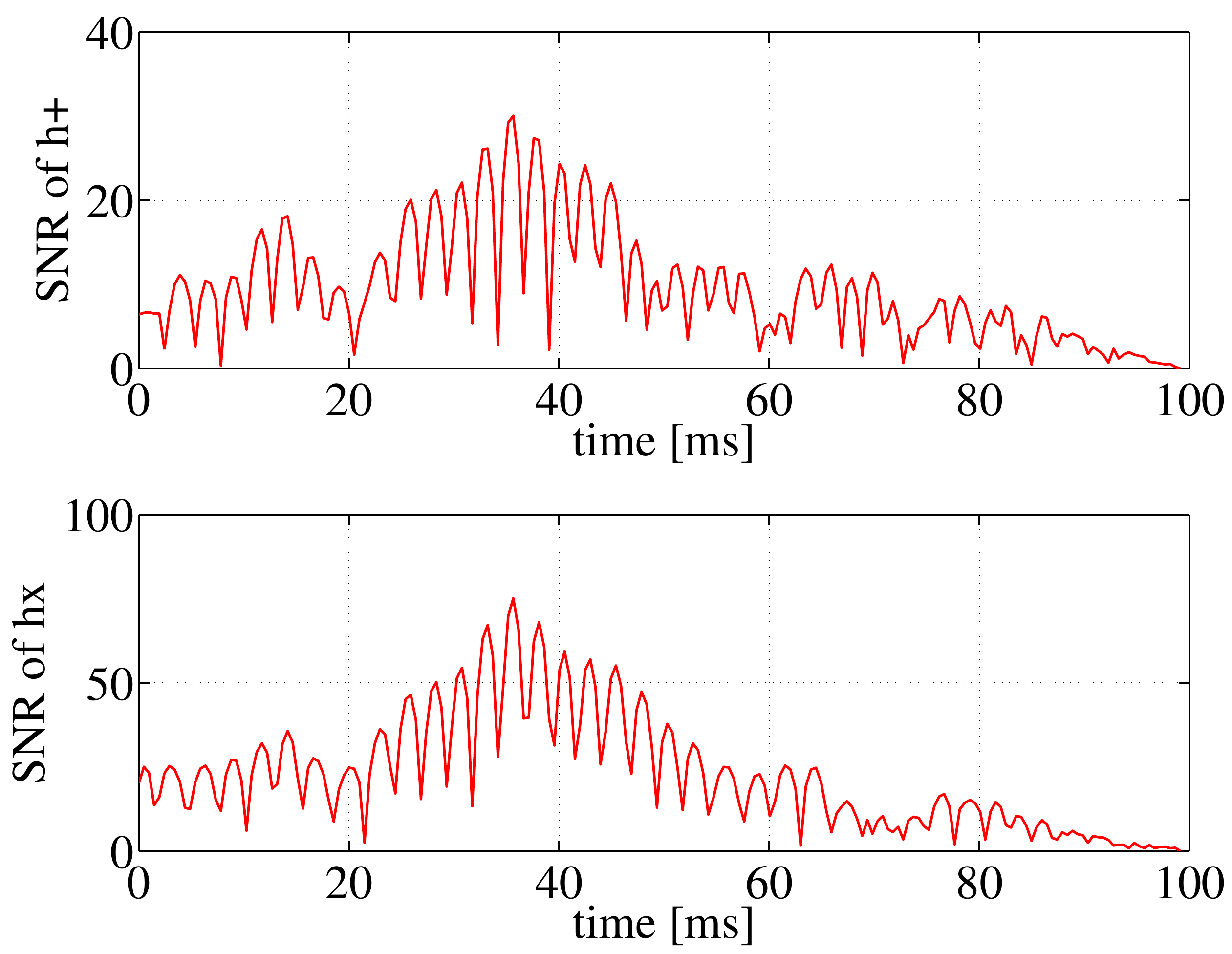}
\includegraphics[width=0.5\linewidth]{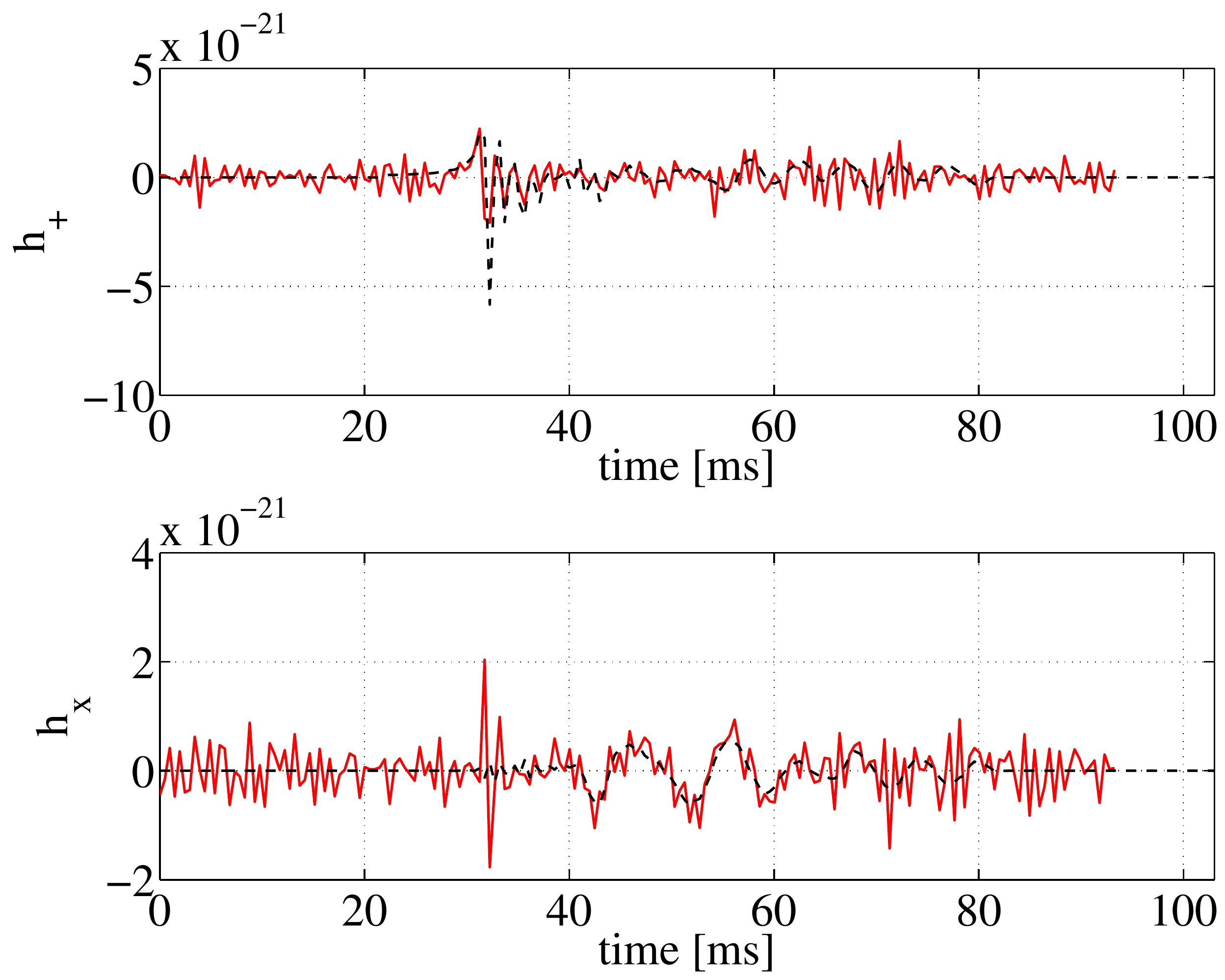}
\includegraphics[width=0.45\linewidth]{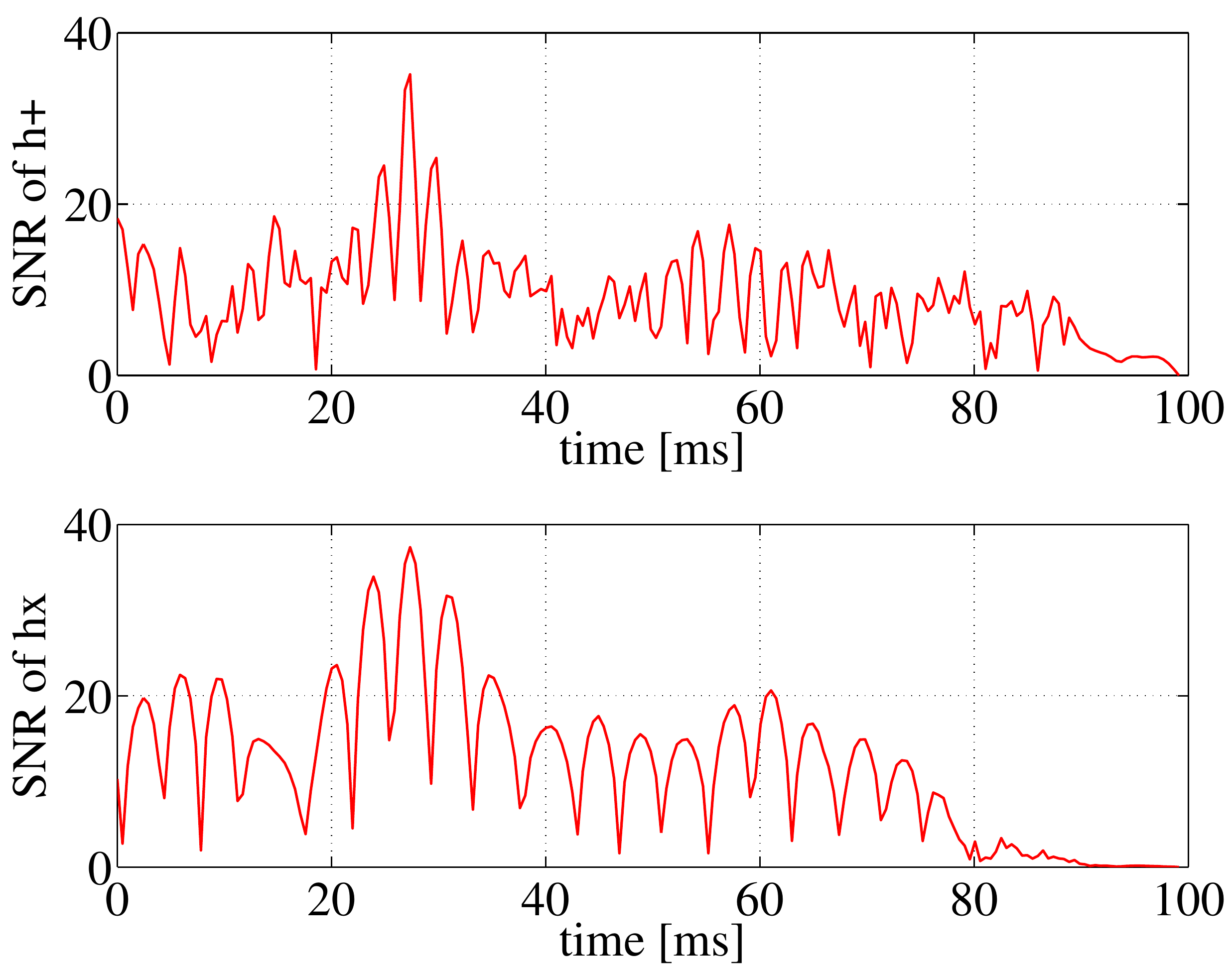}
\caption{Left panels show the reconstructed $h_{+}$ ({\it upper} and $h_{\times}$ 
({\it lower}) waveforms (red lines) of model R3p (top two panels) and R3e 
(bottom two panels) ($d$ = 10 kpc). The black line corresponds to the injected (original) waveform. Right panels show the corresponding output
of matched filtering for the reconstructed signals (compare with the left 
panels), in which the output is the SNR. }
\label{f11}
\end{center}
\end{figure}

\begin{figure}[hbtp]
\begin{center}
\includegraphics[width=0.44\linewidth]{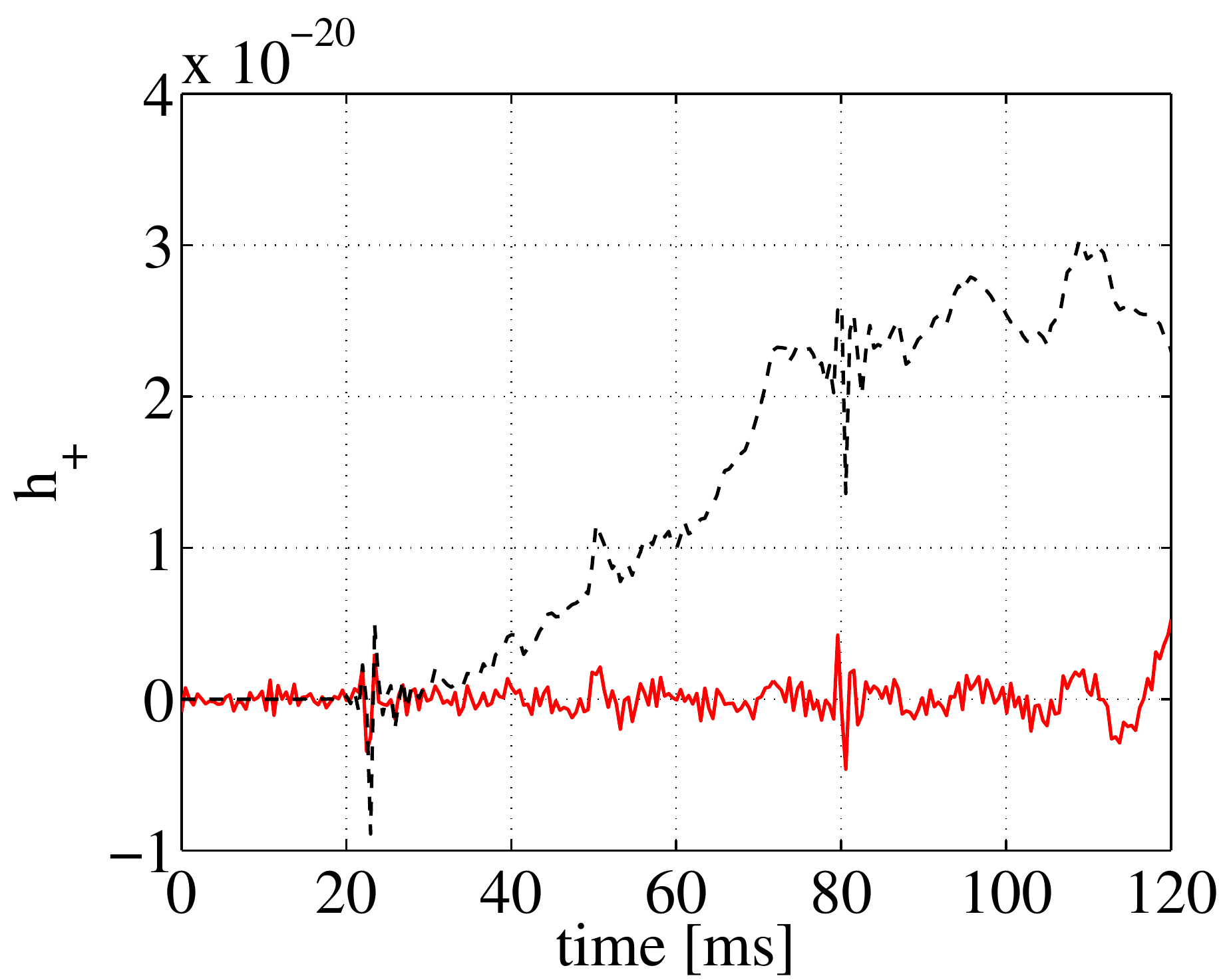}
\includegraphics[width=0.4\linewidth]{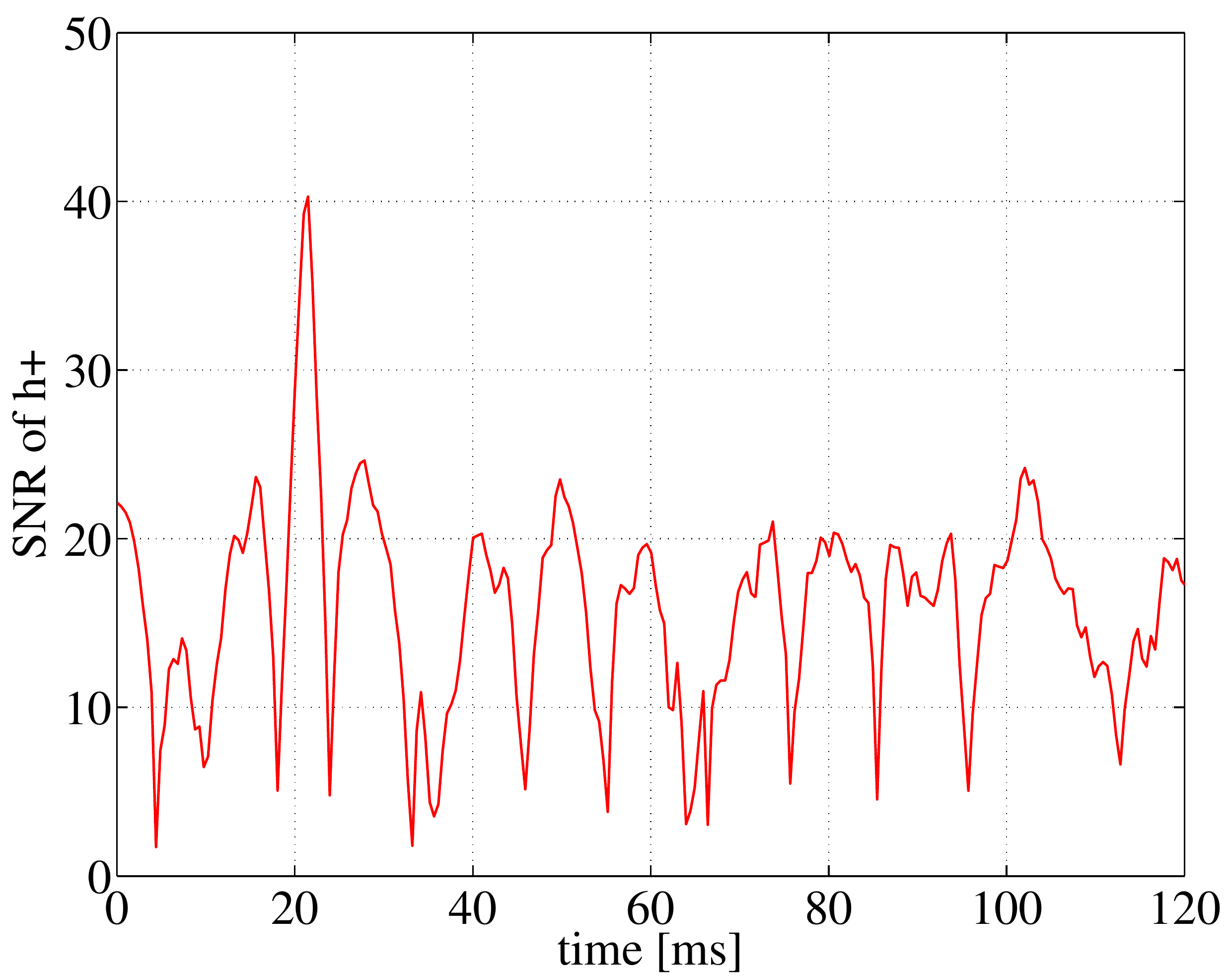}
\caption{Same as Figure \ref{f11} but for model B12X1$\beta$10 from 
 the {\tt TK11} catalogue. }
\label{f12}
\end{center}
\end{figure}

\subsection{Waveform reconstruction}

Figures \ref{f11} and \ref{f12} show comparison between the 
reconstructed (red lines) and the original waveforms of the 
two representative models with rapid rotation (black lines,
 for the $+$ mode ({\it upper}) and for the $\times$ mode ({\it
 lower}), respectively). Seen from the pole of model R3 (i.e., model R3p) 
 in the {\tt KTK14} catalogue (top two pair panels of Figure \ref{f11}), 
the matched-filtering SNR of the reconstructed $h_{+}, h_{\times}$ reach to $\sim 30$ and 60
 ($d = 10$kpc), respectively. Seen from the equator (bottom two pair panels
 of Figure \ref{f11}), the SNR of the 
(conventional) bounce GW signal from rapidly rotating collapse and bounce 
($h_{+}$, upper part of the bottom two panels) 
reaches $\sim 35 -38$.
Remembering that any coherent network analysis 
has bias and one cannot completely correct the bias in general, this result 
is not as bad compared with the ideal SNRs of  
$\sim$ 60  for an optimally oriented and optimally 
located source (Table \ref{table1}).

From Figure \ref{f12}, one can also see that the bounce signal is 
well reconstructed by the {\tt RIDGE} pipeline (with the matched-filtering
 SNR reaching to 40 (bottom panel) for the 2D MHD model from
 the {\tt TK11} catalogue. It is also shown that the 
 quasi-monotonically increasing component in the original waveform 
(black line after $\sim 20$ ms postbounce) disappears in the reconstructed 
 waveform (red line, top panel). This again reflects that 
such low-frequency component is hard to detect due to seismic 
noises. Regarding a 3D (non-rotating) model (not shown) from the 
{\tt KK+} catalogue, the detection efficiency of the high-frequency component 
(with the variation timescale of ms) is not high 
compared to those for Figures \ref{f11} and \ref{f12} 
because of the absence of the distinct waveform morphology. On the other hand, 
 the shape of the waveform with the variation timescale of 50 - 100 ms (which 
 closely corresponds to the SASI modulation) is captured to some extent,
 by which the matched-filtering SNR is in the range
 from $\sim 5$ to 10 for the nearby source ($d = 2$ kpc in this case).


\begin{figure}[htbp]
\begin{center}
\includegraphics[width=0.485\linewidth]{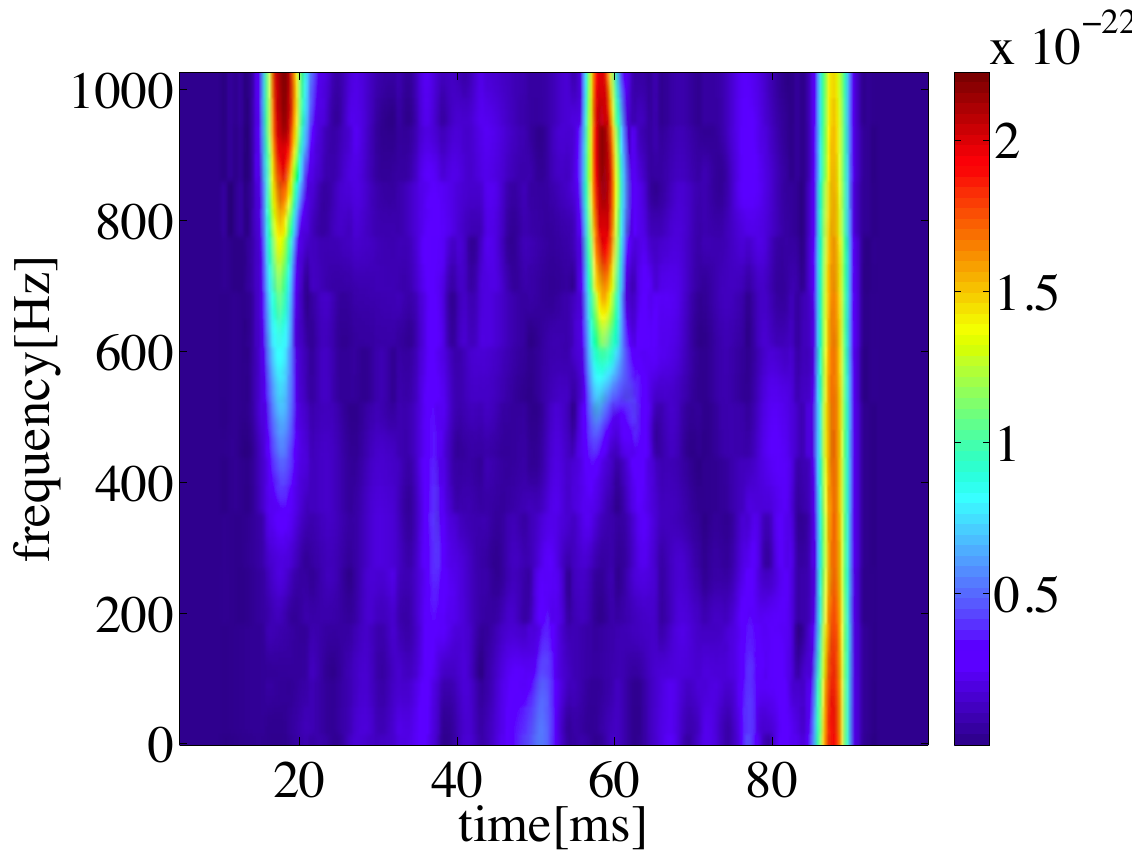}
\includegraphics[width=0.45\linewidth]{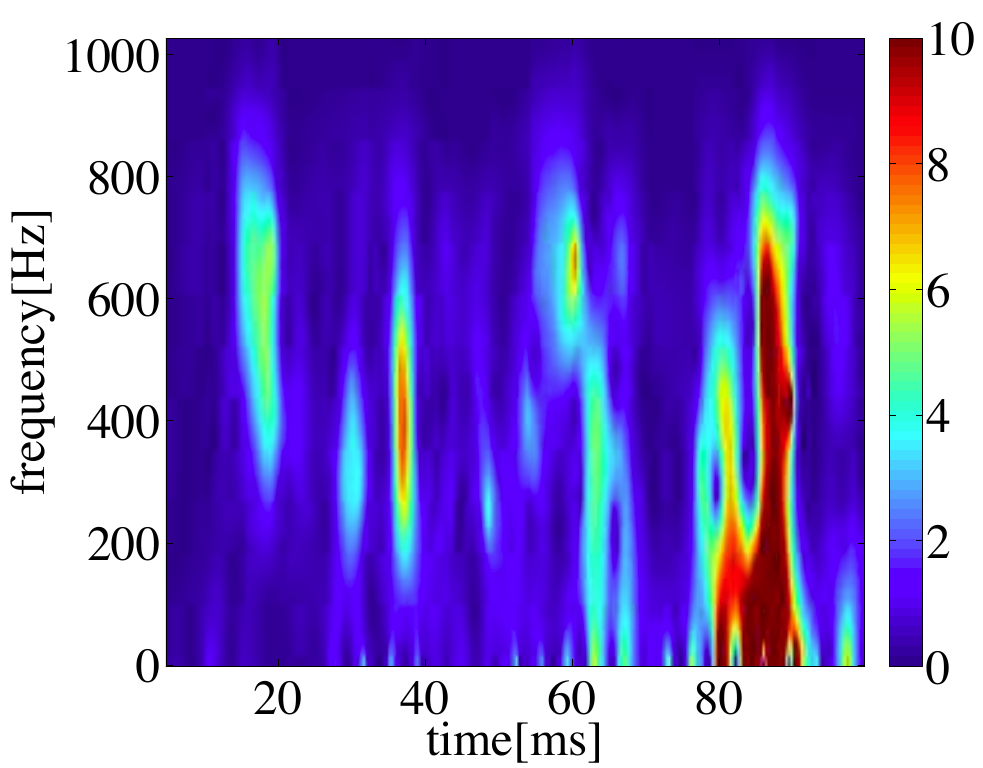}
\includegraphics[width=0.45\linewidth]{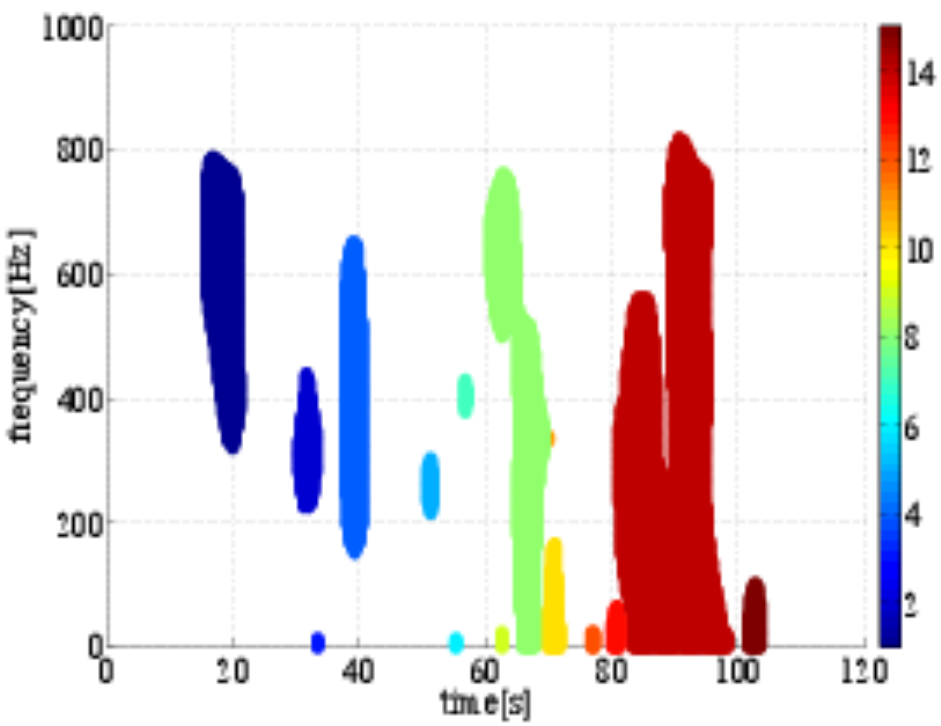}
\includegraphics[width=0.45\linewidth]{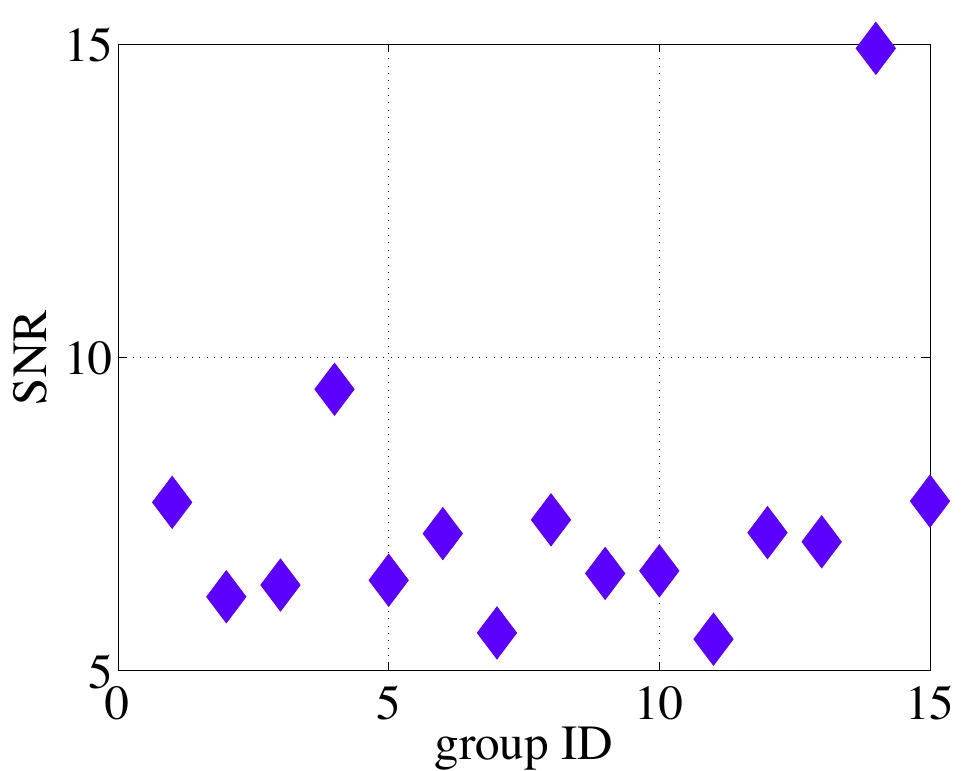}
\caption{Top left panel is the GW spectrogram of the injected signal 
(theoretical prediction) showing the amplitude at $d=10$ kpc
 as a function of simulation time (horizontal axis) and frequency (vertical axis) for model TK11B12X1$\beta$01 from the {\tt TK11} catalogue. Same as the 
 top left panel, but the top right panel is the spectrogram of the reconstructed waveform, in which the color scale represents the time-frequency SNR 
(${\rm SNR}_{\rm TF}$, see text for the definition). 
The bottom left panel shows the groups of the excess with the same ${\rm SNR}_{\rm TF}$ with the color
 scale representing the grouping identification (ID) number.
 The bottom right panel shows the grouping SNR (see text) as a function of the group identification (ID) number.}
\label{f14}
\end{center}
\end{figure}
\section{Inferring the postbounce hydrodynamic episodes}
By performing the GW spectrogram analysis (e.g., 
\cite{EMuller04,Murphy09,EMuller12}), we move on 
to discuss what information we could extract about the hydrodynamic episode 
in the postbounce core. 

For the sake of simplicity, we first focus on the waveform of
 a 2D MHD model from the {\tt TK11} 
catalogue. The top left panel of Figure \ref{f14} is the 
 spectrogram of the original waveform. Colored by red, there are three
 distinct excess at the simulation time $t_{\rm sim} \sim$ 20 ms, 60 ms, and 
90 ms, respectively. The first excess peaking around 1 kHz ($t_{\rm sim} \sim 
20$ ms) comes from the bounce signal, which is followed by the excess 
 at $t_{\rm sim} \sim 60$ ms (we call this as the second excess) 
due to a relatively large ring-down. The third excess ($t_{\rm sim} \sim 90$ ms) which extends to a lower 
 frequency regime ($\lesssim $ 100 Hz) comes from the second core bounce 
(e.g., Figure \ref{f12})
 and the subsequent formation of the secondary MHD jet 
(see, \cite{Takiwaki11} for more details). 

Here it should be noted that these hydrodynamic features imprinted in the spectrogram can be seen 
also in the spectrogram for the reconstructed signals
 (top right panel of Figure \ref{f14}). Comparing the top left with the top 
 right panel, the high frequency domain (colored by red 
in the top left panel from $\sim$ 800 to 1kHz in each of the above three 
excesses) disappear in the top right panel. This is because these frequency 
domain is out of the highest detectors' sensitivity that 
is limited by quantum noises (e.g., Figure \ref{f2}). In the top right 
panel, the color scale represents the ratio of the GW amplitude with 
both the model prediction and the detector's noise to that with the noise only 
(without the model prediction). We call this 
 quantity as the {\it time-frequency SNR} (${\rm SNR}_{\rm TF}$) because 
it is defined in each pixel of the time-frequency ($dt-df$) domain representing 
  the strength of the signal relative to the noise.
 The pixel resolution taken here is 20ms. In this analysis, each 
time-frequency tile is overlapped except for 1 pixel, the group SNR of each 
time-frequency tile is averaged over the overlapped tiles.

By selecting the pixels with the same time-frequency SNR (exceeding 
 5), we can divide the patchy excesses in the spectrogram into several 
disconnected groups (bottom left panel of Figure \ref{f14}). The color scale 
 of the bottom left panel represents the identification (ID) number of each of the 
 groups (which we call as the grouping ID). By summing up the time-frequency 
SNR in the disconnected area ($S$) (having the same ${\rm SNR}_{\rm TF}$)), 
we define the {\it grouping SNR} (i.e., ${\rm SNR}_{\rm group} \equiv \int_{S} dt\,df\,{\rm SNR}_{\rm TF}$).
 The bottom right panel of 
Figure \ref{f14} shows the ${\rm SNR}_{\rm group}$
as a function of the grouping ID. It should be noted that 
  the identification of 
 distinct clusters in the GW spectrograms presented in this work
 is nothing but a very rough and optimistic estimate because we consider 
 only the idealized Gaussian noise. Effects of realistic noise 
 need to be considered, which is one of the most important tasks 
to be studied as a sequel of this work.

The first excess near at bounce (shown as a blue prolate region at 
$t_{\rm sim} \sim 20$ ms, bottom left panel) 
has the group ID = 1 (blue in the color-scale), which is shown to have 
 the grouping SNR 7.5 (e.g., bottom right panel).
The second and third excess ($t_{\rm sim} \sim 60,\, 90$ ms) has the 
group ID= 4 and 14, the ${\rm SNR}_{\rm group}$ of which is 9.6 and 15, 
 respectively. When we set the detection threshold as 8, the bounce 
 signature is expected to be detectable to 9.4 kpc for the H-L-V-K 
observation, while the (strong) ring-down of the PNS (the second excess)
 and the subsequent bounce and the formation of the recurrent MHD
 jets (the third excess) can be detected to 12 kpc and 19 kpc, respectively.

As one would imagine,  
the spectrogram of the {\tt KK+} waveforms,
 as contrary to the {\tt TK11}/{\tt KTK14} waveforms, does not 
possess clear excess. Since it turns out to be difficult to perform the grouping 
analysis as in Figures \ref{f14}, we limit ourselves to 
focus on the {\tt KTK14} waveforms in the following.

\begin{figure}
\begin{center}
\includegraphics[width=0.35\linewidth]{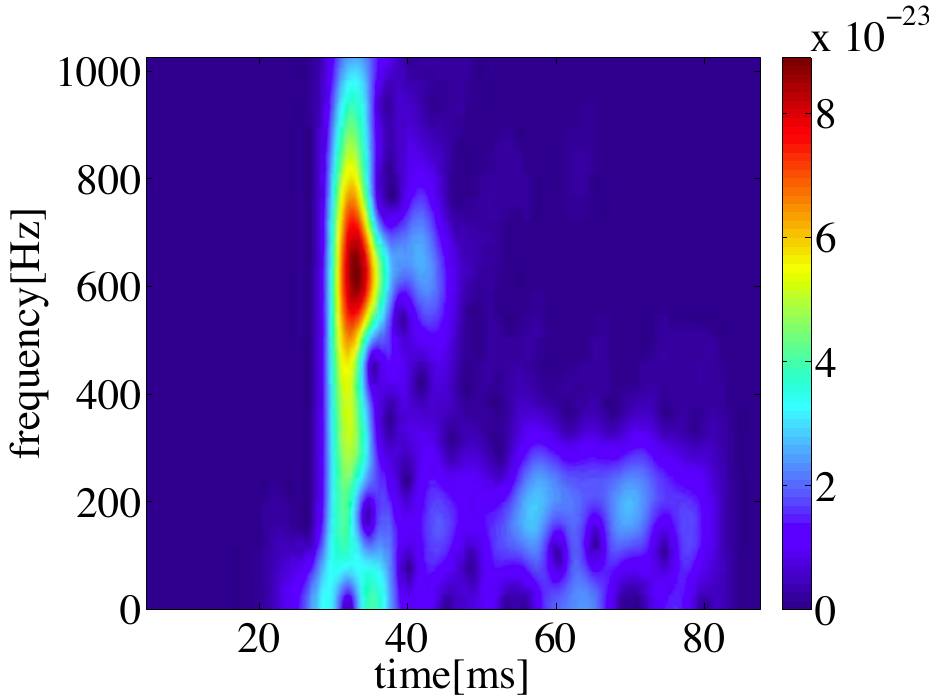}
\includegraphics[width=0.34\linewidth]{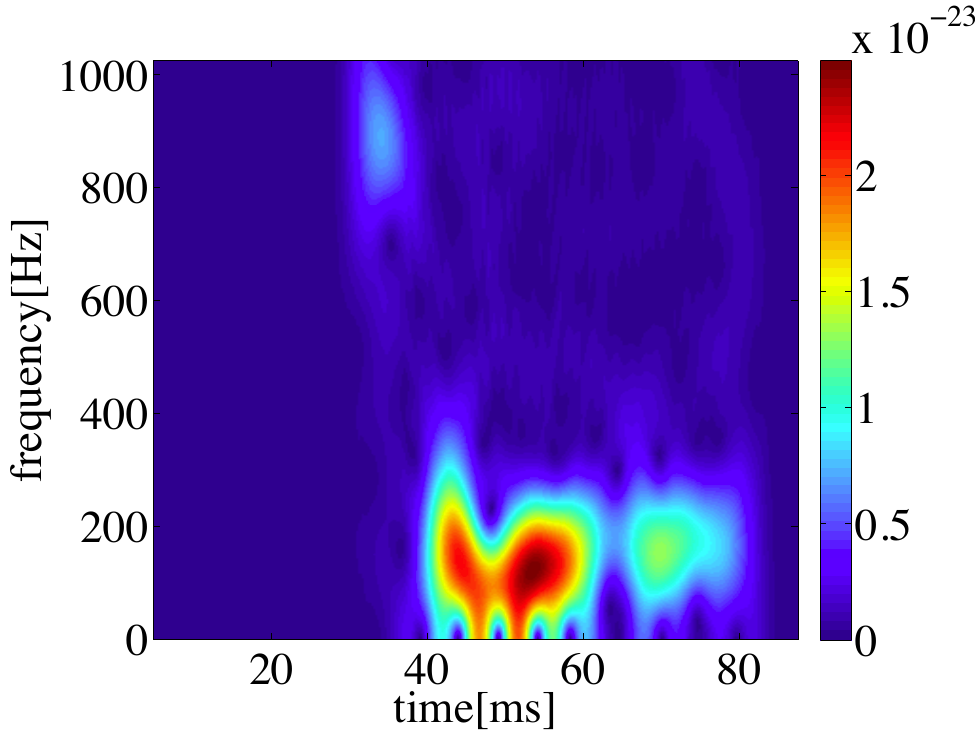}
\includegraphics[width=0.3355\linewidth]{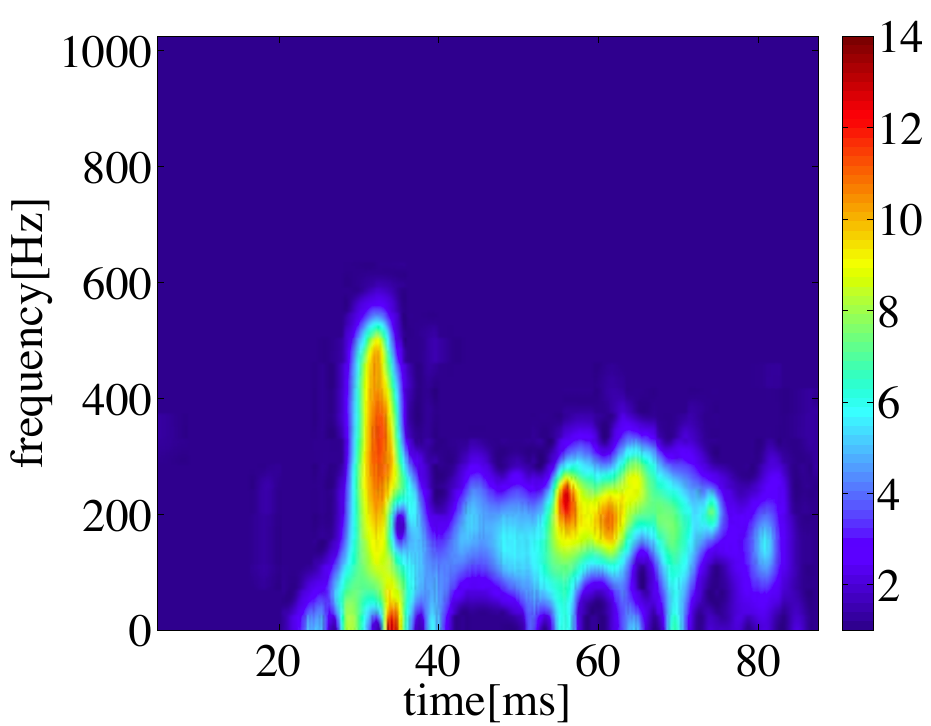}
\includegraphics[width=0.33\linewidth]{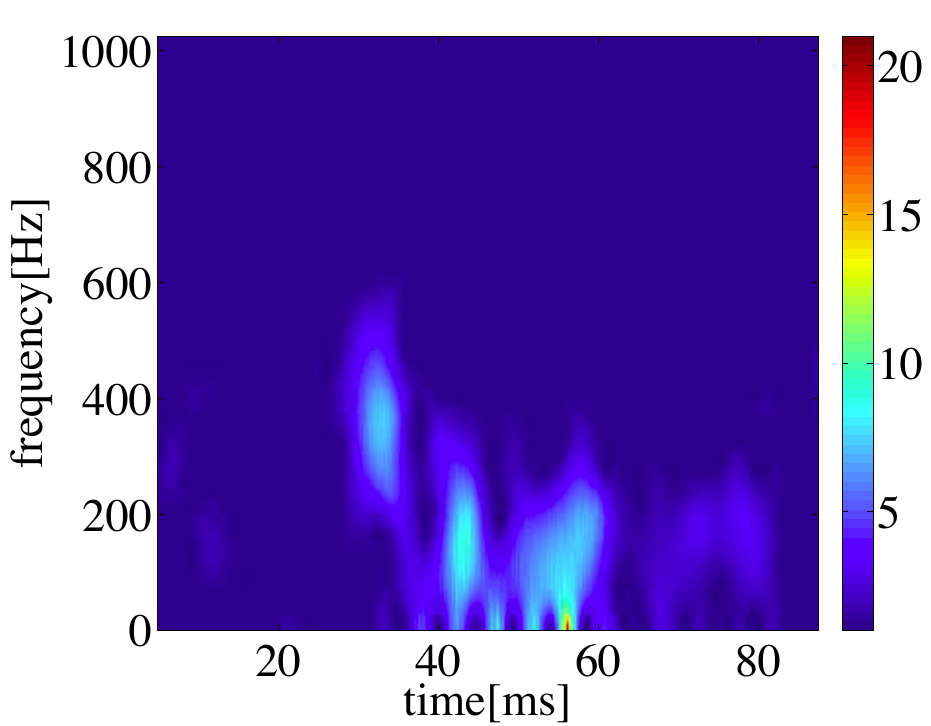}
\includegraphics[width=0.33\linewidth]{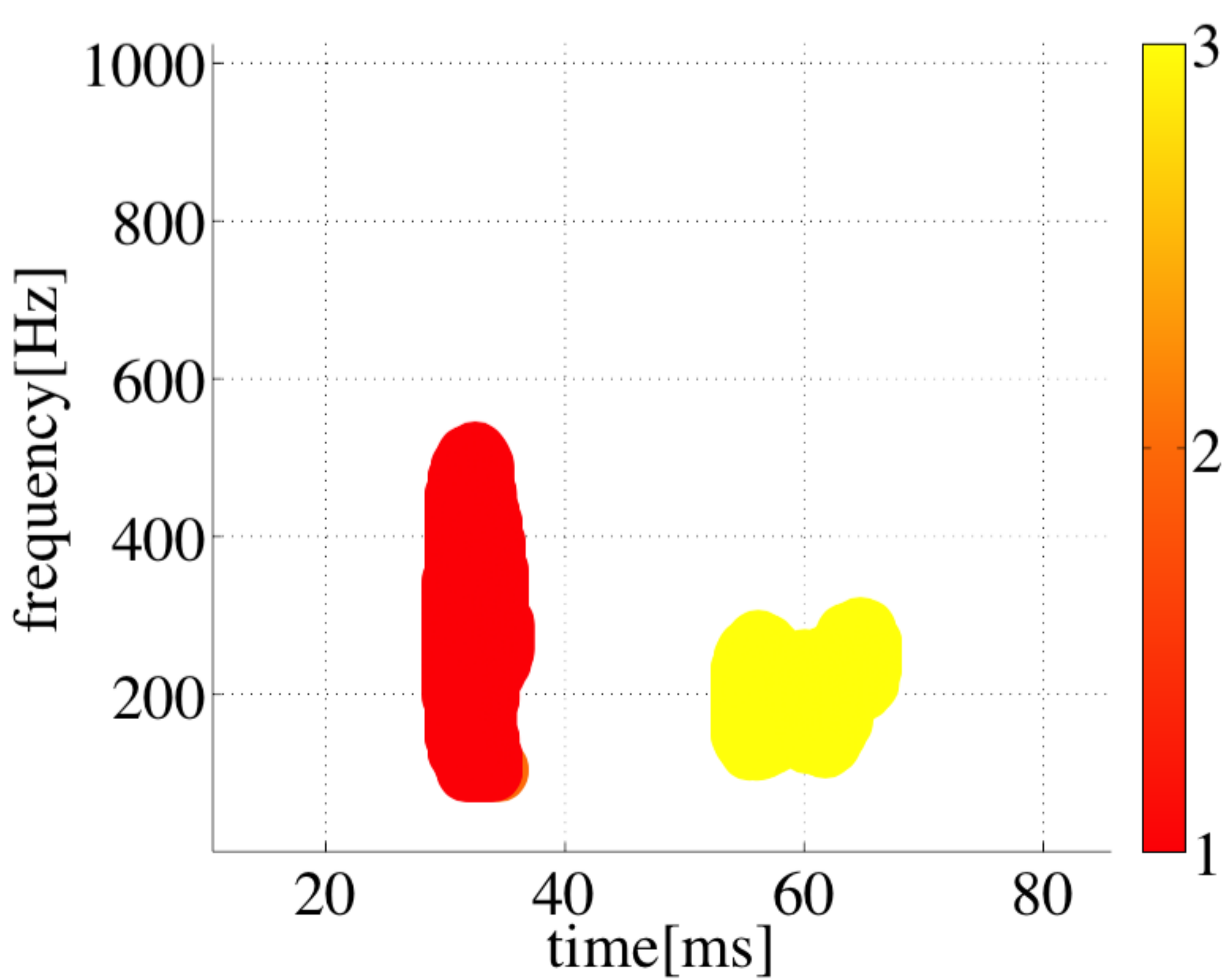}
\includegraphics[width=0.33\linewidth]{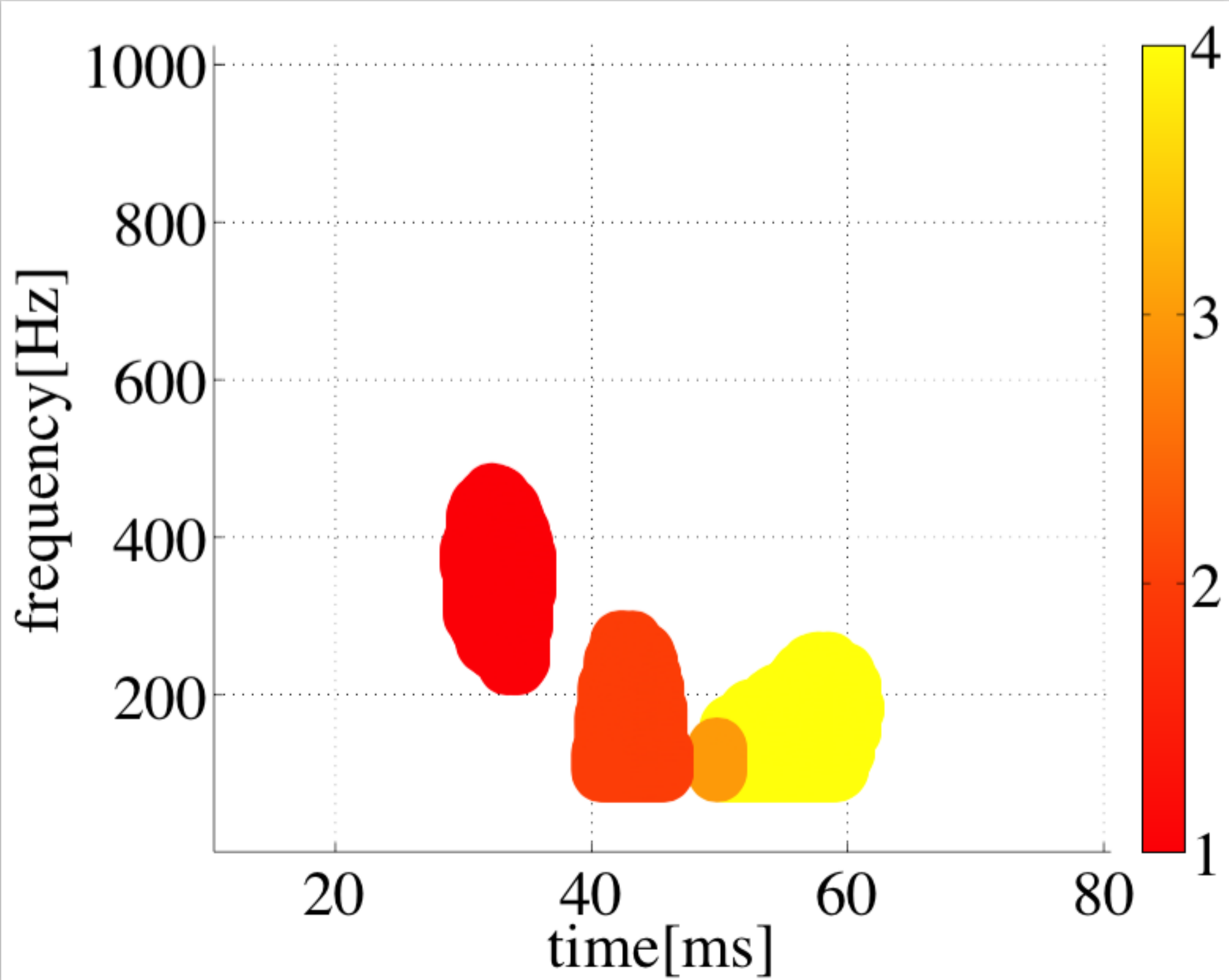}
\includegraphics[width=0.33\linewidth]{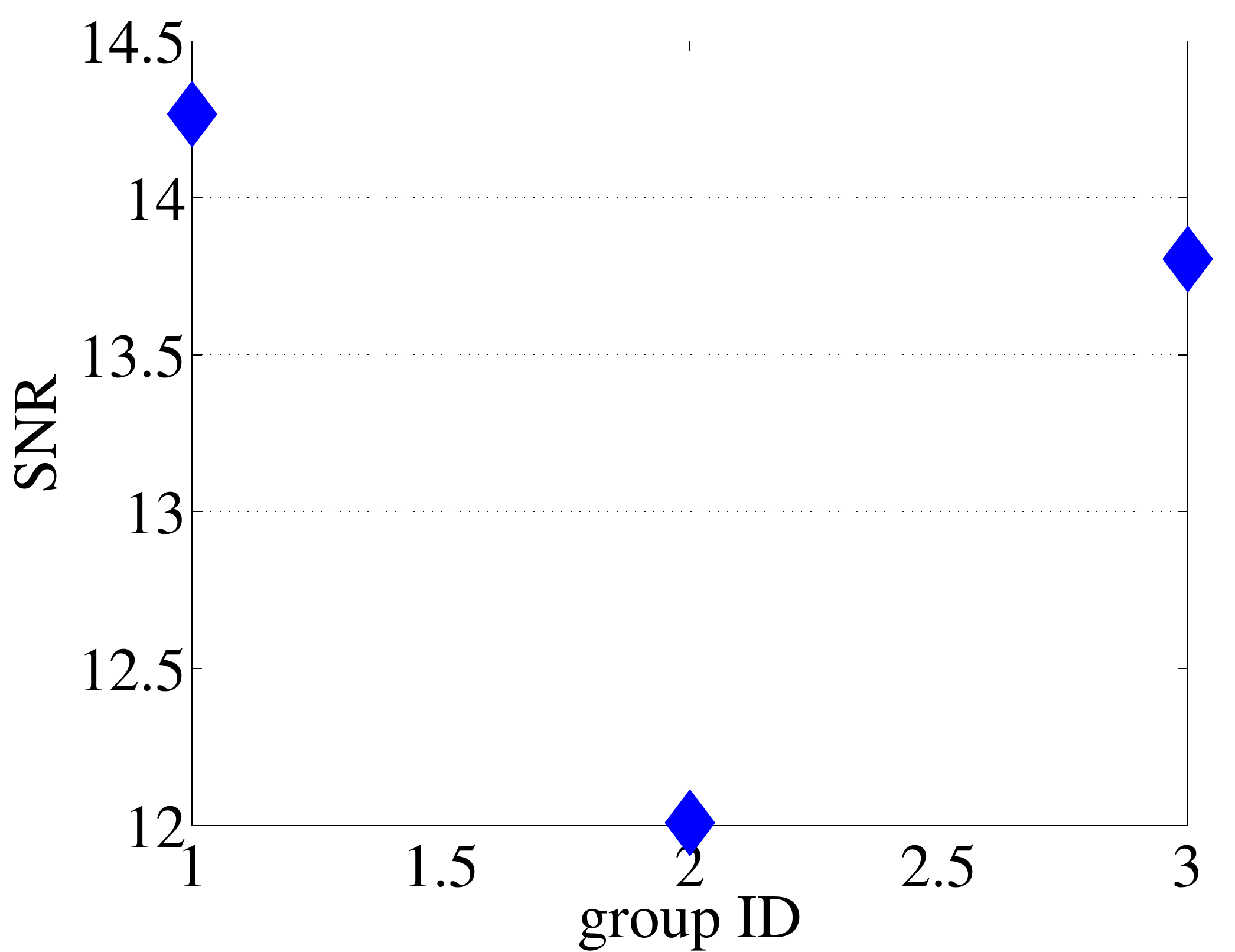}
\includegraphics[width=0.33\linewidth]{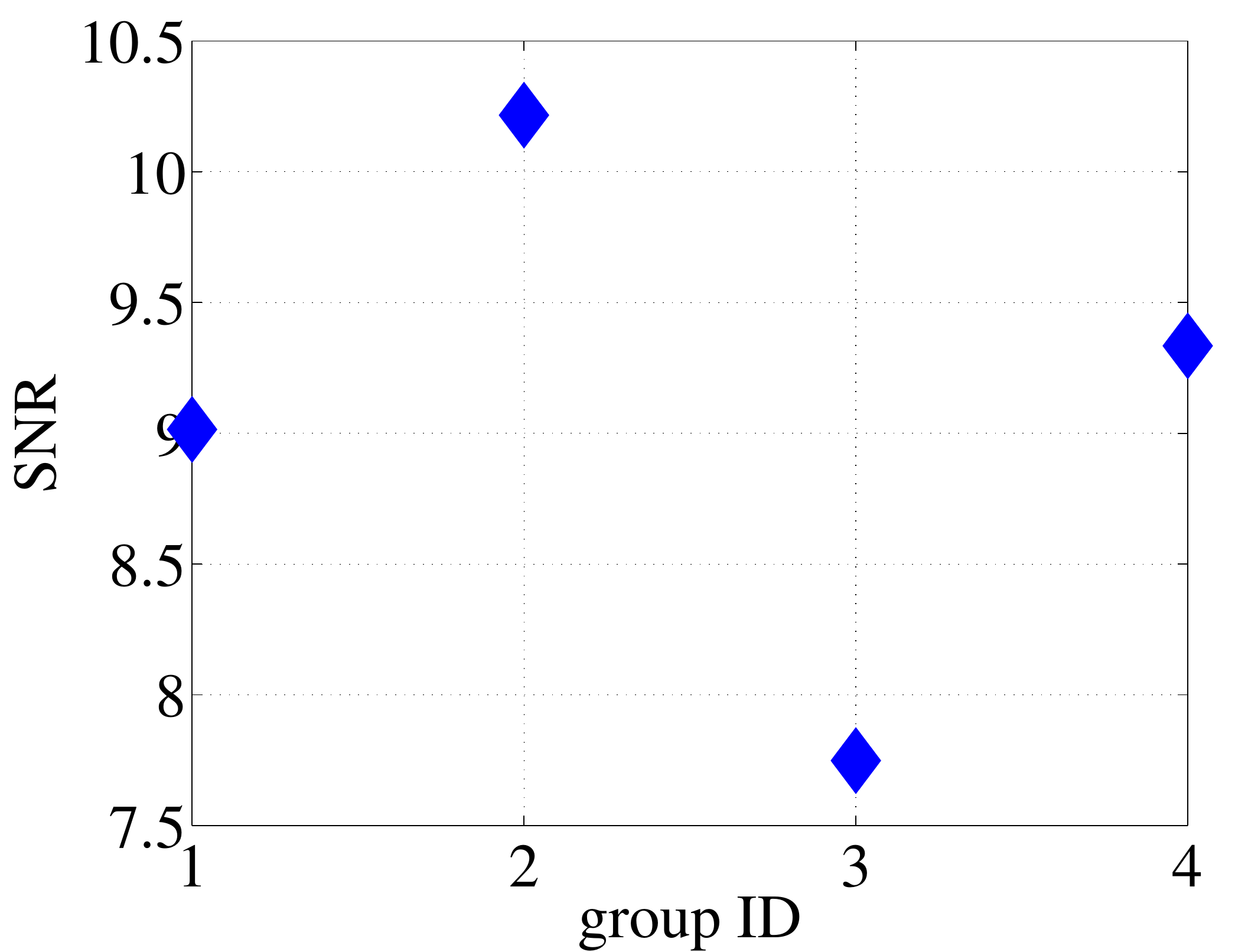}
\caption{Similar to Figure \ref{f14} but for model R3 seen from 
 equator from the 
 {\tt KTK14} catalogue. The left and right panels correspond to the waveform
 with + and $\times$ mode, respectively. 
The first and second column corresponds to the spectrograms of the injected 
 and reconstructed signals, respectively. The third and fourth column show the 
 spectrogram of ${\rm SNR}_{\rm TF}$ and the  ${\rm SNR}_{\rm group}$ with the ID number, respectively.}
\label{f15}
\end{center}
\end{figure}

\begin{figure}
\begin{center}
\includegraphics[width=0.34\linewidth]{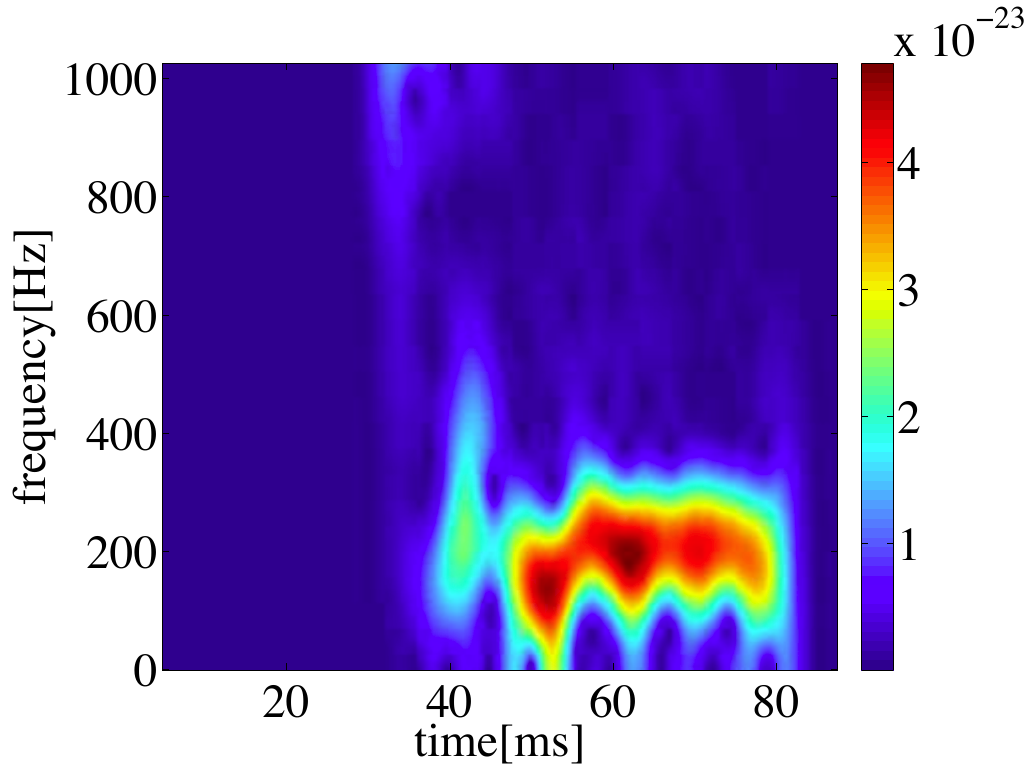}
\includegraphics[width=0.35\linewidth]{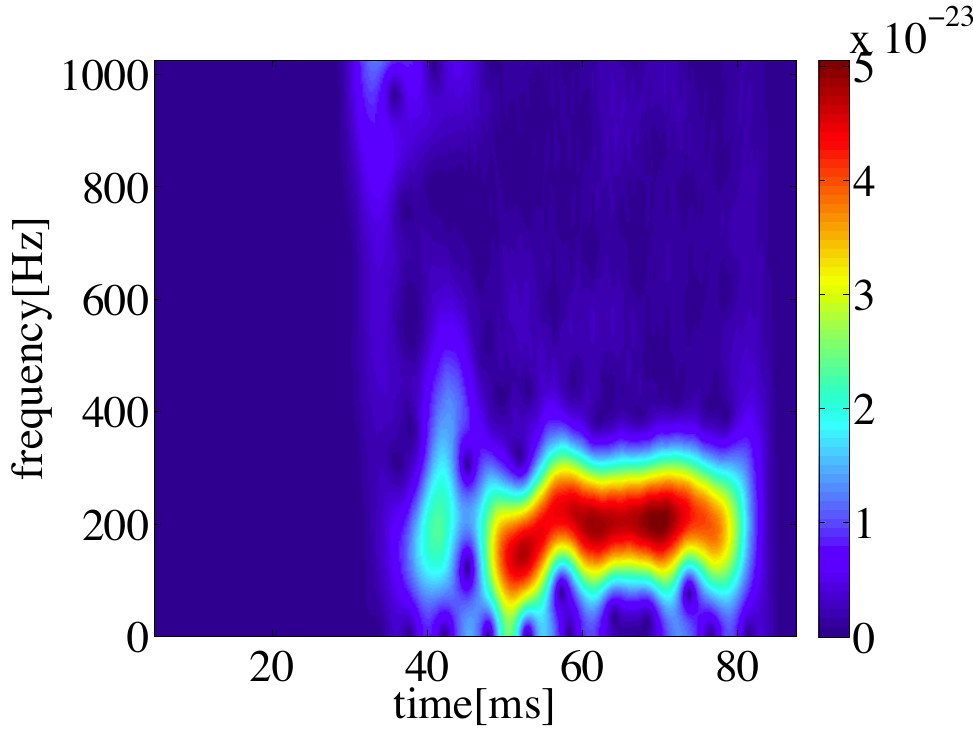}
\includegraphics[width=0.33\linewidth]{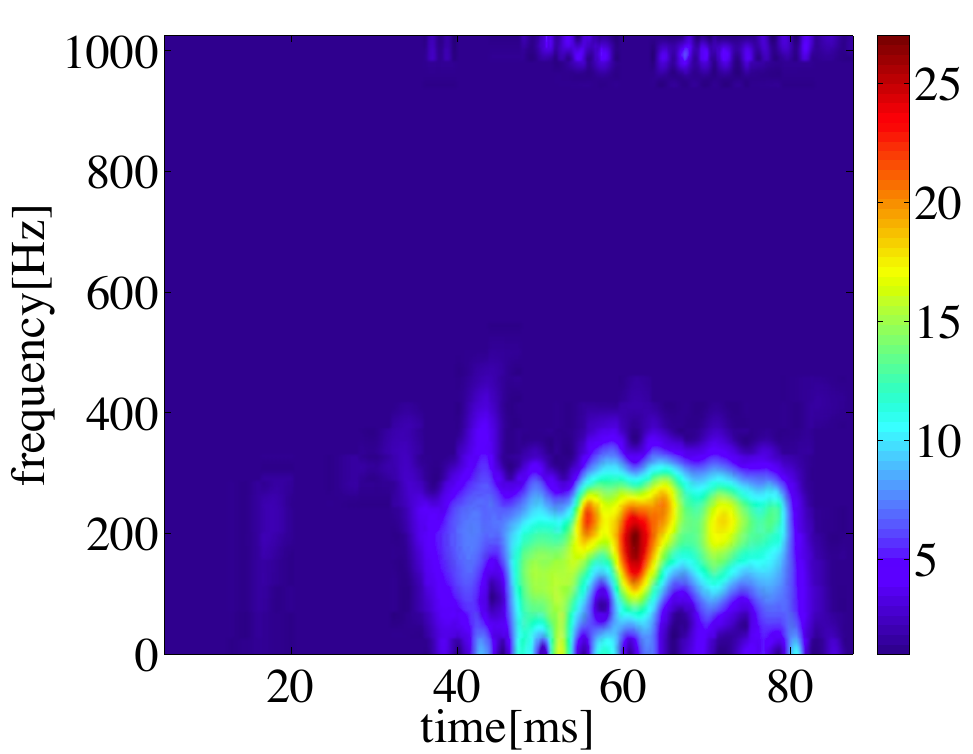}
\includegraphics[width=0.3355\linewidth]{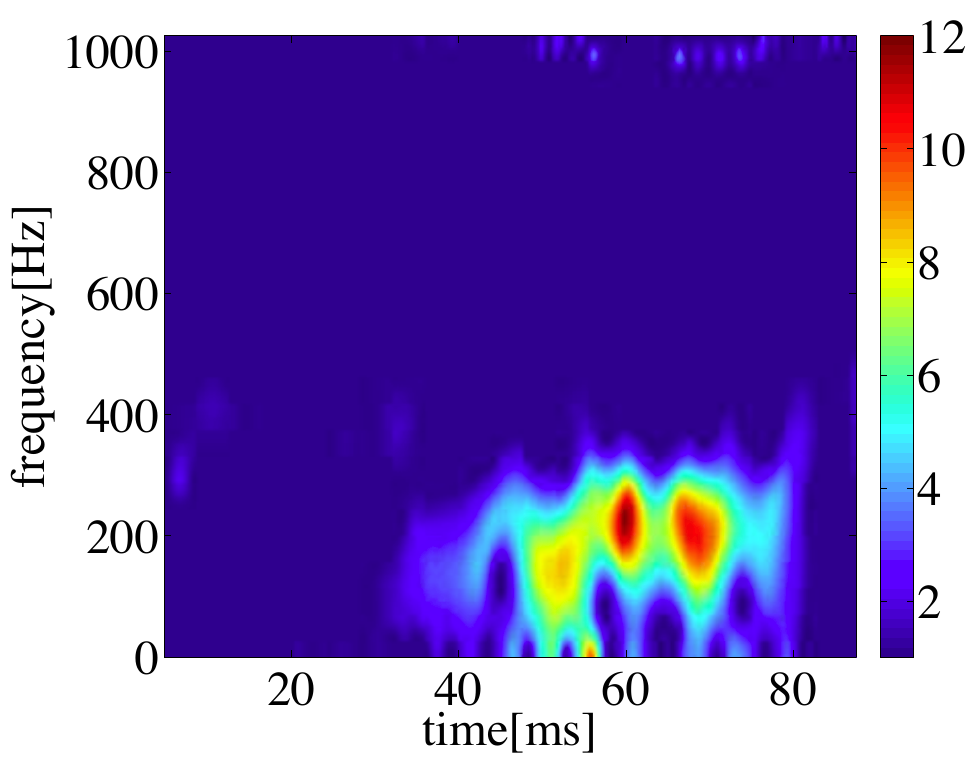}
\includegraphics[width=0.33\linewidth]{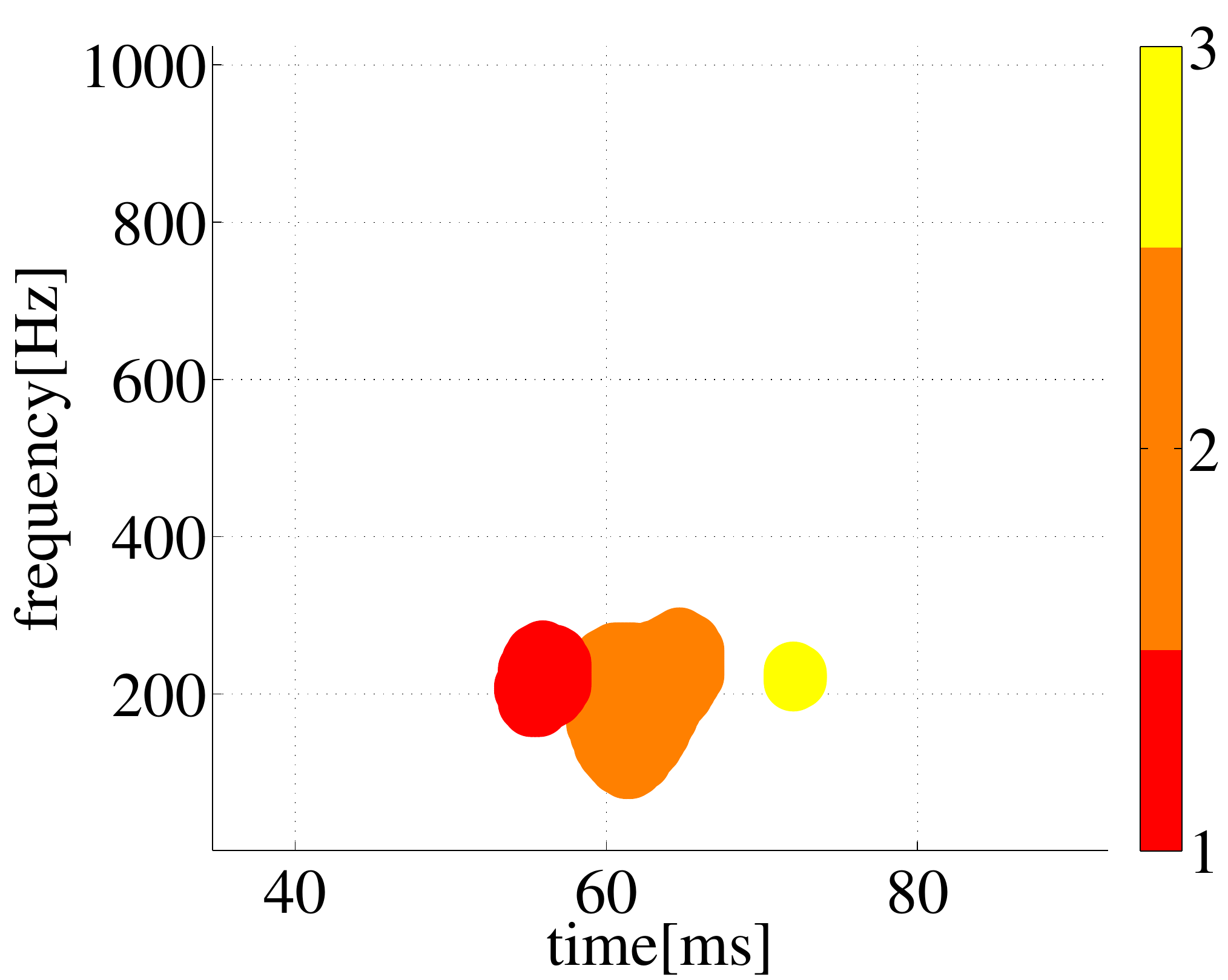}
\includegraphics[width=0.33\linewidth]{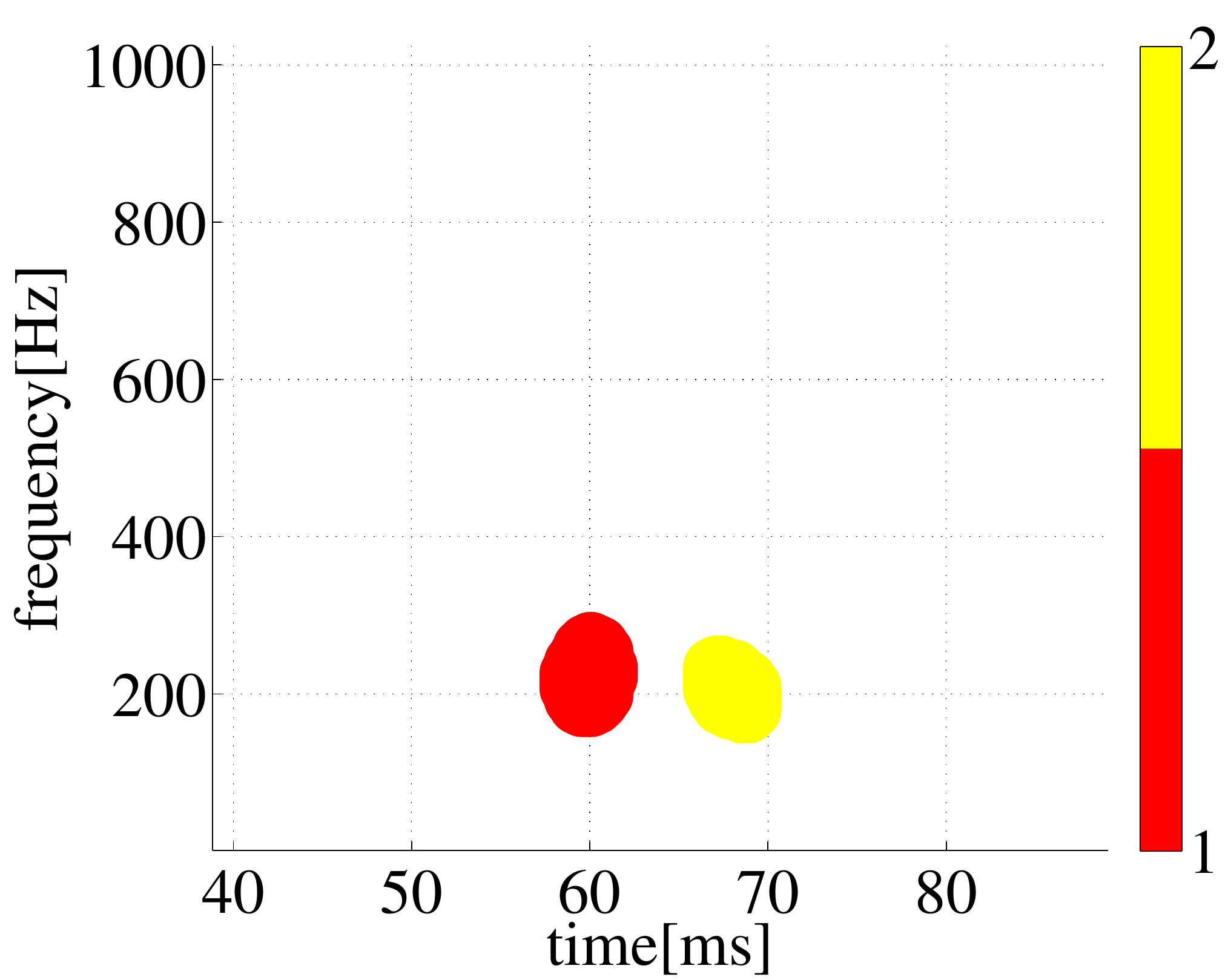}
\includegraphics[width=0.33\linewidth]{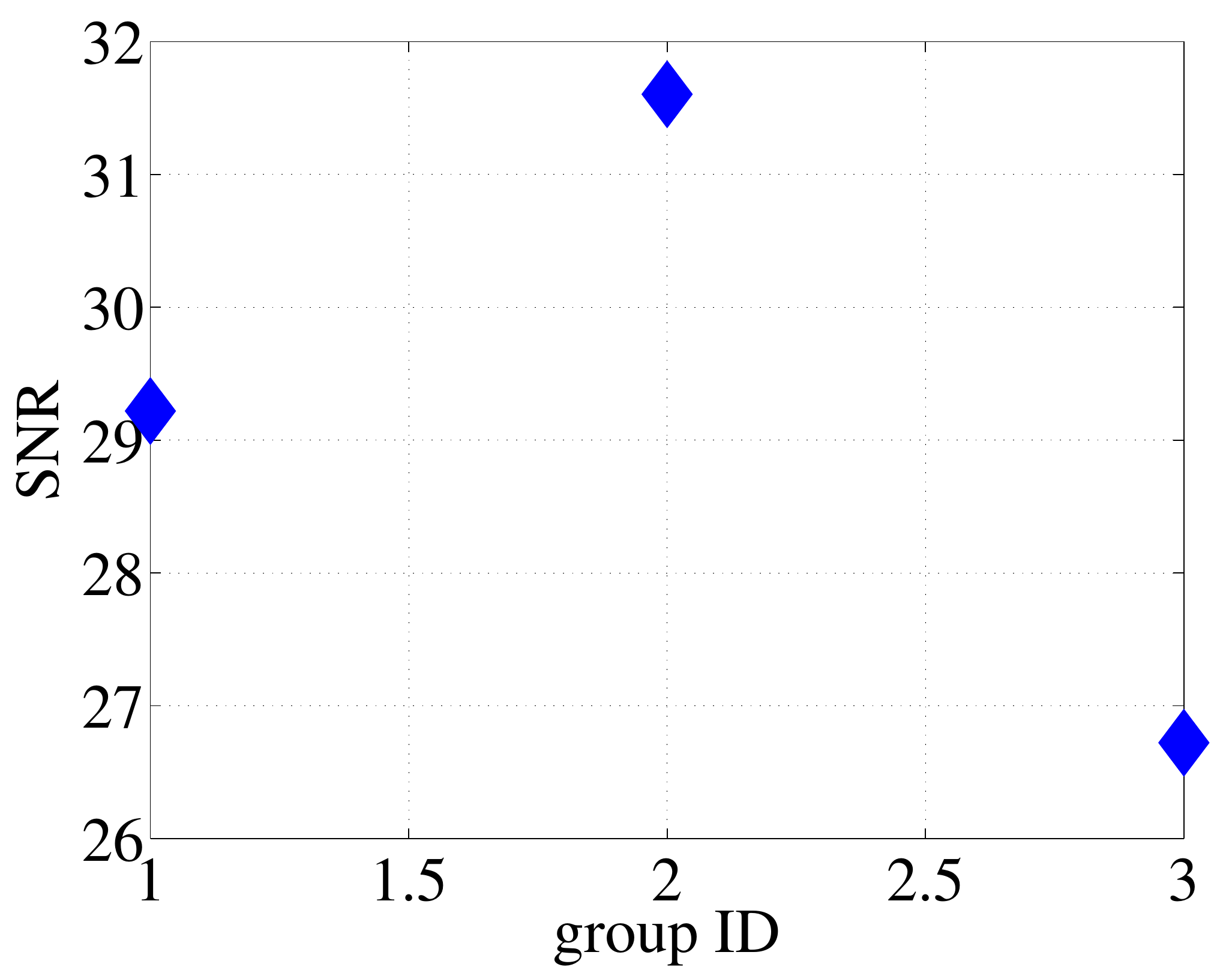}
\includegraphics[width=0.33\linewidth]{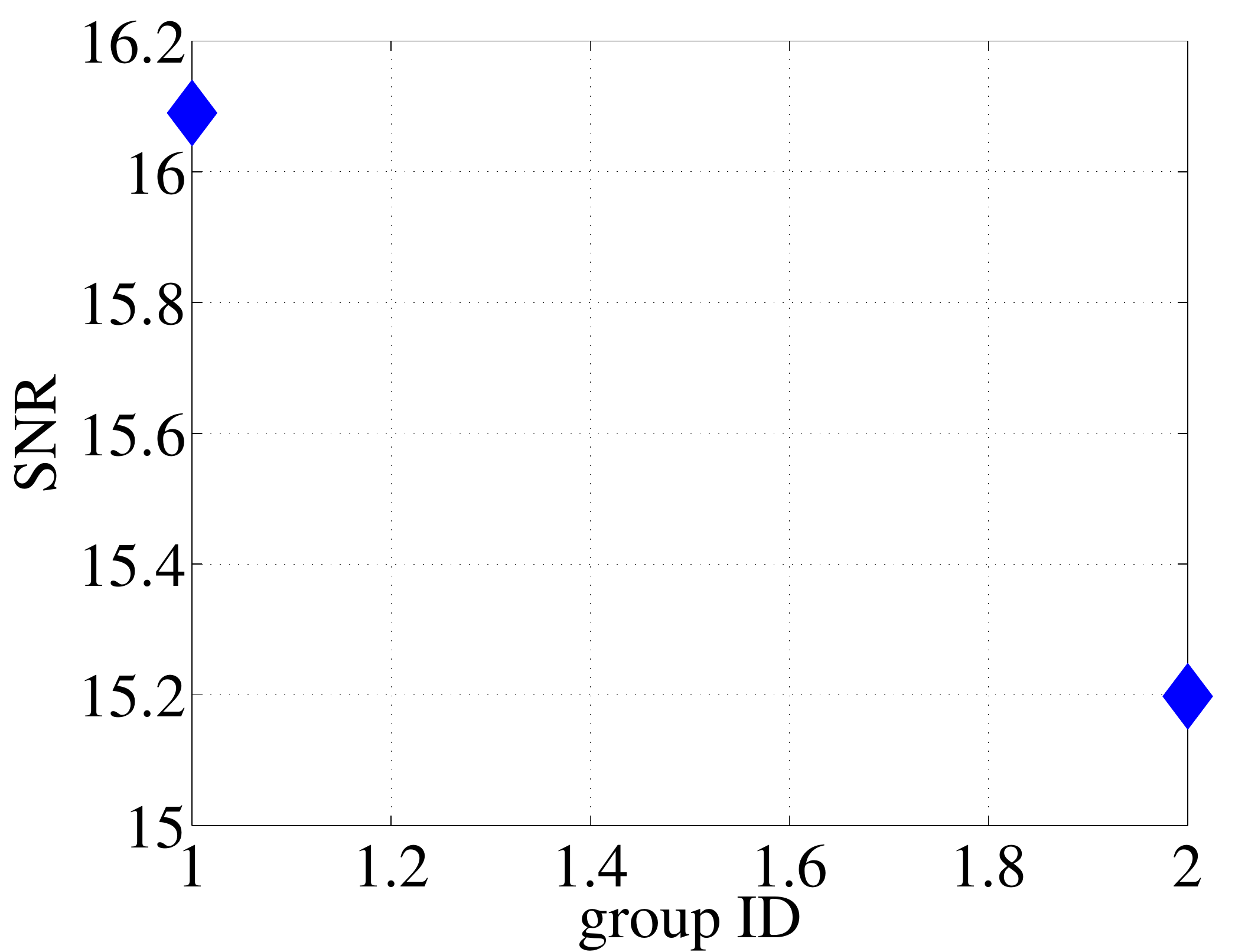}
\caption{Same as Figure \ref{f15} but for model R3 seen from pole.}
\label{f16}
\end{center}
\end{figure}

Figures \ref{f15} and \ref{f16} show the similar analysis for the waveforms 
of model R3, the most rapidly rotating model in the 
{\tt KTK14} catalogue, either seen from equator (R3e) or pole (R3p), 
respectively.

For the equatorial observer (Figure \ref{f15}), a clear excess 
 in the spectrogram is seen in the + mode 
(top left panel) at $t_{\rm sim} \sim 35$ ms (red prolate region),
 which corresponds to the epoch of rotating core bounce. The
 quasi-oscillatory period of the waveform near bounce is 1 $\sim$ 5 ms (e.g., middle 
 panel of Figure \ref{f1}), which accounts for the excess in 
 the frequency range of 200 - 1000 Hz in the spectrogram (top left panel).

This (best-studied) rotating bounce signal is hardly seen in the $\times$ mode 
for the equatorial observer (top right panel) nor for the polar observer 
(top panels of Figure \ref{f16}) with both polarizations 
(+ ({\it left}) or $\times$ ({\it right}) mode, respectively). 
Clearly seen for the polar observer is an another excess that appears 
in the spectrograms between $t_{\rm sim} = 
 40 \sim 80$ ms peaking around the frequency of $200$ Hz (top panels in Figure \ref{f16}). As elaborately
 discussed in \cite{Kuroda14}, this characteristic frequency ($\sim 200$ Hz)
 comes from the growth of one-armed spiral instabilities in the vicinity of 
 the rapidly rotating PNS (see also the periodic waveform patterns 
in the middle panel of Figure \ref{f1}). 

In the reconstructed waveforms (second columns of Figures \ref{f15} and \ref{f16}), it can be seen that the GW signatures seen in the injected signals 
 due to rotating bounce (top left panel 
in Figure \ref{f15}) and the spiral instabilities 
(top panels in Figure \ref{f16}) are still present, although the excess in the 
spectrograms becomes especially weak for the bounce signal 
  at the high frequency regime (compare the left panels in the first and 
second columns of Figure \ref{f15}).

Regarding the $+$ mode seen from the equator (left panels of Figure \ref{f15}), 
${\rm SNR}_{\rm TF}$ of the rotating bounce signal is 14.3 (second panel 
({\it left})), which is assigned to have the ID number 1 (red prolate 
 region in the third column ({\it left})). Setting again 
the detection threshold as 8 with respect to ${\rm SNR}_{\rm group}$, the 
 detection horizon extends to 17.9 kpc. Regarding the GW signature from 
 the spiral instabilities, the corresponding excess 
(${\rm SNR}_{\rm TF} = 13.8$) is assigned to have 
 the ID number 2 and 3 (yellow/orange region in the third column ({\it left})).
 The horizon distance is $\sim 17.3$ kpc. As for the $\times$ mode (right
 panels of Figure \ref{f15}), ${\rm SNR}_{\rm group}$ is 9.0 (the ID number is
 1) and 10.3 (the ID number is 2, which has biggest SNR among the 
 candidate IDs of 2, 3, and 4 (third column ({\it right})), 
the detection horizon of which is 11.3 kpc and 12.9 kpc, respectively.

Seen from pole (Figures \ref{f16}), the excess in the spectrogram 
  of the model R3 waveform (top panels) is categorized into three 
 (${\rm SNR}_{\rm TF}$ = 29.3, 31.7, and 26.7, third column ({\it left}, 
 + mode)) or two (${\rm SNR}_{\rm TF}$ = 16.1 and 15.2 (third column 
({\it right}, $\times$ mode)). Choosing the biggest ${\rm SNR}_{\rm TF}$ among
 the groups, the detection horizon of the + and $\times$ mode turns out to extend up to
 $\sim$ 40 and 20 kpc, respectively. 

Finally, Table \ref{tab1} summarizes the horizon distances of 
 all the rotating models in the {\tt KTK14} catalogue.
Regarding the rotating bounce signals (labelled as "Rotating core-bounce" in the table),
 the horizon distance becomes longest ($\sim$ 18 kpc) seen from equator 
 for the most rapidly rotating model (R3) by the H-L-V-K network considered in this work. 
 In order that the SNR exceeds 8 (to claim detection),
 the initial rotation rate should be higher
 than $\Omega_{\rm ini} = \pi/2$ (rad/s) (model R2) among the 3D-GR models. Regarding the 
 GW signals from non-axisymmetric instabilities, the horizon distances become generally
 longer when seen from pole than seen from equator. This is because the low-mode instabilities 
 characterized by the spiral arms develop most preferentially in the equatorial plane.
 The maximum horizon distance extends up to $\sim$ 40 kpc for the most rapidly rotating model (R3).
 The horizon distance does not decrease monotonically with the initial rotation rate. In fact,
 comparing the initial rotation rate of model R3 ($\Omega_{\rm ini} = \pi$(rad/s)) and that of model R2, 
the maximum horizon distance of model R2 ($\sim $36 kpc) is relatively close to that of model R3
 ($\sim$ 40 kpc).  

From Table \ref{tab1}, we speculate that the chance of detecting GWs from rapidly rotating CCSNe could become quite
 higher for the quasi-periodic signals inherent to the 
non-axisymmetric instabilities than for the short-duration signals emitted 
at rotating collapse and bounce. 
 As repeatedly mentioned before, it should be cautioned again 
that the numbers in Table \ref{tab1} are based on a very optimistic estimate 
using the idealized Gaussian detector noise, and they should be interpreted 
as an upper bound of the horizon distance.
For a more quantitative 
investigation,
one also needs a more accurate waveform prediction based on long-term 3D-GR models
 with sophisticated neutrino transport, toward which we have attempted to make the very first step
 in this study.

\begin{table}[hbt]
\caption{\label{tab1}%
Model summary of optimistic horizon distances based on the spectrogram 
 analysis. In the first column, GW emission for models from the {\tt KTK14} catalogue is 
 categorized either due to rotating core bounce (emitted within 10 ms postbounce) and 
 due to the subsequent growth of the non-axisymmetric instabilities in the vicinity of the PNS.
By setting the detection threshold as ${\rm SNR}_{\rm group} = 8$, the horizon 
 distance is given for each of the 3D-GR models for the equatorial or polar observer
 (labelled such as by R3e and R3p) with $+$ or $\times$ polarization. The blank "---" represents that 
 the ${\rm SNR}_{\rm group}$ does not exceed the threshold. }
\begin{ruledtabular}
\begin{tabular}{lcc}
\textrm{Model}&
\textrm{Rotating core-bounce}&
\textrm{Non-axisymmetric instabilities} \\
\colrule
R3e& 17.9 kpc (+), 11.3 kpc ($\times$) & 17.3 kpc (+), 12.9 kpc ($\times$)\\
R3p& --- & 39.4 kpc (+), 20.1 kpc ($\times$) \\
R2e & 14.0 kpc (+)  & 16.5 kpc (+)\\
R2p &--- & 35.9 kpc (+), 14.0 kpc ($\times$) \\
R1e &--- & 16.8 kpc (+), 7.6 kpc ($\times$)\\
R1p & --- & 5.9 kpc (+), 11.1 kpc ($\times$) \\
\end{tabular}
\end{ruledtabular}
 \end{table}

\section{Summary and Discussions}
\label{summary}

Using predictions from 3D hydrodynamics 
simulations of CCSNe, we presented a coherent network analysis 
to detection, reconstruction, and the source localization of the 
GW signals. The network considered in this work consisted
 of the LIGO Hanford, LIGO Livingston, VIRGO, and KAGRA interferometers.
 We first computed the SNR and the optimistic
detectability of the GW signals with a very idealized situation (i.e.,  
 only a single detector for an optimally oriented and optimally located 
source). Then we considered a more realistic situation using the {\tt RIDGE} coherent network analysis pipeline, in which the 
multiple detectors were
 used for an arbitrary oriented source. By combining with 
the GW spectrogram analysis, it was shown
 that several important hydrodynamics features imprinted in the 
 original waveforms persist in the waveforms of the reconstructed signals. 
The characteristic excess in the GW spectrograms originates not only from 
rotating core-collapse and bounce, the subsequent ring down of 
the PNS as previously identified, but also from the
 formation of MHD
 jets and non-axisymmetric instabilities in the vicinity of the PNS. 
 Regarding the GW signals emitted near at the rotating core bounce, the horizon distance, which we optimistically set by a SNR exceeding 8, extends up to $\sim$ 18 kpc for the most rapidly rotating
 3D model among the employed waveform libraries. Only for models with the
 precollapse angular velocity higher than $\Omega_{0} = \pi/2$ (rad/s), the SNRs of the rotating
 bounce signals exceed the fiducial detection threshold.
Following the rotating core bounce,
the dominant source of the GW emission shifts to the
 non-axisymmetric instabilities that develop in the region between the stalled shock and the 
PNS. It was pointed out 
that the horizon distances from the non-axisymmetric instabilities 
 are generally longer when seen from the direction parallel to the rotational 
axis of the source than seen from the 
 equator. This is because the spiral arms that are inherent to the low-modes instabilities develop more
 preferentially in the equatorial plane. Among the 3D general-relativistic 
models in which the non-axisymmetric 
instabilities set in, the horizon distances extend maximally up 
to $\sim$ 40 kpc seen from the pole and they are rather insensitive 
 to the imposed initial rotation rates. 
In addition to the best studied GW signals due to rotating core-collapse 
 and bounce, it was suggested that 
the quasi-periodic signals due to the non-axisymmetric 
instabilities and the detectability 
should deserve further investigation to elucidate the 
inner-working of the rapidly rotating CCSNe.

While we have shown that the spectrogram analysis is effective 
 for the GW signals from rapidly rotating collapse, which 
 is most likely to be associated with the MHD-driven mechanism, 
the ability on the stochastic waveforms from 
the neutrino mechanism remains to be tested. 
Recently it was demonstrated by
 2D-GR models with elaborate transport scheme 
\cite{Bmueller12a} that a violent
 mass accretion to the PNS leads to an efficient GW emission
 in the late postbounce phase, which can be nicely explained
 by the buoyant frequency near the PNS surface \cite{Murphy09}. Such a 
 high-frequency feature ($\sim$100 to $\sim$1 kHz) are
 generic in 2D explosion models, which is expected 
 to be also the case in 3D \cite{EMuller12}.
With a growing supercomputing power and a rapid development of
 CCSN modeling (e.g., \cite{melson,lentz15,bmueller15,nakamura15,kuroda15}), we speculate that 
the construction of dense waveform catalogues based on self-consistent 
3D-GR models is becoming a reality in the decade to come 
\cite{Kotake12_ptep,tony14}.
We hope that in combination of the refined waveform predictions 
 the GW analysis schemes will be also updated such as by taking a 
 coincidence with neutrino signals (\cite{yokozawa,tamborra14}, e.g.,
 \cite{mirizzi15,Kotake12} for a review), which should be indispensable 
 to decipher the CCSN mechanism from the multi-messengers observables
 in the NEXT nearby event.

\begin{acknowledgments}
KH would like to thank B. Allen for warm hospitality during his stay in 
Hannover and S.D. Mohanty for valuable comments and encouragement. 
We are grateful to N. Kanda and the members in his lab in Osaka-city university 
for helpful discussions. KK and TT are thankful to K. Sato and S. Yamada for 
continuing encouragements.
Numerical computations were carried out in part on 
XC30 and general common use computer system at the center for 
Computational Astrophysics, CfCA, 
the National Astronomical Observatory of Japan,
Oakleaf FX10 at Supercomputing Division in University of Tokyo, and on SR16000 at YITP in Kyoto University.
 This study was supported in part by the Grants-in-Aid for the Scientific 
Research from the Ministry of Education, Science and Culture of Japan 
(Nos. 24103006, 24244036, 26707013, and 26870823) 
and by HPCI Strategic Program of Japanese MEXT.
\end{acknowledgments}
\appendix 
\section*{Appendix A, Waveform Catalogues}
\label{catalogues}

\subsection{The KK+ waveforms}

The top panels of Figure \ref{f1} show an example waveform
 (left panel) and a snapshot (right panel) of a 3D model that
 is trending towards an explosion by the neutrino mechanism 
 \cite{kotake_ray,Kotake09,kotake11}.
In the 3D model, the neutrino luminosity ($L_{\nu}$) from 
 the PNS was treated as a parameter to trigger explosions 
and the initial conditions were derived from a steady-state 
approximation of the postshock structure and the Newtonian 
hydrodynamics outside an 
inner boundary at 50 km was solved. The core neutrino luminosity is taken as 
$L_{\nu} = 6.8 \times 10^{52}$ and $ 6.4 \times 10^{52}\,{\rm erg}/{\rm s}$
for the {\tt KK+09} (corresponding to model A in \cite{Kotake09}) and 
 the {\tt KK+11} waveform (model C2 in \cite{kotake11}, top panels in Figure \ref{f1}),
 respectively. 
In model {\tt KK+11}, stellar rotation was taken into account by 
adding a uniform rotation to the flow at the outer 
boundary of the computational domain (e.g., \cite{iwakami2}), the angular 
 momentum of which is assumed to agree with recent stellar evolution models 
\cite{hege05}.

The top right panel is a snapshot (at $t_{\rm sim} = 513$ ms) seen from the rotational
 axis for model C2. Note that the time ($t_{\rm sim}$) is measured from the epoch 
 when simulations are started. The first and third quadrants of the panel show
 the profiles of the high-entropy bubbles (colored by red) 
inside the surface of the standing shock wave 
(the second and fourth quadrants). The side length of the panel is 1000 km.
The high-entropy bubbles are seen to develop like a spiral arm, which is a signature of the spiral SASI modes. 
Under the influence of the spiral and 
sloshing SASI modes and neutrino-driven convection, the 3D model starts to 
be exploding at $t_{\rm sim} \sim 200$ ms after the stalled 
 shock comes to a steady state. In the top left panel,
 the waveforms only $t_{\rm sim} \gtrsim 200$ ms is shown because the amplitudes 
 are zero (or very small) before the non-spherical hydrodynamical instabilities
 enter to the non-linear phase 
(see also the inset of the top right panel). As shown, 
 the wave amplitudes change stochastically with time, because the
 non-sphericities in the postshock region are essentially governed by
  turbulent flows. Note that the wave amplitudes of our simplistic 3D models 
are qualitatively in agreement with those obtained in more realistic 
3D models \cite{ewald11}.

 The inset of the top right panel shows
 the waveform contributed only from anisotropic
 neutrino emission (pink line (seen from pole, $+$ mode),
green line (seen from pole, $\times$ mode), 
light blue line (seen from equator, $+$ mode), and 
red line (seen from equator, $\times$ mode, respectively). As already 
discussed in \cite{muyan97,mueller04,kotake_ray,Kotake09,kotake11,ott_rev}, the 
time variability of the neutrino(-originated) GWs is much longer ($\gtrsim O(10)$ ms) 
 due to the memory effect \cite{Braginskii87} than that of the matter GWs
 ($\lesssim O(10)$ ms (top left panel)).
As a result, the peak frequencies of the neutrino GWs are typically 
 below $\sim$ 100 Hz. These low frequency GWs are very difficult to 
 detect by ground-based detectors whose sensitivity is limited 
mainly by seismic noises. In the case of KAGRA, the sensitivity at $20$ Hz is 
$5 \times10^{-23}$, which is about $20$-times 
 worse than the most sensitive frequency domain around $80 - 200$ Hz. 
In the following, 
 we thus focus on the matter GW signals that 
are more important in discussing the detectability. 

\subsection{The KTK14 waveforms}

The middle panels of Figure \ref{f1} show the waveform (left panels) and 
 the 3D entropy plot ($49$ ms postbounce) of a rapidly rotating model 
 (R3) in \citet{Kuroda14}. For this model, an
 angular velocity of $\Omega_0 = \pi$ rad/s is added to 
 the non-rotating $15 M_{\odot}$ progenitor of \cite{WW95} 
with a quadratic cut-off parameter at the radius of $X_0 = 1000$ km.
 Shortly after bounce ($\sim 15$ ms postbounce), 
 one-armed spiral modes were observed to develop in the postshock 
region for this rapidly rotating model.
As a result, the waveforms show narrow-band and highly quasi-periodic 
 signals (regardless of the GW polarizations, see the left middle panels), 
which persist until the end of simulations ($t_{\rm sim} \sim 60$ ms). Since the typical frequency of the 
 quasi-periodic waveform can be well explained by the propagating 
 acoustic waves between the stalled shock and the rotating PNS 
surface, the waveforms are most likely to be associated with the 
 appearance of the spiral SASI (see \citet{Kuroda14} for more detail).
 Regarding the $+$ mode of the signal seen from equator
 (not shown in Figure \ref{f1}), typical GW features of the so-called 
type I waveforms
 (e.g., \cite{Dimmelmeier02}) were clearly seen \cite{Kuroda14}, i.e., 
a first positive peak just before bounce precedes the deep
  negative signal at bounce, which is followed by the
 subsequent ring-down phase. 

 In addition to the rapidly rotating model (R3), we use three waveforms 
  of more slowly rotating
 models from \cite{Kuroda14}, which 
correspond to models R0 ($\Omega_0 $ = 0 rad/s (non-rotating)), 
R1 ($\Omega_0 $ = $\pi/6$ rad/s), and R2 
($\Omega_0 $ = $\pi/2$ rad/s), respectively. 
All of the 3D models are based on full GR
hydrodynamic simulations,
 in which an approximate three-flavour neutrino transport was solved 
 with the use of an analytical closure scheme (e.g., \cite{kuroda12}).
The wave amplitudes for the non-rotating (model R0) and slow-rotating (model
 R1) stay much smaller ($\alt 3 \times 10^{-22}$) during the simulation 
time ($\lesssim 50$ ms postbounce). These GW amplitudes and frequencies 
 are consistent with 3D (post-)Newtonian \cite{Scheidegger10} 
or GR models \cite{Ott07_prl,Ott12a} with
 more idealized transport scheme and 2D GR models with more 
detailed neutrino transport \cite{BMuller13}.

\subsection{The TK11 waveforms}

To discuss GWs from models that produce MHD explosions, we use four 
 waveforms from \cite{Takiwaki11}. The authors performed 
2D special-relativistic (SR) MHD simulations
 with the use of an approximate GR potential
 \cite{ober06a}, in which a neutrino leakage scheme was employed 
 to take into account neutrino cooling \cite{taki09}. The computed 
models were named with 
 the first part, B12, representing the strength of the initial magnetic
 field parallel to the spin axis ($10^{12}$ G), the second part, X1,
 X5, or X20, indicating the degree of differential rotation
 ($X_0$ = 100, 500, 2000 km, respectively), and the third part, 
 $\beta = 0.1$ or $1$, showing the rotation parameter (the ratio 
 of the rotational energy to the absolute value of the gravitational
 energy prior to core collapse). 

The bottom left panel of Figure \ref{f1} shows that 
 the waveform from MHD explosions tends to have a quasi-monotonically 
increasing component, which is followed by the 
typical type I GW signature near at bounce 
($0.02 \lesssim t_{\rm sim} \lesssim 0.03$ s). Such feature
 was only observed for models with strong precollapse 
magnetic field and with rapid rotation initially imposed
 (e.g., model B12X1$\beta$0.1). The increasing trend comes from bipolar flows 
(bottom right panel of Figure \ref{f1}) as shown by \cite{Shibata06,Obergaulinger06}.
Again the low frequency waveform ($\lesssim 100$ Hz) are hard to detect, but it may be
 worth mentioning here that future space interferometers like Fabry-Perot type DECIGO are
 designed to be sensitive in the frequency regimes \cite{fpdecigo,kudoh}.

The GW amplitudes and frequencies of the {\tt TK11} catalogue are 
 consistent with those obtained in previous 2D \cite{Shibata06,Obergaulinger06}
 and 3D \cite{Scheidegger10} MHD models.
  We chose to take the signal predictions from 2D models in order 
to compare the wave amplitudes with those in 3D models (i.e., {\tt KTK14}
 catalogue), and also to discuss how difficult it is to detect the low 
frequency components for ground-based interferometers even by performing 
 the coherent network analysis.

\section*{Appendix B, Coherent Network Analysis}
\label{coherent}

In order to compute the signal detection, 
reconstruction, and 
source localization of the model waveforms in the last section, 
 we perform a coherent network analysis using a pipeline called {\tt RIDGE} 
(see \cite{hayama07} for details).
In the algorithm, one combines information from
 multiple GW detectors coherently to perform a maximum likelihood analysis,
 taking into account the antenna patterns, geographical locations of 
 the detectors, and the sky direction to the source
(\cite{guersel89,klimenko05,mohanty05,rakhmanov,hayama07,klimenko08,sutton10} 
and therein). The {\tt RIDGE}
 pipeline takes full advantage of the global network of 
 currently working and future interferometers 
(LIGO Hanford (H), LIGO Livingston (V), VIRGO (V), and KAGRA (K)), 
resulting in enhanced detection efficiency.
For the detailed description of the pipeline,
 see Refs. \cite{hayama07,hayama2013}.

Using the {\tt RIDGE} pipeline, we perform Monte Carlo simulations 
(see \cite{hayama07,hayama2013}
 for more details) to investigate the detectability
 of the model waveforms in section \ref{catalogues}.
For the detector noise spectrum densities
 of the four detectors (H,L), (V), and (K), 
 we use the ones in \cite{ligodesign,advv,KAGRA} and keep the locations and 
orientations the same as the real detectors. Gaussian, stationary noise 
was generated by first generating four independent realizations of white 
noise with the sampling frequency of $2048$Hz and then passing them 
through the FIR filters having transfer functions that approximately 
match the design curves.

Signals placed at a distance of $10$kpc from the earth were added to the 
simulated noise at regular intervals. The sky locations where signals
 are injected in are (longitude,latitude) is $(-180^\circ,-90^\circ)$ 
to $(180^\circ,90^\circ)$ with resolution of $(10^\circ,10^\circ)$.
 The time window for the data analysis is $100$ms, $300$ms, 1 s,
 for waveforms in the {\tt KTK14}, {\tt TK11}, and {\tt KK+09, KK+11}
 catalogues, respectively. These time windows are much larger 
than the signal durations, so the detection performance is not optimized. 
It is possible to obtain higher detection efficiency by optimizing search 
algorithm, but the optimization is beyond the scope of this paper. 
The value of the likelihood of the multiple detector data 
is calculated by changing over the possible sky 
locations $\hat{\Omega}=(\theta,\phi)$, and the maximum of the likelihoods is 
chosen. If the maximum likelihood value is above a given threshold, the chosen 
event candidate is recorded in a detection list. Since the likelihood
 values are obtained as a function of $\theta$ and $\phi$, this 
two-dimensional output, $\boldsymbol{S}(\theta,\phi)$, is called 
{\it skymap}.


\bibliography{ref}

\end{document}